\documentclass[fleqn,10pt]{wlscirep}
\usepackage[utf8]{inputenc}
\usepackage[T1]{fontenc}
\usepackage{float}
\usepackage{bm}
\usepackage{amssymb}
\usepackage{braket}
\usepackage{multirow}
\usepackage{enumitem}
\usepackage{fix-cm}
\usepackage{calligra}
\usepackage{yfonts}
\usepackage{tgothic}
\usepackage{tcolorbox}
\usepackage[table]{xcolor}
\usepackage{colortbl}
\usepackage{makecell}
\usepackage{wrapfig}
\usepackage[normalem]{ulem}
\usepackage{xcolor}
\usepackage{rotating}

\usepackage{chngcntr}

\title{Quantum Simulation of Ligand-like Molecules through Sample-based Quantum Diagonalization in Density Matrix Embedding Framework}

\author[1]{\small Ashish Kumar Patra}
\author[1]{\small Anurag K. S. V.}
\author[1]{\small Sai Shankar P.}
\author[1, 2]{\small Ruchika Bhat}
\author[3]{\small Raghavendra V.}
\author[4]{\small Rahul Maitra}
\author[, 1]{\small Jaiganesh G.\thanks{(Corresponding Author) email: jaiganesh@qclairvoyance.in, drjaiganesh15@gmail.com}}
\affil[1]{\small Qclairvoyance Quantum Labs, Secunderabad, TG 500094, India.}
\affil[2]{\small The University of Arizona, Tucson, AZ 85721, USA.}
\affil[3]{\small SRM Institute of Science and Technology, Chennai, TN 603203, India.}
\affil[4]{\small Indian Institute of Technology Bombay, Mumbai, MH 400076, India.}

\begin{abstract}
 
    The accurate treatment of electron correlation in extended molecular systems remains computationally challenging using classical electronic structure methods. Hybrid quantum-classical algorithms offer a potential route to overcome these limitations; however, their practical deployment on existing quantum computers requires strategies that both reduce problem size and mitigate hardware noise. In this work, we investigate ground-state energy calculations of ligand-like molecules using Sample-based Quantum Diagonalization (SQD) within the Density Matrix Embedding Theory (DMET) framework, focusing on low-symmetry systems with diverse bonding motifs that exhibit subsystem-dependent variations in fragment-environment entanglement. These entanglement-based variations directly influence bath orbital construction, impurity sizes, and the structure of the embedded Hamiltonians, posing nontrivial challenges for both embedding and quantum sampling. By combining DMET fragmentation with SQD-based construction of reduced configuration spaces through quantum sampling and iterative configuration recovery, we perform quantum simulations on IBM’s Eagle R3 (\texttt{ibm\_sherbrooke}) and IBM's Heron R3 (\texttt{ibm\_boston}) superconducting quantum hardware thereby, showing that the entanglement structure across embedding subsystems plays a central role in determining the efficiency and accuracy of the simulations. Despite these complexities, we show that the DMET-SQD framework yields ground-state energies in strong agreement with DMET-FCI benchmarks, achieving chemical accuracy (1 kcal/mol) across all systems studied. These results demonstrate that SQD-based quantum simulations can be robustly extended to low-symmetry, chemically realistic, industry relevant molecules, and highlight the importance of entanglement-aware embedding strategies for scalable quantum electronic structure calculations.
\end{abstract}
\begin{document}

\flushbottom
\maketitle

\thispagestyle{empty}


\section{Introduction}\label{sec: intro}

\quad In quantum chemistry, the accurate simulation of many-electron systems remains a central goal. However, the increasing complexity of molecular systems renders such endeavours computationally demanding, often necessitating varying levels of approximation~\cite{Levine2014Ch16}. Current classical computational resources are insufficient to fully capture electron correlation in larger molecular systems, such as proteins. 
More recently, distributed computing and an efficient flavor of Full Configuration Interaction (FCI) known as the Small-Tensor Product Distributed Active Space (STP-DAS) has been used to simulate the $HBrTe$ molecule leading to the diagonalization of a Hilbert space exceeding $10^{15}$ determinants~\cite{Shayit2025_HBrTe} involving 88 electrons and 100 spin orbitals in the x2c-TZVPall basis set~\cite{x2c_TZV_basis_set}. For comparison, one of the largest conventional FCI simulations carried out using classical methods involved propane (\( \mathrm{C_3H_8} \)), requiring the diagonalization of a Hilbert space spanning approximately \( 1.31 \times 10^{12} \) determinants~\cite{propane_fci_Gao2024}, even when employing a minimal STO-3G basis set~\cite{Hehre1969_GaussianSTO}.
In quantum chemistry, Hartree-Fock (HF) theory continues to serve as the foundational approximation for all subsequent methods~\cite{Fock1930}.

With the advent of quantum computers, the accurate simulation of chemically and
biologically relevant systems are expected to become increasingly tractable.
Quantum algorithms promise the capability to model and evaluate protein-ligand interactions~\cite{wiley_protein_ligand_interaction, sapt_pro_lig, Bowling2025_ProteinLigand_FragQC} and binding energetics, explore conformational
landscapes of flexible bio-molecules~\cite{barton_1956_pes, pes_scan_qchem_ml, Anurag2025_ButanePES, hao2025largescaleefficientmoleculegeometry, dmet_sqd, Kawashima2021_dmet_vqe_ion, scalable_qc_qpe_vqe}-tasks that rapidly become intractable on classical computers~\cite{schuch2009_qmahard}.This has motivated increasing interest in quantum computing, which offers alternative algorithmic frameworks capable of addressing many-electron correlations more efficiently~\cite{cao2019_quantumchemistry, aspuruguzik2005_simulated}.

Fault-tolerant Application-Scale Quantum (FASQ)~\cite{Eisert2025_QuantumAdvantage} algorithms, such as Quantum Phase Estimation (QPE)~\cite{kitaev1995abelian_qpe} and Quantum Singular Value Transformation (QSVT)~\cite{Gily_n_2019_qsvt, wang2025randomizedQSVT}, face significant challenges due to their requirement for deep quantum circuits, which remain impractical on current Noisy Intermediate-Scale Quantum~(NISQ) hardware~\cite{Preskill_2018}. To facilitate execution on near-term devices, variational approaches such as the Variational Quantum Eigensolver (VQE)~\cite{Peruzzo2014_vqe} were proposed. VQE evaluates the expectation value of the Hamiltonian directly through a hybrid classical-quantum optimization loop, iteratively improving the overlap between the parametrized quantum state and the true ground state. Recent developments~\cite{Belaloui2025_current_vqe} have produced more hardware-efficient variants, yet key limitations persist, including statistical fluctuations and physical noise inherent to NISQ devices, as well as the requirement for prohibitively large numbers of samples and extensive error mitigation strategies~\cite{Tilly2022_VQE_Review}.

Early hardware demonstrations of molecular problems using VQE, such as protein-ligand interactions, have been limited to active spaces involving around four qubits~\cite{wiley_protein_ligand_interaction}, primarily restricted to small organic systems.  
To overcome these limitations, sample-based approaches such as Sample-Based Quantum Diagonalization (SQD)~\cite{sqd_first_paper} have emerged. SQD integrates Quantum Selected Configuration Interaction (QSCI)~\cite{kanno2023qsci} with an error-mitigation scheme known as Iterative Self-Consistent Configuration Recovery (S-CoRe)~\cite{sqd_first_paper} and employs classical Selected Configuration Interaction (SCI)~\cite{sci_evangelista_1} for post-processing diagonalization. These methods have enabled the quantum sampling of molecules with up to 36 spatial orbitals, including the [4Fe-4S] complex in a 77-qubit experiment, using 6400 nodes of the \textit{Fugaku} super-computer~\cite{sqd_first_paper}. We note that SQD and VQE are complementary rather than competing paradigms: whereas VQE performs iterative variational optimization of a parametrized ansatz, SQD leverages quantum sampling to construct a compact configuration subspace for subsequent classical diagonalization, with the dominant computational burden borne by the classical post-processing step.

While quantum computers are anticipated to handle strongly correlated systems more effectively, providing a compelling rationale for their adoption, counterarguments emphasize potential limitations~\cite{critical_lims_sqd_2025}. Nonetheless, within the emerging paradigm of Quantum-Centric Super-Computing (QCSC)~\cite{sqd_first_paper}, where the samples obtained from the quantum computers are further processed in classical super-computers in a hybrid methodological framework, SQD has facilitated a series of further algorithmic advancements.

Recent developments in sample-based quantum algorithms include applications to materials simulation~\cite{Alexeev2024_QCSMaterials_sqd}, excited-state energy estimation~\cite{Barison2025_QCExcitedStates_SQD}, hybrid methods combining SQD with Krylov Subspace Diagonalization~\cite{Yoshioka2025KrylovDiagonalization} and QDrift~\cite{Campbell2019RandomCompiler}, leading to techniques such as Sample-based Krylov Quantum Diagonalization (SKQD)~\cite{Yu2025_SBKrylov}, Quantum Krylov using Unitary Decomposition (QKUD)~\cite{asthana2025quantumkrylovalgorithmusing} and SqDRIFT~\cite{piccinelli2025_sqdrift}. Additionally, methods have been developed to iteratively expand the configuration space, achieving chemical accuracy while sampling only a fraction of the symmetry space (configurations obeying number and spin symmetries)~\cite{hi_vqe_2024, Yoo2026extending_hi_vqe, Patra2026_PIGen, Patra2026mlcsg}, thereby reducing the classical computational costs as compared to SQD~\cite{sqd_first_paper}.

Classical fragmentation techniques offer a practical route to handling the exponential scaling of electronic structure methods by dividing large molecular systems into smaller subsystems. By decomposing the global wavefunction or Hamiltonian into fragment-level contributions, these approaches enable the study of chemically realistic systems without treating the entire system at once, while retaining a controlled description of inter-fragment interactions. Prominent examples include Density Matrix Embedding Theory (DMET)~\cite{wouters2016dmets_guide, Knizia2012_DMET, negre2025_new_dmet, cances_2025_analysis_dmet} and its multi-reference variants~\cite{Pham2018DMET_laura, Verma2025DMETPDFT_laura}, Divide and Conquer (DC) method~\cite{Nakai2023DC, DCStatePrep_2021_Araujo}, and the Fragment Molecular Orbital (FMO) framework~\cite{fmo_fedorov_2012}, all of which have demonstrated practical applicability across a wide range of molecular and materials problems~\cite{yamazaki2018_practical_appl}. These techniques provide a systematic route for reducing the effective dimensionality of quantum chemical calculations by localizing electronic structure in real space or orbital subspaces, while incorporating environmental effects through different mechanisms: embedding potentials in density-based approaches such as DFT embedding, density matrix matching in DMET, and effective Hamiltonians in wavefunction-based embedding frameworks~\cite{Sun_2016_quantum_embedding_theories, wouters2016dmets_guide}.

Beyond these foundational approaches, several modern fragmentation schemes aim to preserve higher-order correlation effects while maintaining computational efficiency~\cite{kowalski_cc_downfolding, bauman2025coupledclusterdownfoldingtheory,Hermes2020VLASSCF, Otten2022Localized_laura,ewf_embedding_2022,jinlong_2023_mbe, Xu2024MBE_VQE_Deflation}. Coupled-cluster downfolding methods construct effective low-dimensional Hamiltonians by integrating out inactive degrees of freedom using classical CC theory, yielding size-consistent and systematically improvable reduced models~\cite{kowalski_cc_downfolding, bauman2025coupledclusterdownfoldingtheory, kowalski2024resourceadaptivequantumflowalgorithms}. Variational Localized Active Space methods, such as LASSCF~\cite{Hermes2020VLASSCF, Otten2022Localized_laura}, explicitly partition the active space into weakly entangled local subspaces that can be optimized independently, providing a natural interface for hybrid classical-quantum workflows. Related embedding paradigms include wavefunction-based embedding methods (EWF)~\cite{ewf_embedding_2022}, Many-Body Expansion (MBE)~\cite{jinlong_2023_mbe, Xu2024MBE_VQE_Deflation}, and Dynamical Mean-Field Theory (DMFT)~\cite{ Georges1996_DMFT, blumenthal2025_dmft_intuition}, which employ different strategies for capturing non-local correlation and dynamical effects. More recently, Bootstrap Embedding (BE)~\cite{ye2020bootstrap, Liu2023Bootstrap, Hardikar2024, cho2025quemb} has been introduced as a hybrid quantum-classical embedding framework that builds upon DMET by enforcing consistency of reduced density matrices across fragment boundaries, matching the density matrices of the central region of a fragment with those of its neighboring fragments.

In parallel, several approaches have focused on adapting classical fragmentation techniques for use with quantum simulation algorithms. Notable examples include DMET-VQE implementations on superconducting and trapped-ion platforms~\cite{wiley_protein_ligand_interaction, Kawashima2021_dmet_vqe_ion, iijima2023accuratequantumchemicalcalculations}, as well as DMET-SQD~\cite{dmet_sqd} and EWF-SQD~\cite{shajan2026_ewf_protein} approaches demonstrated on IBM quantum hardware. Related efforts such as divide-and-conquer VQE (DC-VQE)~\cite{fujii_2022_deep_vqe} and fragment molecular orbital extensions to quantum solvers, including Effective FMO-VQE~\cite{FMO_VQE_Lim_2024}, further illustrate the breadth of strategies combining classical fragmentation with quantum sampling or variational techniques. Collectively, these works define a growing ecosystem of hybrid methods that aim to balance accuracy, scalability, and near-term hardware constraints in quantum-enabled electronic structure calculations.

SQD has gained significant interest in recent times~\cite{sqd_first_paper, hi_vqe_2024, Patra2026_PIGen, Yoo2026extending_hi_vqe} and is subject to further exploration to develop and assess its performance within fragmentation frameworks~\cite{dmet_sqd, shajan2026_ewf_protein, Bierman_2026_QBE_SQD}. Towards this, we assess the feasibility of SQD within the DMET framework for ligand-like molecules, extending upon the recently introduced DMET-SQD framework~\cite{dmet_sqd} for strongly correlated symmetric systems. Herein, we extend this framework to accommodate a wider class of molecules that are biologically relevant, unlike the previously studied systems~\cite{dmet_sqd}, which had high point group symmetry; these ligand-like molecules belong to low-symmetry classes (typically $C_1$). This makes the universal thresholding criteria to pick the dominant configurations over individual embedding problems a challenging task, a difficulty that provides a rigorous test of the robustness and practical applicability of the proposed approach.

The ligand-like molecules considered for the study include: Cyanic Acid($HOCN$), Formaldehyde Oxime ($CH_3NO$), O-methyl-hydroxylamine ($CH_5NO$), Methyl Isocyanate ($C_2H_3NO$), Acetaldehyde Oxime ($C_2H_5NO$),
Carbamide/Urea ($CH_4N_2O$), Nitrosyl Chloride ($NOCl$), and Hydroxythiocyanate ($HOSCN$). All of these compounds have molecular weights below 76~Da (see Fig.~\ref{fig:dmet_sqd_mols}) and together span a chemically diverse set of small, pharmacologically motivated molecular systems. Several members of this set, particularly urea derivatives and methoxyamine-related systems, have established relevance in medicinal chemistry
\cite{Caimi2017_Methoxyamine_Fludarabine, Eads2021_Temozolomide_Methoxyamine,
Ghosh2020_Urea_DrugDiscovery, Listro2022_UreaAnticancerReview}. Furthermore, due to the 
heterogeneous nature of individual embedding problems, unlike 
previous studies, the variation in the fragment-environment 
entanglement across the studied low-symmetry molecules 
manifests directly in the bath orbital construction and 
impurity sizes, posing nontrivial challenges for both 
embedding and quantum sampling. This will be further 
elaborated in the Results Section (Section~\ref{sec: results_disc}), and makes the chosen set of molecules an appropriate testbed for assessing 
DMET-SQD in the context of biologically and industrially 
relevant molecular systems. In particular, the virtual orbital occupation threshold $\varepsilon_{\mathrm{occ}}$ governing bath orbital inclusion emerges as a critical, system-dependent parameter whose selection directly determines the balance between embedding accuracy and quantum hardware feasibility, and we therefore subject it to a systematic sensitivity analysis.

Concisely detailing the work, in \textbf{Section~\ref{sec: theory}} an overview of the methods used are described briefly, in \textbf{Section~\ref{sec: methodology}} the experimental details and methodology are discussed where we emphasize that the variable entanglement structure of the embedded fragments may vary the performance of SQD due to the lack of symmetry in the molecules, in \textbf{Section~\ref{sec: results_disc}} the results obtained are discussed and finally in \textbf{Section~\ref{sec: conclusion}} we conclude the work and mention future prospects. 

\section{Theoretical Background}
\label{sec: theory}

\subsection{Hamiltonian Formulation and Local Unitary Cluster Jastrow ansatz}\label{sec: ham_ansatz}

Under the Born-Oppenheimer approximation~\cite{born_oppie_1927}, the electronic structure problem is most conveniently expressed in second quantization, where the Hamiltonian is written in terms of fermionic creation and annihilation operators acting on a basis of spin orbitals~\cite{szabo1996_modernqc}:

\begin{equation}
    \hat{H}_{sq} = \sum_{i, j} h_{ij} \hat{a}^{\dagger}_i \hat{a}_j + \frac{1}{2}\sum_{p,q,r,s} h_{pqrs} \hat{a}^{\dagger}_p \hat{a}^{\dagger}_q \hat{a}_r \hat{a}_s.
\end{equation}

This representation separates one-electron and two-electron contributions and forms the natural interface for quantum simulation, since it can be systematically mapped onto qubit operators using standard fermion-to-qubit encodings such as the Jordan-Wigner transformation~\cite{Jordan1928}. In this work, we initialize the quantum register in the restricted Hartree-Fock (HF) reference state, which is a single Slater determinant that is straightforward to prepare on quantum hardware and typically exhibits a significant overlap with the true ground state.

Starting from this reference, we construct a correlated variational wavefunction using the Local Unitary Cluster Jastrow (LUCJ) ansatz~\cite{lucj_ansatz}, which is designed to balance chemical expressivity with circuit efficiency on near-term quantum devices. The ansatz parameters are initialized using amplitudes inspired by coupled-cluster singles and doubles (CCSD) theory~\cite{bartlett_cc_theory_2007,jensen2006introduction}, providing a chemically motivated starting point for optimization. The LUCJ ansatz consists of multiple layers, each comprising an orbital-rotation operator, a diagonal Jastrow interaction that captures electron-electron correlations, and the inverse orbital rotation, thereby incorporating correlation effects in a structured and hardware-aware manner~\cite{Matsuzawa2020_JastrowLowDepth,AbuGhanem2025_heavy_hex_eagle}. The resulting variational state is given by

\begin{equation}
    \ket{\psi}_{UCJ} = \prod_{\mu=1}^{L} e^{\hat{K}_\mu} e^{i\hat{J}_{\mu}} e^{-\hat{K}_\mu} \ket{\psi}_{HF}.
\end{equation}

Here, $\ket{\psi}_{HF}$ denotes the Hartree-Fock reference state and $L$ is the total number of LUCJ layers. The operators $\hat{K}_\mu$ and $\hat{J}_\mu$ respectively generate orbital rotations and two-body correlation effects within each layer, enabling the ansatz to systematically incorporate electronic correlation beyond the mean-field level while maintaining a circuit structure amenable to near-term quantum hardware. For closed-shell systems, a spin-balanced form of the ansatz is employed to enforce spin symmetry and reduce the number of independent variational parameters, improving optimization stability without sacrificing expressivity.

\subsection{Density Matrix Embedding Theory}\label{sec: dmet}

Density Matrix Embedding Theory (DMET)~\cite{Knizia2012_DMET} is a frequency-independent quantum embedding method conceptually related to Dynamical Mean-Field Theory (DMFT)~\cite{Georges1996_DMFT}, in which the environment of a chosen fragment is represented through a set of entangled orbitals, referred to as bath orbitals. In DMET, a set of user-defined fragment orbitals $A_y$ is selected from a localized orbital basis, and the remaining orbital space constitutes the environment $Env_y$. This environment is decomposed into entangled bath orbitals and unentangled core and virtual orbitals according to
\begin{equation}
Env_y = B_y \cup Cor_y \cup Vir_y,
\end{equation}
where $B_y$ denotes the bath orbitals capturing fragment-environment entanglement, while $Cor_y$ and $Vir_y$ correspond to unentangled occupied and virtual orbitals, respectively.

\begin{figure}[htbp]
    \centering
    \includegraphics[width=0.9\linewidth]{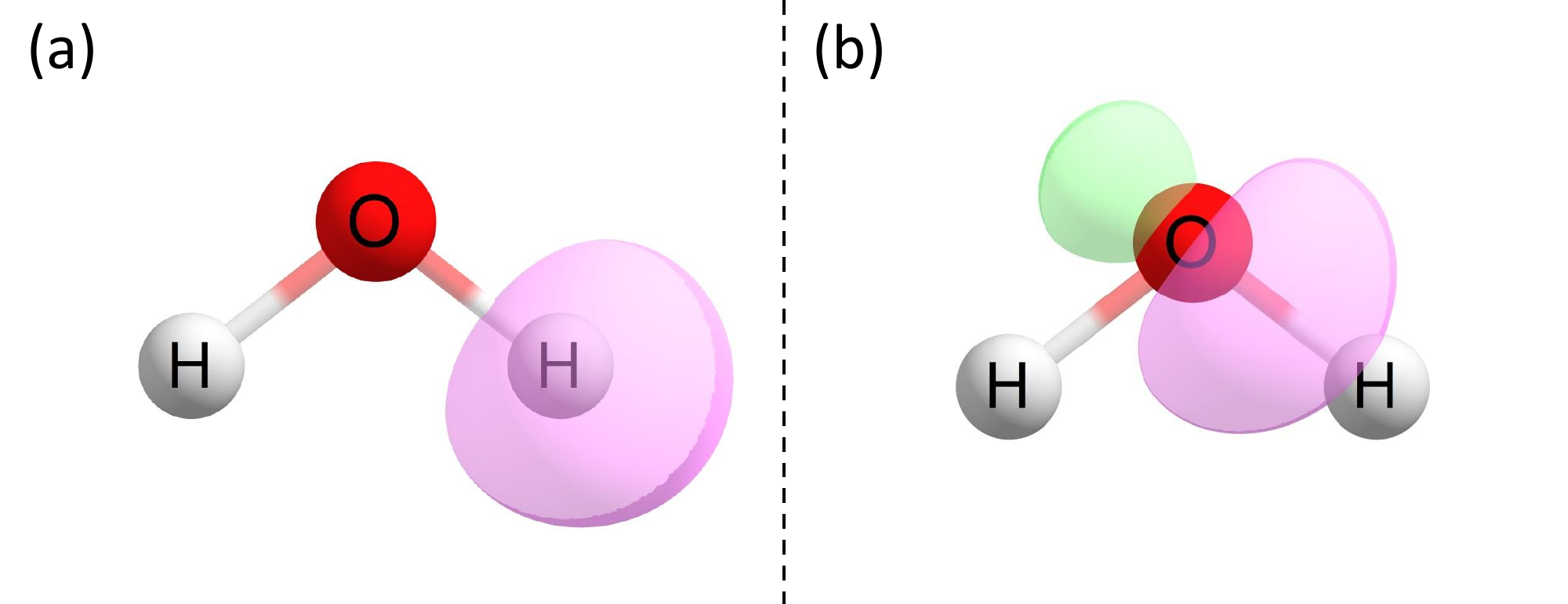}
    \caption{Illustrative example of bath-orbital construction for the $H_2O$ molecule in the STO-3G basis. The system is fragmented so that the orbital localized on a single hydrogen atom constitutes the fragment. \textbf{(a)} shows an iso-surface of this localized fragment orbital; \textbf{(b)} shows the corresponding bath orbital obtained via DMET, which is a linear combination of the remaining localized spatial orbitals in the molecular orbital basis. Together, \textbf{(a)} and \textbf{(b) }span the impurity subspace for the H-fragment.}
    \label{fig:water_bath}
\end{figure}

The bath orbitals are constructed from a mean-field reference (typically Hartree-Fock) by projecting the one-particle reduced density matrix (1-RDM) onto the fragment subspace and diagonalizing the resulting matrix. The eigenvectors with fractional occupations define the bath orbitals, and their number is bounded by the number of fragment orbitals~\cite{horn2012matrix,MacDonald1933_RayleighRitz}. An illustrative example of bath construction for a hydrogen fragment in a water molecule is shown in Fig.~\ref{fig:water_bath}. The fragment and bath orbitals together form the impurity space on which an embedded Hamiltonian is solved to obtain the correlated wavefunction. A global chemical potential $\mu_{glob}$ is adjusted to ensure electron number consistency across all fragments, corresponding to a one-shot interacting bath formulation~\cite{wouters2016dmets_guide}.

The fragment energy $E_{A_y}$ is given by
\begin{equation}\label{eqn: e_frag}
E_{A_y} \approx \sum_{p\in A_y} \left( 
\sum_{q}^{L_{A_y}+L_{B_y}} 
\left( \frac{t_{pq} + \overline{h}_{pq}^y}{2} \right) D_{qp}^y  
+ \frac{1}{2} \sum_{qrs}^{L_{A_y}+L_{B_y}} (pq|rs) P_{qp|sr}^y 
\right),
\end{equation}
where $y$ indexes the fragments in the partitioning of the molecular 
system, and $E_{A_y}$ denotes the energy contribution associated with fragment $A_y$. The indices $p,q,r,s$ label spatial orbitals belonging to the impurity space, defined as the union of fragment and bath orbitals with total dimension $L_{A_y}+L_{B_y}$. The matrix elements $t_{pq}$ denote the one-electron integrals in the molecular orbital basis, while $\overline{h}_{pq}^y$ are effective one-body terms incorporating the mean-field contribution of the core orbitals. The quantities $D_{qp}^y = \langle \hat a_p^\dagger \hat a_q \rangle$ and $P_{qp|sr}^y = \langle \hat a_p^\dagger \hat a_r^\dagger \hat a_s \hat a_q \rangle$ denote the fragment one- and two-particle reduced density matrices (1-RDM and 2-RDM), respectively. The two-electron integrals $(pq|rs)$ are Coulomb integrals in the molecular orbital basis. The prefactors of $1/2$ avoid double counting when summing over all orbital indices, and the total energy of the system is obtained by summing the contributions from all fragments.

\subsection{Sample-based Quantum Diagonalization} \label{sec: qsampling}

SQD~\cite{sqd_first_paper} is a combination of the quantum sampling method QSCI~\cite{kanno2023qsci} and an error-mitigation technique introduced in the work~\cite{sqd_first_paper}, S-CoRe, which ``recovers'' the noisy configurations. These configurations thus obtained create a sub-space, which is further expanded, and then the projection sub-space is obtained over which SCI~\cite{sci_evangelista_1} is performed to obtain the ground state energy.

\subsubsection{Quantum State Preparation and Sampling}

\begin{figure}[htbp]
    \centering
    \includegraphics[width=0.7\linewidth]{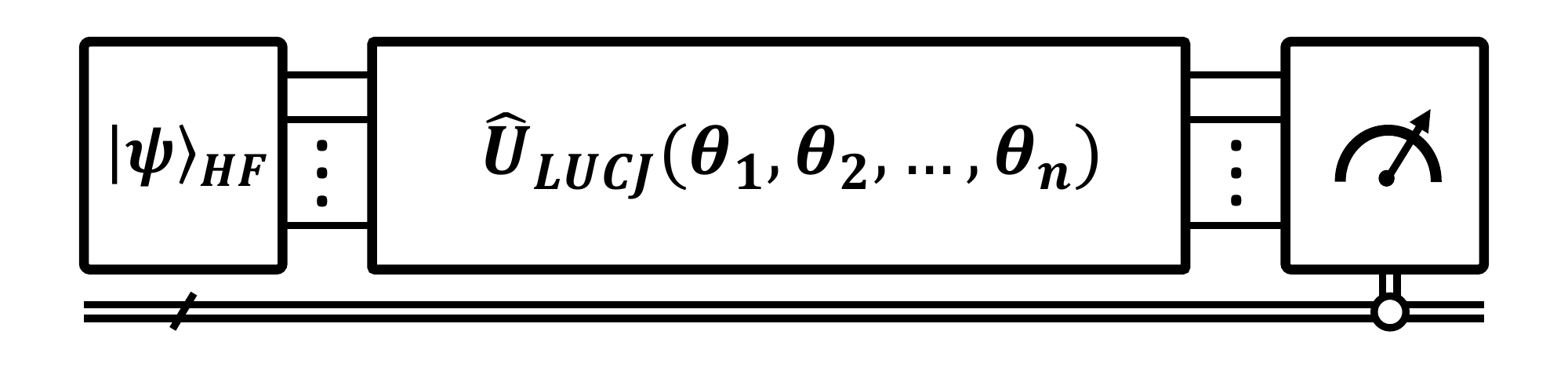}
    \caption{Quantum Circuit Diagram for performing Sample-based Quantum Diagonalization. The initial state is prepared as the Hartree-Fock configuration, and is acted upon by an LUCJ ansatz which is pre-initialized with $t_i^a$ and $t_{ij}^{ab}$ parameters obtained through CCSD calculation. Finally, measurement is performed in the $\hat{\sigma}_z$ basis.}
    \label{fig: qsampling_cartoon}
\end{figure}

We begin with the goal of approximating the ground state energy of a many-body Hamiltonian by constructing and (classically) diagonalizing a reduced Hilbert space that contains the \textit{dominant configurations} contributing to the ground state configuration, \textit{sampled using a quantum computer}. Starting from an approximate ground state, like $\ket{\psi}_{HF}$, a variational ansatz is initialized with suitable parameters: $    \ket{\psi}_{init} = \hat{U}(\theta_1, \theta_2,...,\theta_n) \ket{\psi}_{HF}$ (in our case, the LUCJ ansatz with CCSD parameters) to increase the overlap between the trial state with the true ground state, as provided in Fig.~\ref{fig: qsampling_cartoon}.

Quantum sampling is performed in the computational basis, and a set $\mathcal{S}_{samp}$ is obtained with \textit{d} configurations:
\begin{equation}
    \mathcal{S}_{samp} = \{ \ket{\phi_1}, \ket{\phi_2},...,\ket{\phi_d} \}
\end{equation}
 where every $\ket{\phi_i}$ can be represented as a tensor product of the $\alpha$ spin and $\beta$ spin configurations: $\ket{\phi_i} = \ket{\alpha_i} \otimes \ket{\beta_i}$.
The total number of electrons in the molecular systems is designated as $N_{elec}$, and correspondingly the number of electrons in $\alpha$ configuration and $\beta$ configurations are designated as $N_{\alpha}$ and $N_{\beta}$ respectively. For molecular systems which contain fully-filled spatial orbitals (closed-shell) and which possess zero spin, the following conditions must be satisfied for all the computational basis states contributing to the eigen-states of the system:
\begin{equation}\label{eqn: number_spin_symm}
    N_{\alpha} + N_{\beta} = N_{elec};\quad N_{\alpha} - N_{\beta} = 0.
\end{equation}

Due to the noisy nature of quantum computers, we end up also sampling configurations which violate the particle-number and spin-$z$ symmetries mentioned in Eq.(\ref{eqn: number_spin_symm}). These ``noisy'' configurations are recovered using configuration recovery~\cite{sqd_first_paper}.

\subsubsection{Self-Consistent Iterative Configuration Recovery} \label{sec: score}

S-CoRe, or Configuration Recovery, is an iterative procedure designed to restore the sampled noisy configurations. Given an approximate ground-state occupation number distribution for the state-vector $\ket{\Psi}$:
\begin{equation}
\begin{split}
n_{p\sigma} = \langle \Psi | \hat{a}^\dagger_{p\sigma}\hat{a}_{p\sigma} | \Psi \rangle,
\end{split}
\end{equation}
where $n_{p\sigma}$ is the occupancy corresponding to the $p^{th}$ spatial orbital with $\sigma$ spin configuration, S-CoRe updates each sampled configuration, over multiple batches (the $b$-th batch indexed by $b \in \{1,\ldots,K\}$, where $K$ is 
the total number of batches) by probabilistically flipping occupations of the spin-orbitals based on the values of $n_{p\sigma}$ until the total electron number and spin projection match the target values. 
This produces a set of corrected configuration sub-spaces $\mathcal{S}_{CR}^{(b)}$, ${\forall\,b\in \{1,\ldots,K\}}$ that respect the intended symmetries while remaining close to the original sampled distribution.

For projection and diagonalization, the sub-spaces $\mathcal{S}_{CR}^{(b)}$ are converted to $\mathcal{S}_{proj}^{(b)}$ as detailed in Section \ref{sec: sci}. With the $E^{(b)}$ and $\ket{\Psi^{(b)}}$ obtained post the classical diagonalization for all the K batches, the lowest energy $min_b E^{(b)}$ is taken as the current best estimate, and the average orbital occupation is thus calculated as:
\begin{equation}\label{eqn: avg_occ_cr_2}
n_{p\sigma} = \frac{1}{K} \sum_{b=1}^K \langle\psi^{(b)}| \hat{a}^\dagger_{p\sigma}\hat{a}_{p\sigma}|\psi^{(b)}\rangle,
\end{equation}

which is again used in the next iteration and so on. This self-consistent iteration ensures that the symmetry constraints are enforced, and the recovered sub-space contains configurations closer to the desired orbital occupation distribution. From these recovered configurations at each iteration, K batches of subspaces $\{\mathcal{S}_{CR}^{(b)}\}_{b \in K}$ are obtained.

\subsubsection{Selected Configuration Interaction} \label{sec: sci}

For the set of configurations obtained in the $b^{th}$ batch, the sub-space is designated as $\mathcal{S}_{CR}^{(b)}$. As the subsequent steps remain the same for all the batches, we shall drop the super-script and denote the sub-space as $\mathcal{S}_{CR}$ without loss of generality. 

In SQD, the configurations $\mathcal{S}_{CR} = \{\ket{\alpha_1}\otimes\ket{\beta_1}, \ket{\alpha_2}\otimes \ket{\beta_2},...,\ket{\alpha_d}\otimes\ket{\beta_d} \}$ do not represent the sub-space onto which the Hamiltonian is projected. Instead in the case of \textbf{closed-shell systems} all the unique alpha configurations $\mathcal{A}_{unique} = \{\ket{\alpha_1}, \ket{\alpha_2}, ...,\ket{\alpha_d}\}$ and unique beta configurations $\mathcal{B}_{unique} = \{\ket{\beta_1}, \ket{\beta_2},...,\ket{\beta_d}\}$ are again made into a final set:
\begin{equation}\label{eqn: ab_unique}
    \mathcal{P}_{unique} = \mathcal{A}_{unique} \cup \mathcal{B}_{unique}
\end{equation}
Now, a new configuration space is obtained from the tensor product of $\mathcal{P}_{unique}$ with itself, which we define as $\mathcal{S}_{proj}$:
\begin{equation}\label{eqn: s_proj_creation}
\mathcal{P}_{unique} \otimes \mathcal{P}_{unique}
    = \mathcal{S}_{proj} = \{\, \ket{\lambda}\otimes\ket{\phi} : \forall \,\,\ket{\lambda} \in \mathcal{P}_{unique},\ \forall \,\,\ket{\phi} \in \mathcal{P}_{unique} \,\}.
\end{equation}

\begin{tcolorbox}[
  width=\textwidth,
  colback=gray!10,
  colframe=black,
  halign=center,
  top=8pt, bottom=8pt,
  before skip=1em, after skip=1em,
  sharp corners
]

Hence, $\mathcal{S}_{proj}$ is the\textbf{ sub-space onto which the Hamiltonian is projected onto, and not the} $\mathcal{S}_{CR}$ obtained after configuration recovery. $\mathcal{S}_{proj}$ is subsequently diagonalized using Davidson's Algorithm~\cite{Davidson1975_iterative_eigensolver}. The construction in which the unique configurations from $\mathcal{A}_{unique}$ and $\mathcal{B}_{unique}$ are merged into $\mathcal{P}_{unique}$, followed by forming the tensor-product space $\mathcal{P}_{unique} \otimes \mathcal{P}_{unique}$, is motivated by the observation that the ground-state wavefunction  generally exhibits correlated contributions from multiple combinations of spin-resolved configurations.In particular, if a configuration of the form $\ket{\alpha}\otimes\ket{\beta}$ carries a non-zero amplitude in the ground state, then configurations formed from the same $\alpha$ and $\beta$ sectors in different pairings may also contribute, as observed in correlated wavefunction structures~\cite{hi_vqe_2024}. This construction is also consistent with the original SQD formulation, where the subspace is generated by combining unique spin configurations to form all possible pairs~\cite{sqd_first_paper}.

\end{tcolorbox}

The Hamiltonian is projected (for every $b^{th}$ sub-space) into the sub-space $\mathcal{S}_{proj}$,
\begin{equation}
\hat{H}_{\mathcal{S}_{proj}} = \hat{P}_{\mathcal{S}_{proj}} \hat{H} \hat{P}_{\mathcal{S}_{proj}}, \qquad
\hat{P}_{\mathcal{S}_{proj}} = \sum_{\mathbf{x}\in \mathcal{S}_{proj}} |\mathbf{x}\rangle\langle \mathbf{x}|,
\end{equation}

and the solution is obtained by performing diagonalization using Davidson's method~\cite{Davidson1975_iterative_eigensolver}, finally providing us $E^{(b)}$ and $\ket{\Psi^{(b)}}$, which are again utilized in Eq.~(\ref{eqn: avg_occ_cr_2}) to obtain the average orbital occupancies.

Because the diagonalization cost scales polynomially in \(|\mathcal{S}_{proj}|\), the method remains tractable as long as the number of selected configurations does not grow exponentially with system size~\cite{Davidson1975_iterative_eigensolver}.

\section{Methodology}\label{sec: methodology}
This section describes the DMET-SQD framework, integrating fragmentation, impurity solving, and self-consistent quantum procedures via global chemical potential.
Section~\ref{sec: dmet_sqd_workflow} details the workflow; Section~\ref{sec: exp_details} covers hardware, software, and settings.
\subsection{DMET-SQD Workflow}\label{sec: dmet_sqd_workflow}
The workflow\cite{Kawashima2021_dmet_vqe_ion} used to perform DMET-SQD is as follows:
\begin{enumerate}[itemsep=0pt, topsep=6pt, parsep=2pt, partopsep=3pt]
    \item \textbf{Full System Low-level Reference:} The molecular co-efficient matrix $C$, and the HF determinant (in the case of single-reference systems) of the whole system are obtained using Restricted HF method (here we only encounter molecular systems with fully filled spatial orbitals). 
    \item \textbf{Orbital Localization and Fragmentation: } Then, the molecular spatial orbitals are localized (using Meta-L$\ddot{o}$wdin localization\cite{sun2017pyscf}) and based on the user-defined criteria (in this case, one atom per fragment), the orbitals are fragmented.

    \item \textbf{Bath Orbital Construction:} The bath orbitals are constructed with the remaining environment orbitals for the fragment $A_y$. Here, the occupation threshold $\epsilon_{occ}$ is set to $10^{-13}$ for discarding virtual orbitals ~\cite{wouters2016dmets_guide}.
    \item \textbf{Impurity Fock Matrix Updation:} The Fock matrix $\hat{F}_{imp}$ (where $imp=A_y + B_y$) for each of the impurities is obtained, and the chemical potential $\mu_{glob}$ is subtracted from the diagonal entries corresponding to the fragment orbitals: 
    \begin{equation}\label{eqn: fock_update}
        \hat{F}_{imp}^{new}(\mu_{glob}) = \hat{F}_{imp} - \mu_{glob} \hat{P}_{frag}, \quad \hat{P}_{frag} =   \begin{cases}
    1 & \text{if $i = j$ and $i \in A_y $} \\
    0 & \text{otherwise}
  \end{cases} 
    \end{equation}
    \item \textbf{Impurity SCF Updation:} Using $\hat{F}_{imp}^{\mathrm{new}}(\mu_{\mathrm{glob}})$ from Eq.~\eqref{eqn: fock_update}, an SCF calculation is carried out in the impurity space to obtain updated molecular properties, such as the impurity coefficient matrix $C_{imp}^{\mathrm{new}}(\mu_{\mathrm{glob}})$ and the corresponding one-particle density matrix $D^{y}(\mu_{\mathrm{glob}})$. The standard convention is to initialize $\mu_{glob}^{init} = 0$.
    \item \textbf{High-level Quantum Solver (SQD):} Now, we use SQD, \textit{as the high-level solver} to estimate the ground state energy and eigen-vector of these "impurities". The one-electron ($h_{imp}^{pq}(\mu_{glob})$) and two-electron ($h_{imp}^{pqrs}(\mu_{glob})$) Hamiltonian terms (where $pqrs \in imp$) obtained after the localization, serve as the $\hat{H}_{emb}(\mu_{glob})$, and the configurations obtained through quantum sampling (and subsequent Configuration Recovery) are used to form the subspace onto which $\hat{H}_{emb}$ is projected. \textit{In the case of calculating the reference energies, this step would be replaced by FCI, without any loss of generalization.}
    \item \textbf{Fragment Energy and RDM Extraction:} Using the eigen-vector obtained in the previous step, the following quantities are calculated: 
    \begin{itemize}[itemsep=4pt, topsep=2pt, parsep=0pt, partopsep=0pt]
        \item fragment energy $E_{A_y}(\mu_{glob})$ using Eq. (\ref{eqn: e_frag}),
        \item number of electrons in the fragment~\cite{Kawashima2021_dmet_vqe_ion} $N_{A_y}^{ele}$, where $N_{A_y}^{ele}(\mu_{glob}) = \sum_{p\in A_y} D_{pp}^{y}(\mu_{glob})$
        \item 1-RDM $D_{pq}^y(\mu_{glob})$ and
        \item 2-RDM $P_{qp|sr}^y(\mu_{glob})$ of the impurities, where $pqrs \in imp$.
    \end{itemize}
    \item \textbf{Aggregation of Fragment Contributions:} Now similar to the fragment $A_y$, the steps 3 to 7 are repeated for all the remaining $N_{frag} - 1$ fragments, where $N_{frag}$ represents the number of fragments in the system. The total energy $E_{tot}(\mu_{glob})$, the number of electrons of the system $N_{tot}$, and the error in electron number $N^{ele}_{err}$ can be thus obtained by:
    \begin{equation} \label{eqn: dmet_workflow_N_ele}
    E_{tot}(\mu_{glob}) = E_{nuc} + \sum_y^{N_{frag}} E_{A_y}(\mu_{glob}),\quad
        N_{tot}(\mu_{glob}) = \sum_{y}^{N_{frag}} N_{A_y}^{ele}(\mu_{glob}),\quad N_{err}^{ele}(\mu_{glob}) = N_{tot}(\mu_{glob}) - N_{true} 
    \end{equation}
    where $E_{nuc}$ is the nuclear repulsion energy of the total system, and $N_{true}$ is the exact number of electrons in the system (which we know a priori).
    \item \textbf{Chemical Potential Optimization:} Next, the root of the function $N_{err}^{ele}(\mu_{glob})$ is computed to within the convergence threshold $\epsilon_{conv}$, by iteratively updating $\mu_{glob}$, and repeating the steps 3 to 8, using a root finding method (in this case we use Newton-Secant method~\cite{papakonstantinou2013secant}). In other words, we wish to obtain a $\mu_{glob}^{\star}$ such that:
\begin{equation}\label{eqn: chem_pot_condition}
\left| N_{tot}(\mu_{glob}^{\star}) - N_{true} \right| < \epsilon_{conv}.
\end{equation}
\end{enumerate}

\begin{figure}[htbp]
    \centering
    \includegraphics[width=\linewidth]{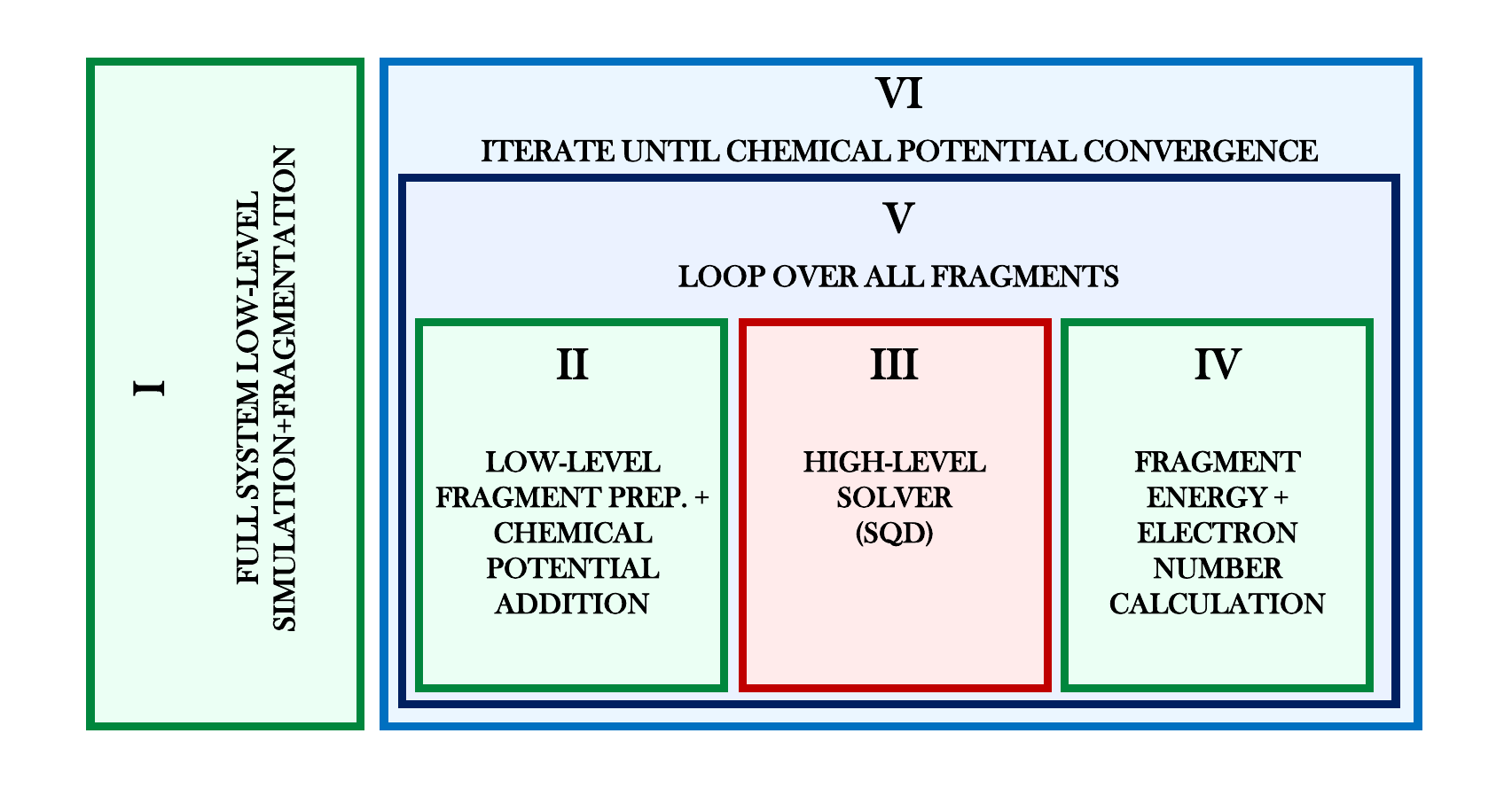}
    \caption{A simplified workflow for DMET-SQD}
    \label{fig:dmet_workflow}
\end{figure}

The nine workflow steps described above are grouped into six broader conceptual categories, designated as \textbf{Parts~I-VI}, to provide a higher-level organizational structure for the DMET-SQD algorithm, as illustrated in Fig.~\ref{fig:dmet_workflow}. \textbf{Step 1}, \textbf{Step 2} can be considered to fall under a common section, designated as \textbf{Part I}, which deals with initial low-level solving of the entire system to obtain certain pre-requisites, \textbf{Steps 3, 4} and \textbf{5}, deal with the creation of bath orbitals, and the inclusion of $\mu_{glob}$ into the hamiltonians of the impurities, which can be designated as \textbf{Part II}. \textbf{Step 6} falls under \textbf{Part III}, where the previously obtained Hamiltonians are solved using a high-level solver, such as SQD. \textbf{Step 7} involves obtaining the relevant information from the state-vector obtained from high-level solving. This crucial step allows us to perform further iterations, and can be designated as \textbf{Part IV}. \textbf{Step 8} and \textbf{Step 9} involve looping over the fragments to obtain the total energy and the number of electrons in the system, and subsequently tweaking $\mu_{glob}$ to satisfy Eq.~(\ref{eqn: chem_pot_condition}), respectively. These steps are designated as \textbf{Part V} and \textbf{Part VI}.

\subsection{Experimental Details}\label{sec: exp_details}

The sampling part of the quantum simulation experiments in this work was executed on the Eagle R3 quantum processor (IBM Sherbrooke, containing 127 qubits)~\cite{AbuGhanem2025_heavy_hex_eagle}. The corresponding hardware calibration metrics recorded immediately prior to the computation are provided in Section~\ref{sec: sherbrooke_calib} of the Supplementary Material. These values represent the initial calibration state of the device and should not be interpreted as a faithful reflection of the noise characteristics during execution. This distinction is important, as the full set of molecular simulations required approximately two days to complete, largely due to queueing delays between successive quantum-sampling jobs, during which hardware performance may drift inevitably due to queue wait times on shared quantum hardware, with drift bounded by the $\sim$24-hour automated recalibration cycle as noted in Section~\ref{sec: sherbrooke_calib} of the Supplementary Material.

For DMET-SQD and DMET-FCI, \verb|Tangelo v0.4.3|\cite{senicourt2022tangeloopensourcepythonpackage} was used.  LUCJ circuits were generated using \verb|ffsim v0.10.0|\cite{ffsim}, and quantum circuits were constructed and transpiled with \verb|Qiskit v1.4.2|\cite{javadiabhari2024quantumcomputingqiskit} and \verb|Qiskit IBM Runtime v0.36.1|. All quantum-chemistry tasks (HF, CCSD, and SCI) were carried out using \verb|PySCF v2.10.0|\cite{sun2017pyscf}. The root-finding method used was the Newton–Secant method from \verb|scipy v1.15.3|~\cite{2020SciPy-NMeth}.

The molecules considered in this study are listed in Section~\ref{sec: tab:nat_mols_energy} of the Supplementary Material. The STO–3G basis set was employed to enable controlled benchmarking of the DMET-SQD framework against DMET-FCI, allowing for direct and systematic comparison in a regime where exact classical reference solutions remain tractable. In particular, the use of a minimal basis allows us to avoid additional approximations such as active space truncation in the DMET-FCI calculations, thereby ensuring a consistent and unbiased reference. Furthermore, the reduced Hilbert space complexity makes this setting well suited for studying the behavior of the embedding and sampling procedures under NISQ-era hardware constraints. Fragmentation was performed such that each atom is treated as an individual fragment for all molecules.

While ROHF-based DMET formulations for open-shell systems are well-established~\cite{Ai_2022_DMET_ROHF_SIM, Mitra_2021_DMET_ROHF_Defects, Ai_2025_DMET_ROHF_Lanthanide}, the ligand-like molecules considered in this study are closed-shell systems, and a restricted Hartree-Fock reference is therefore employed for constructing the DMET bath. The fragmentation scheme adopted is one-atom-per-fragment as a stringent test of bath orbital construction, including chemically nontrivial cases where covalent bonds, such as double bonds, are cut across fragments. This aggressive partitioning may introduce significant errors, but those errors will be identical in both DMET-SQD and DMET-FCI and can hence be safely ignored, while simultaneously keeping the sample space as small as possible to reduce shot budget for SQD sampling and also to reduce the classical post-processing overhead. We note that DMET is not rigorously size-consistent in the formal sense; however, since both DMET-SQD and DMET-FCI are evaluated under identical embedding assumptions, this limitation does not affect the conclusions drawn from their comparison.

Since the SQD solver relies on SCI for diagonalization, it is formally neither size-consistent nor size-extensive, and its accuracy depends explicitly on the configurations retained. However, the hardware-driven sampling and subsequent configuration recovery ensure that the most dominant determinants across the full Hilbert space are naturally picked, and this dominant subspace can be systematically enlarged to improve the accuracy. This ensures that despite the formal lack of size-consistency, the numerical accuracy is maintained even for larger basis sets and more extended systems within the scope of the approach considered here. 

For benchmarking, FCI is employed as it accesses the full configuration space and is both size-consistent and size-extensive. This ensures that any deviation in correlation energy arises solely from the embedding and sampling procedure, thereby providing a rigorous validation of the ability of DMET-SQD to reproduce DMET-FCI correlation within the scope of the present study.

The number of shots used for quantum sampling was $10^4$ and was kept consistent across all the impurities, regardless of the number of spin orbitals within them. For each impurity, 5 iterations of S-CoRe were performed. The maximum allowed samples per batch was set to $10^5$, which is the quantum sub-sampling threshold that is applied to the configurations obtained post configuration recovery. The circuits were transpiled~\cite{Li2019SABRE} with optimization set to level 3, and no additional error handling techniques were employed in these series of experiments. The convergence threshold $\epsilon_{conv}$ for $\mu_{glob}$ was set to $1.48\times10^{-8}$, which is the default parameter in \verb|scipy.optimize.newton| method. The default virtual orbital threshold $\epsilon_{occ}$ was $10^{-13}$ for bath orbital construction. This threshold determines the inclusion of the non-negligibly entangled environment orbitals into the impurity space, thereby directly controlling the impurity size. Consequently, it influences both the amount of correlation captured and the quantum resource requirements of the corresponding SQD simulations.

\subsection{Bath-Orbital Threshold Sensitivity Analysis}

The construction of the DMET bath orbitals depends explicitly on the occupation threshold $\varepsilon_{\mathrm{occ}}$, which determines whether a given environment orbital is classified as entangled and therefore included in the impurity space. Since the size of the resulting impurity Hamiltonian directly affects both the classical and quantum computational costs, it is important to assess the sensitivity of the embedding procedure to the choice of $\varepsilon_{\mathrm{occ}}$.

To investigate this dependence, additional calculations were performed using a range of
occupation thresholds,
\begin{equation}
\varepsilon_{\mathrm{occ}} \in
\left\{
10^{-5},
10^{-7},
10^{-9},
10^{-11},
10^{-13},
10^{-15}
\right\},
\end{equation}
for representative embedding problem, particularly for the $HOSCN$ molecule, in the Heron R3 quantum processor (IBM Boston), the hardware calibration details of which are provided in Section~\ref{sec: boston_calib} of the Supplementary Material. For each threshold, the corresponding bath-orbital dimensions, impurity sizes, quantum resources, and the resulting energetics were analyzed. Particular attention was paid to the variation in fragment-environment entanglement across the studied low-symmetry molecules, since weakly entangled orbitals near the threshold can lead to significant changes in the impurity dimension.

The threshold sensitivity analysis was primarily performed to quantify the relationship between fragment-environment entanglement and impurity construction, and to determine whether the inclusion of weakly entangled bath orbitals substantially affects the resulting embedded Hamiltonians. Unless otherwise stated, all production DMET-SQD and DMET-FCI calculations reported in this work employ the value $\varepsilon_{\mathrm{occ}} = 10^{-13}$,
which was chosen to ensure that all non-negligibly entangled environment orbitals are
retained in the impurity space while maintaining a tractable computational cost.

\section{Results}\label{sec: results_disc}

\begingroup
\begin{table}[htbp]
\centering
\caption{Quantum resource estimation summary for the molecular fragments. Beginning from the left: Chemical formula of the molecule, Fragment, number of spatial orbitals and number of electrons represented as (o, e), number of qubits used, depth of the quantum circuit, number of $R_z$, $\sqrt{X}(S_x)$, $X$, $ECR$ gates used respectively for the specific instance of execution, dimension of the Symmetry Space(|S.S|), dimension of the Hilbert Space(|H.S|), the QPU time taken for performing DMET-SQD, and the number of iterations for $\mu_{glob}$ convergence ($N_{it}$).}
\label{tab1:resource_dmet_sqd}
\begin{tabular}{|c|c|c|c|c|c|c|c|c|c|c|c|c|}
\hline
Molecule & Frag & (o, e) & Qubits & Depth & $R_z$ & $S_x$ & $X$ & $ECR$ & |S.S| & |H.S| & T(s) & $N_{it}$ \\
\hline
\multirow{4}{*}{$HCNO$}
 & [H] & (2, 2) & 4 & 34 & 46 & 38 & 0 & 12 & 4 & 16 & \multirow{4}{*}{68} & \multirow{4}{*}{4} \\
 & [C] & (10, 10) & 20 & 389 & 1939 & 1524 & 108 & 584 & 63{,}504 & 1{,}048{,}576 &  &  \\
 & [N] & (10, 10) & 20 & 365 & 1833 & 1548 & 88 & 550 & 63{,}504 & 1{,}048{,}576 &  &  \\
 & [O] & (10, 10) & 20 & 343 & 1804 & 1410 & 126 & 550 & 63{,}504 & 1{,}048{,}576 &  &  \\
\hline
\multirow{6}{*}{$CH_3NO$}
 & [C] & (10, 10) & 20 & 366 & 1847 & 1454 & 134 & 580 & 63{,}504 & 1{,}048{,}576 & \multirow{6}{*}{107} & \multirow{6}{*}{4} \\
 & [H] & (2, 2) & 4 & 34 & 43 & 38 & 0 & 12 & 4 & 16 &  &  \\
 & [H] & (2, 2) & 4 & 34 & 44 & 38 & 0 & 12 & 4 & 16 &  &  \\
 & [H] & (2, 2) & 4 & 35 & 47 & 38 & 0 & 12 & 4 & 16 &  &  \\
 & [N] & (10, 10) & 20 & 341 & 1814 & 1511 & 88 & 542 & 63{,}504 & 1{,}048{,}576 &  &  \\
 & [O] & (10, 10) & 20 & 378 & 1893 & 1479 & 138 & 586 & 63{,}504 & 1{,}048{,}576 &  &  \\
\hline
\multirow{8}{*}{$CH_5 NO$}
 & [C] & (10, 10) & 20 & 369 & 1946 & 1578 & 109 & 586 & 63{,}504 & 1{,}048{,}576 & \multirow{8}{*}{130} & \multirow{8}{*}{4} \\
 & [H] & (2, 2) & 4 & 35 & 45 & 38 & 0 & 12 & 4 & 16 &  &  \\
 & [H] & (2, 2) & 4 & 35 & 48 & 38 & 0 & 12 & 4 & 16 &  &  \\
 & [H] & (2, 2) & 4 & 34 & 44 & 38 & 0 & 12 & 4 & 16 &  &  \\
 & [H] & (2, 2) & 4 & 35 & 46 & 38 & 0 & 12 & 4 & 16 &  &  \\
 & [H] & (2, 2) & 4 & 34 & 46 & 38 & 0 & 12 & 4 & 16 &  &  \\
 & [O] & (10, 10) & 20 & 369 & 1796 & 1463 & 120 & 552 & 63{,}504 & 1{,}048{,}576 &  &  \\
 & [N] & (10, 10) & 20 & 410 & 1950 & 1543 & 130 & 590 & 63{,}504 & 1{,}048{,}576 &  &  \\
\hline
\multirow{7}{*}{$C_2H_3NO$}
 & [C] & (10, 10) & 20 & 591 & 1929 & 1597 & 126 & 604 & 63{,}504 & 1{,}048{,}576 & \multirow{7}{*}{130} & \multirow{7}{*}{4} \\
 & [C] & (10, 10) & 20 & 396 & 1871 & 1503 & 125 & 578 & 63{,}504 & 1{,}048{,}576 &  &  \\
 & [H] & (2, 2) & 4 & 35 & 46 & 38 & 0 & 12 & 4 & 16 &  &  \\
 & [H] & (2, 2) & 4 & 35 & 46 & 38 & 0 & 12 & 4 & 16 &  &  \\
 & [H] & (2, 2) & 4 & 34 & 43 & 38 & 0 & 12 & 4 & 16 &  &  \\
 & [N] & (10, 10) & 20 & 371 & 1901 & 1532 & 104 & 568 & 63{,}504 & 1{,}048{,}576 &  &  \\
 & [O] & (10, 10) & 20 & 383 & 1868 & 1544 & 116 & 572 & 63{,}504 & 1{,}048{,}576 &  &  \\
\hline
\multirow{9}{*}{$C_2H_5NO$}
 & [C] & (10, 10) & 20 & 361 & 1927 & 1589 & 107 & 580 & 63{,}504 & 1{,}048{,}576 & \multirow{9}{*}{165} & \multirow{9}{*}{4} \\
 & [C] & (10, 10) & 20 & 343 & 1813 & 1432 & 134 & 568 & 63{,}504 & 1{,}048{,}576 &  &  \\
 & [H] & (2, 2) & 4 & 35 & 47 & 38 & 0 & 12 & 4 & 16 &  &  \\
 & [H] & (2, 2) & 4 & 35 & 47 & 38 & 0 & 12 & 4 & 16 &  &  \\
 & [H] & (2, 2) & 4 & 35 & 46 & 38 & 0 & 12 & 4 & 16 &  &  \\
 & [H] & (2, 2) & 4 & 35 & 48 & 38 & 0 & 12 & 4 & 16 &  &  \\
 & [H] & (2, 2) & 4 & 35 & 48 & 38 & 0 & 12 & 4 & 16 &  &  \\
 & [N] & (10, 10) & 20 & 375 & 1906 & 1545 & 94 & 568 & 63{,}504 & 1{,}048{,}576 &  &  \\
 & [O] & (10, 10) & 20 & 635 & 1865 & 1424 & 139 & 572 & 63{,}504 & 1{,}048{,}576 &  &  \\
\hline
\multirow{8}{*}{$CH_4 N_2 O$}
 & [C] & (10, 10) & 20 & 633 & 1809 & 1405 & 142 & 568 & 63{,}504 & 1{,}048{,}576 & \multirow{8}{*}{119} & \multirow{8}{*}{4} \\
 & [N] & (10, 10) & 20 & 655 & 1877 & 1389 & 131 & 588 & 63{,}504 & 1{,}048{,}576 &  &  \\
 & [H] & (2, 2) & 4 & 34 & 43 & 38 & 0 & 12 & 4 & 16 &  &  \\
 & [H] & (2, 2) & 4 & 35 & 47 & 38 & 0 & 12 & 4 & 16 &  &  \\
 & [H] & (2, 2) & 4 & 34 & 44 & 38 & 0 & 12 & 4 & 16 &  &  \\
 & [H] & (2, 2) & 4 & 33 & 42 & 38 & 0 & 12 & 4 & 16 &  &  \\
 & [N] & (10, 10) & 20 & 408 & 1925 & 1500 & 117 & 578 & 63{,}504 & 1{,}048{,}576 &  &  \\
 & [O] & (10, 10) & 20 & 444 & 1811 & 1437 & 122 & 564 & 63{,}504 & 1{,}048{,}576 &  &  \\
\hline
\multirow{3}{*}{$NOCl$}
 &  [N] & (8, 10) & 16 & 223 & 897 & 628 & 88 & 300 & 3{,}136 & 65{,}536 & \multirow{3}{*}{62} & \multirow{3}{*}{4} \\
 & [O] & (8, 10) & 16 & 219 & 891 & 636 & 80 & 300 & 3{,}136 & 65{,}536  &  &  \\
 & [Cl] & (12, 18) & 24 & 342 & 2441 & 2117 & 109 & 738 & 48{,}400 & 16{,}777{,}216 &  &  \\
\hline
\multirow{5}{*}{$HOSCN$}
 & [H] & (2, 2) & 4 & 34 & 43 & 38 & 0 & 12 & 4 & 16 & \multirow{5}{*}{98} & \multirow{5}{*}{4} \\
 & [O] & (10, 10) & 20 & 454 & 1838 & 1427 & 199 & 619 & 63{,}504 & 1{,}048{,}576 &  &  \\
 & [S] & (15, 18) & \textbf{30} & \textbf{1,081} & 4384 & 4018 & 205 & 1351 & \textbf{25{,}050{,}025} & \textbf{1{,}073{,}741{,}824} &  &  \\
 & [C] & (10, 10) & 20 & 651 & 1911 & 1534 & 146 & 600 & 63{,}504 & 1{,}048{,}576 &  &  \\
 & [N] & (10, 10) & 20 & 558 & 2021 & 1622 & 143 & 646 & 63{,}504 & 1{,}048{,}576 &  &  \\
\hline
\end{tabular}
\end{table}
\endgroup

\begin{figure}[ht]
    \centering
    \includegraphics[width=1.0\linewidth]{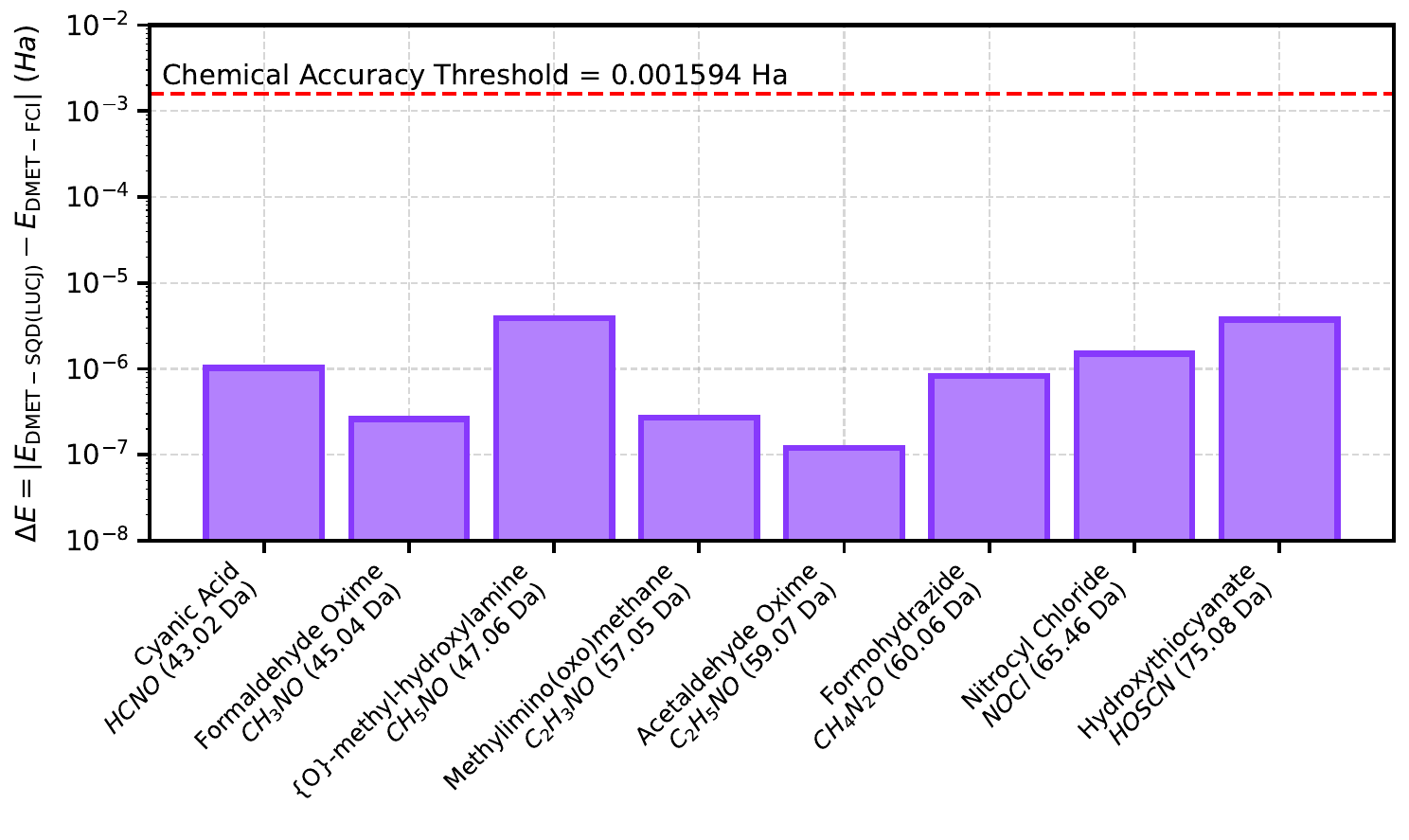}
    \caption{Comparison of the DMET-SQD energies obtained for the set of studied molecules on IBM Sherbrooke Quantum Hardware ($4^{th}-5^{th}$ June, 2025) with DMET-FCI. All the energies obtained are within the chemical accuracy criterion (1 kcal/mol $\approx$ 0.001594 Ha), which is marked with the red-dashed line. The x-axis denotes the name of the studied molecule, the molecular formula, and the molecular weight. The y-axis denotes the absolute energy difference between the energy obtained through quantum hardware using DMET-SQD ($E_{DMET-SQD}$) and the reference energy $E_{DMET-FCI}$, which is defined as $\triangle E$. Since DMET-SQD and DMET-FCI employ identical fragmentation settings, basis sets, and occupation thresholds, $\Delta E$ isolates the accuracy of the SQD impurity solver relative to FCI within the same embedded active space.}
    \label{fig:dmet_sqd_mols}
\end{figure}

\begin{figure}[htbp]
    \centering
    \includegraphics[width=1.0\linewidth]{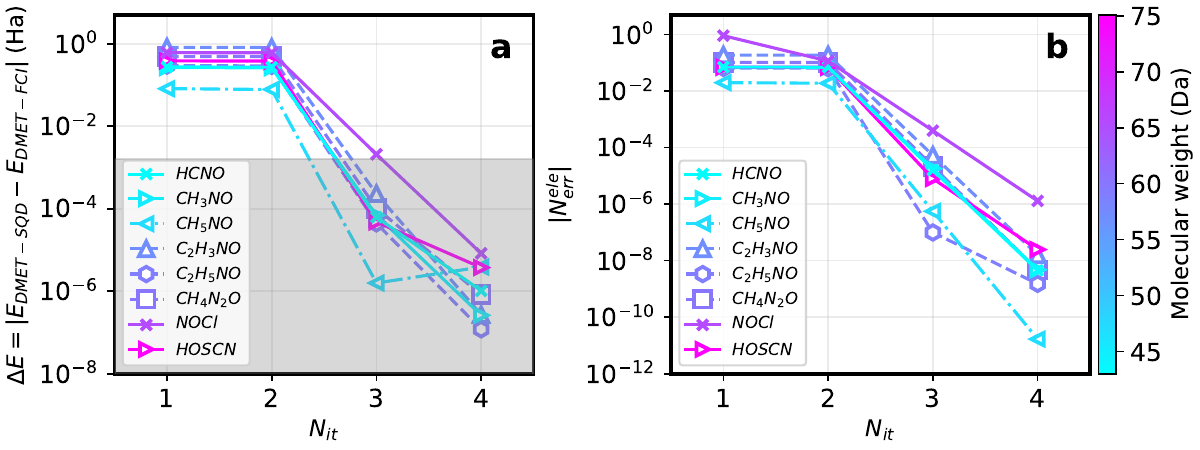}
    \caption{Convergence behavior of the DMET-SQD energies for a set of molecular systems. (a) Absolute energy deviation with respect to  the final energy obtained through DMET-FCI, $\Delta E = \lvert E_{\mathrm{DMET\text{-}SQD}} - E_{\mathrm{DMET\text{-}FCI}} \rvert$, (b) Corresponding absolute electron-number error, $|N_{err}^{ele}| = |\sum_{y}^{N_{frag}} N_{A_y}^{ele} - N_{true}|$ as given in Eq.~(\ref{eqn: dmet_workflow_N_ele}), as a function of the iteration number $N_{\mathrm{it}}$ of the chemical potential $\mu_{\mathrm{glob}}$ 
    convergence loop. Different molecular systems are distinguished by marker style, while the color scale denotes molecular weight.}
    \label{fig:dmet_sqd_ibm_molecules}
\end{figure}

This section is organized as follows: Section~\ref{sec:accuracy} presents the main accuracy and convergence results across all eight molecules; Section~\ref{sec:epsocc} examines the effect of the occupation threshold $\varepsilon_{\mathrm{occ}}$ on embedding and quantum resources for HOSCN; and broader implications of the results.

\subsection{DMET-SQD Accuracy and Convergence}\label{sec:accuracy}

 The absolute energy differences between the DMET-SQD energies and DMET-FCI energies of the molecules are presented in Fig.~\ref{fig:dmet_sqd_mols}. As is evident, the energy differences obtained are consistently lower than $10^{-5}$ Ha, and in some cases are within micro-Hartree precision in comparison to DMET-FCI. We note that since DMET-SQD and DMET-FCI employ identical fragmentation settings, basis sets, and occupation thresholds, all systematic errors originating from basis set incompleteness, one-atom-per-fragment partitioning, and bath truncation cancel exactly in $\Delta E$, which therefore isolates the accuracy of the SQD impurity solver relative to the FCI reference within an identical embedding framework, making this a stringent comparison of the chosen high-level quantum computing solver against the exact classical reference. Two additional factors that may influence the reported accuracy are the SQD subspace thresholding and hardware noise. The maximum number of samples per batch was set to $10^5$ across all fragments, 
ensuring that SQD's quantum sub-sampling threshold does not constitute a limiting factor in the reported accuracy. The hardware calibration metrics recorded immediately prior to execution are provided in Appendix~C.

The details pertaining to the fragmentation settings, quantum resource requirements, and the dimensions of the symmetry space and Hilbert space are summarized in Table~\ref{tab1:resource_dmet_sqd}. The QPU time `T(s)' in the table is measured via the \texttt{IBM Qiskit Runtime} API's provided metadata and represents the cumulative quantum sampling time across all fragments and DMET iterations, excluding classical post-processing (S-CoRe and SCI diagonalization). The resource estimation is presented in the IBM Sherbrooke superconducting hardware basis gate set \{$X$, $R_z$, $\sqrt{X}(S_x)$, $ECR$\}. The largest quantum-sampling experiment employed 30 qubits, with a symmetry-space dimension of 25,050,025 and a Hilbert-space dimension of 1,073,741,824. The largest quantum circuit depth was 1081. These entries have been highlighted in the table to indicate the maximal values observed in our work. In contrast, the complete (unfragmented) quantum simulation of the molecules with SQD would require the following number of qubits in the STO-3G basis: $HOCN$ (32 qubits), $CH_3NO$ (36 qubits), $CH_5NO$ (40 qubits), $C_2H_3NO$ (46 qubits), $C_2H_5NO$ (50 qubits), $CH_4N_2O$ (48 qubits), $NOCl$ (38 qubits), and $HOSCN$ (50 qubits).

Section~\ref{sec: orb_lists} in Supplementary Material provides the detailed orbital space decomposition for each fragment. We note that in DMET, only the impurity space comprising the fragment orbitals $A_y$ and bath orbitals $B_y$ is solved on the quantum computer, requiring $2(|A_y|+|B_y|)$ qubits per fragment at most. The core orbitals $\mathrm{Cor}_y$ and virtual orbitals $\mathrm{Vir}_y$ are excluded from the quantum simulation entirely - their contributions enter only as a classical mean-field correction to the embedded Hamiltonian. Consequently, there is no additional qubit overhead associated with $\mathrm{Cor}_y$ or $\mathrm{Vir}_y$. The largest fragment in this work amounts to a reduction from 50 qubits (full HOSCN) to at most 30 qubits ([S] fragment).

Fig.~\ref{fig:dmet_sqd_ibm_molecules} illustrates the convergence behavior of the DMET-SQD workflow with respect to the number of iterations $N_{it}$ (required for $\mu_{glob}$ convergence) for a set of ligand-like molecules. Panel (a) reports the absolute energy difference $\Delta E = |E_{\mathrm{DMET\text{-}SQD}} - E_{\mathrm{DMET\text{-}FCI}}|$ on a logarithmic scale, while panel (b) shows the corresponding electron number deviation $N_e^{\mathrm{err}}$ for each impurity. A systematic and monotonic reduction of errors is observed as the number of iterations increases, with all systems achieving energy deviations below $10^{-5}$~Ha and electron number errors below $10^{-7}$ by $N_{it}=4$. Notably, this convergence behavior is consistent across molecules with varying chemical composition and molecular weight, indicating that the DMET-SQD workflow is robust with respect to system size within the studied regime. These results demonstrate that, despite aggressive fragmentation (one-atom-per-fragment) and minimal basis sets, DMET-SQD is able to recover correlation energies consistent with the reference DMET-FCI solver to within chemical accuracy.

The simulation results showing the energies obtained through DMET-SQD on a quantum computer, DMET-FCI as reference, and the absolute energy difference between them are shown in Section~\ref{sec: tab:nat_mols_energy} in Supplementary Material. The energy differences lie along the range of micro-Hartree, indicating strong agreement with the reference. The geometric co-ordinates of the molecules used for simulation were geometry-optimized (force-field level)~\cite{RDKit_2025_09_1} and are given in the Section~\ref{sec:coords_supp} of the Supplementary Material.

These results indicate that the quantum sampling, configuration recovery, and subsequent creation of projection sub-space through symmetrized expansion of the unique spin configurations (Eq.~\ref{eqn: ab_unique} and Eq.~\ref{eqn: s_proj_creation}), is leading to diagonalization over a sub-space which is a sizable fraction of the \textit{symmetry space}~\cite{hi_vqe_2024}.

\begin{sidewaysfigure}
    \centering
    \includegraphics[width=\textwidth]{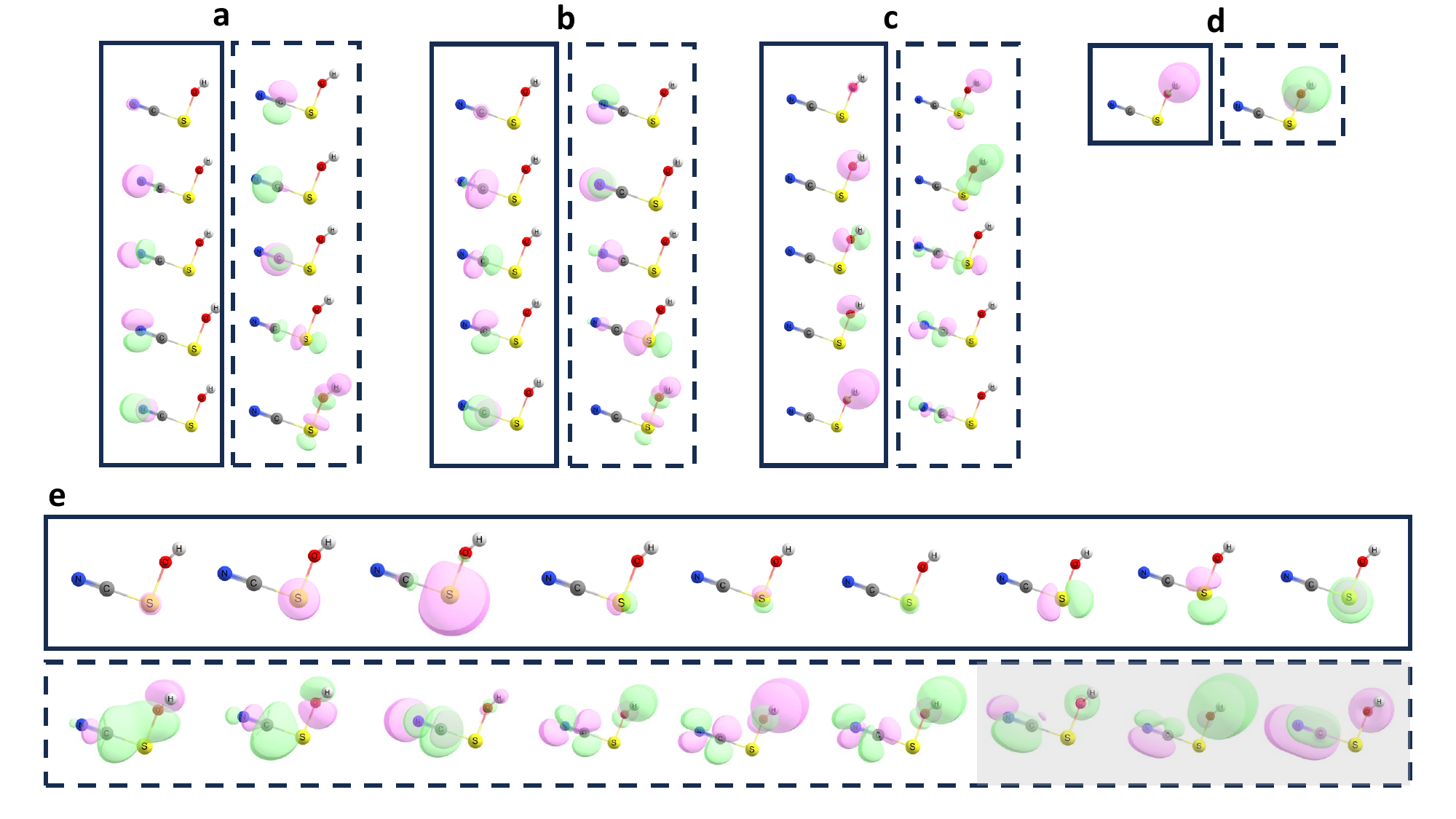}
    \caption{The localized fragment and bath orbitals for the HSOCN molecule for the fragments (a), (b), (c), (d) and (e) denoting the fragments [N], [C], [O], [H], and [S] respectively. The boxes with a solid line denote the localized fragment orbitals, and the boxes with a dashed line denote the corresponding bath orbitals. The last three bath orbitals in subplot (e) are not considered as part of the impurity orbitals based on the value of $\epsilon_{occ} = 10^{-13}$.}
    \label{fig:hsocn_bath}
\end{sidewaysfigure}

\subsection{Entanglement-aware Embedding via Occupation Threshold \texorpdfstring{$\varepsilon_{\mathrm{occ}}$}{epsilon\_occ}}\label{sec:epsocc}

From a quantum computing perspective, the choice of $\epsilon_{occ}$ plays a critical role in balancing correlation accuracy and hardware feasibility. A smaller cutoff retains more bath orbitals, thereby increasing the impurity size, which in turn raises the number of qubits and circuit depth required for SQD. This leads to a degradation in sampling quality due to increased noise accumulation on NISQ hardware, which requires more robust configuration recovery. Conversely, a larger cutoff reduces the impurity size, improving sampling efficiency and robustness, but at the cost of neglecting weaker correlations. This trade-off is particularly pronounced in low-symmetry ($C_1$) molecular systems, where geometric details strongly influence fragment-environment entanglement patterns. Our observations indicate that $\epsilon_{occ}$ is therefore a physically and computationally significant control knob that governs both the expressibility of the embedding and the viability of quantum sampling. Notably, this sensitivity is particularly consequential in the DMET-SQD context, where impurity size directly governs qubit count, circuit depth, and shot budget - quantum hardware constraints that are inconsequential for classical impurity solvers such as FCI.

\subsubsection{Bath Orbital Construction for various Occupation Thresholds}

The construction of bath orbitals in practical DMET implementations relies on the spectrum of the environment block of the 1-RDM, where only a subset of eigenvectors corresponding to nontrivial (fractional) eigenvalues are retained as entangled bath orbitals. In realistic quantum chemistry calculations, several of these eigenvalues may lie numerically very close to $0$ or $1$, making it difficult to distinguish between genuinely entangled orbitals and effectively unentangled environment orbitals. Therefore, this leads to a practically feasible and efficient construction of the impurity, consisting of the localized fragment orbitals and the entangled bath orbitals with eigenvalues adhering to $\epsilon_{occ}$, as discussed in the methodology, to truncate negligible contributions and define the bath space~\cite{wouters2016dmets_guide}.

In our DMET-SQD simulations with atom-wise fragmentation, this truncation directly impacts the resulting impurity sizes. For instance, in the cases of NOCl and HOSCN, although the nominal impurity size for the [Cl] and [S] fragments would correspond to $18$ orbitals, the application of the eigenvalue cutoff reduces the effective bath to only $3$ and $6$ orbitals, respectively, yielding smaller impurity spaces consisting of $12$ and $15$ orbitals, respectively. This phenomenon has been critically analyzed and observed for the HOSCN molecule, as shown in Fig.~\ref{fig:hsocn_bath}. This reflects that only a limited subset of orbitals significantly contributes to the fragment–environment entanglement, as determined by the chosen $\epsilon_{occ}$ threshold. This behaviour is consistent with the underlying fragment-environment entanglement structure of these molecules, and its implications for quantum resource requirements and embedding accuracy are examined subsequently.

To quantify the impact of $\varepsilon_\text{occ}$ on the embedding and the resulting quantum resource requirements, we performed a systematic threshold sweep on HOSCN using six values $\varepsilon_\text{occ} \in \{10^{-5}, 10^{-7}, 10^{-9}, 10^{-11}, 10^{-13}, 10^{-15}\}$, with all other simulation parameters held fixed. By MacDonald's theorem~\cite{MacDonald1933_RayleighRitz}, the rank-$|A_y|$ coupling between the fragment and its environment bounds the number of eigenvalues of the environment block $D^{\mathrm{env}}_y$ that can be displaced from 0 or 2, so at most $|A_y|$ fractional (entangled) eigenvalues exist per fragment. The 
fractionality of each eigenvalue $\varepsilon$ of $D^{\mathrm{env}}_y$ is measured by $\delta = \min(\varepsilon, 2-\varepsilon)$: for $\varepsilon$ close to 0 (empty orbital) or 2 (filled orbital), $\delta \approx 0$ and the orbital is unentangled, whereas $\delta \approx 1$ signals a maximally entangled bath orbital. The MacDonald bound therefore directly limits the number of eigenvalues with $\delta > 0$ to at most $|A_y|$ per fragment. For HOSCN, this yields an upper bound of 1 for [H], 5 for each of [O], [C], and [N], and 9 for [S], consistent with their respective fragment orbital counts $|A_y|$ given in Section~\ref{sec: orb_lists} in Supplementary 
Material.

\begin{figure}[htbp]
    \centering
    \includegraphics[width=\linewidth]{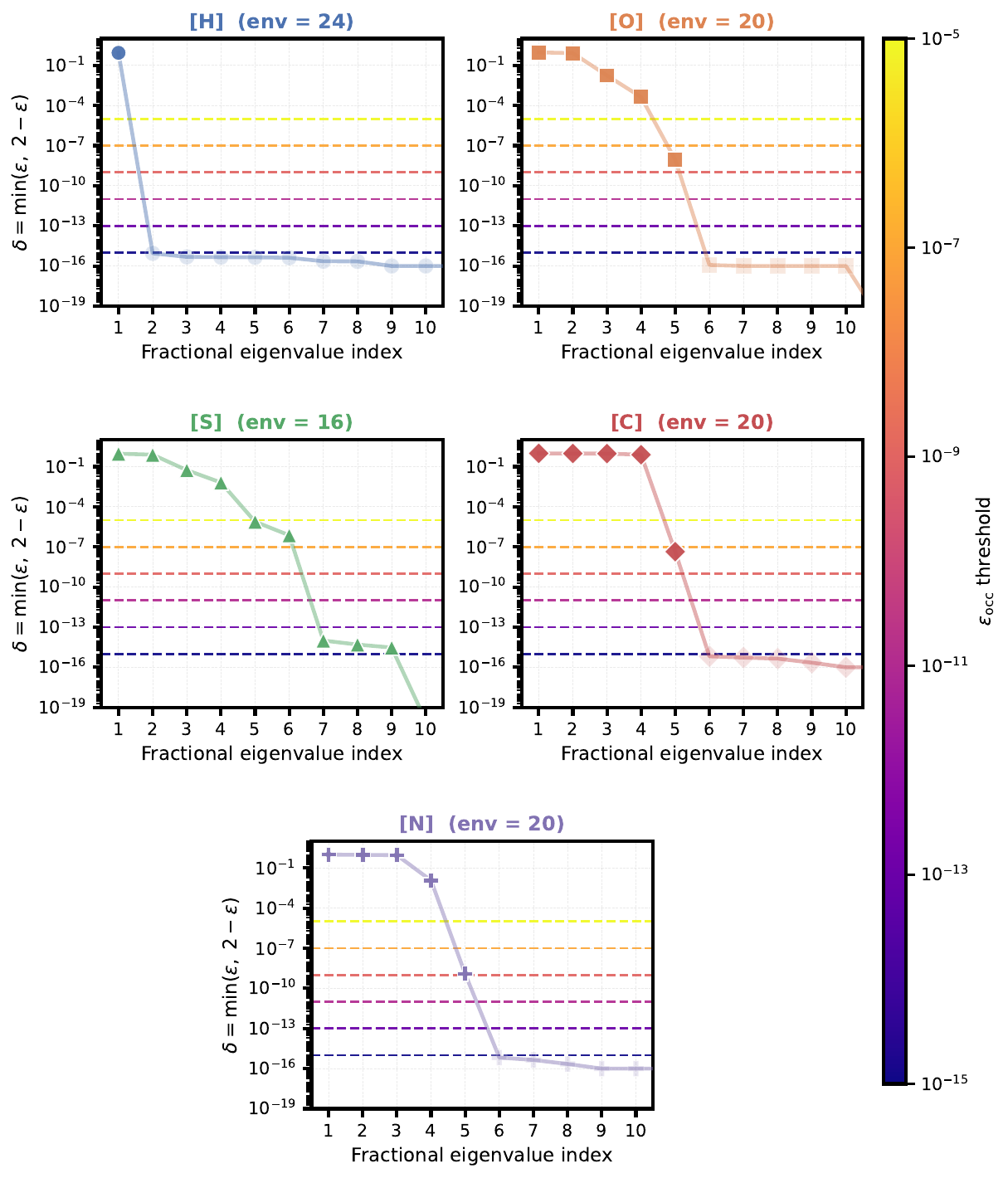}
    \caption{Environment 1-RDM eigenvalue fractionality spectrum $\delta = \min(\varepsilon, 2-\varepsilon)$ for all fragments of HOSCN in the STO-3G basis (\textbf{25 total spatial orbitals}), at six occupation thresholds $\varepsilon_\text{occ} \in \{10^{-5}, 10^{-7}, 10^{-9}, 10^{-11}, 10^{-13}, 10^{-15}\}$. The environment size per fragment follows directly from $|Env_y| = N_\text{orb} - |A_y|$: [H] has $|Env_y| = 24$ ($|A_y|=1$), [O], [C], and [N] each have $|Env_y| = 20$ ($|A_y|=5$), 
    and [S] has $|Env_y| = 16$ ($|A_y|=9$). Dashed horizontal lines mark each threshold, and eigenvectors corresponding to eigenvalues above a given line are selected as bath orbitals at that $\varepsilon_\text{occ}$.}
    \label{fig:hoscn_eigenvalue_spectrum}
\end{figure}

The environment 1-RDM eigenvalue fractionality spectra per fragment, shown in Fig.~\ref{fig:hoscn_eigenvalue_spectrum}, directly confirm this bound: each panel contains at most $|A_y|$ eigenvalues with $\delta > 0$, while all remaining environment eigenvalues collapse to the noise floor, corresponding to exactly unentangled core or virtual orbitals. The entanglement structure is, however, highly fragment-dependent in how those $|A_y|$ fractional eigenvalues are distributed: [H], [O], [C], and [N] exhibit sharp spectral gaps, with their few fractional eigenvalues well-separated from the noise floor, so that $|B_y|$ is stable across all thresholds. In contrast, [S] shows a gradual spectrum spanning several orders of magnitude — its 9 fractional eigenvalues decay continuously from $\delta \sim 10^{-1}$ down to $\delta \sim 10^{-14}$ without a clean gap, so that $|B_y|$ grows from 3 to 9 as $\varepsilon_{\mathrm{occ}}$ is tightened from $10^{-5}$ to $10^{-15}$.

Since $\varepsilon_\text{occ}$ directly controls which of these MacDonald-bounded fractional eigenvalues are admitted as bath orbitals, it serves as a physically grounded, entanglement-aware embedding control parameter: a tighter threshold retains 
more of the fragment--environment entanglement at the cost of larger impurity spaces, while a looser threshold discards weakly entangled orbitals to reduce quantum resource requirements. This sensitivity is particularly consequential for [S], whose absence 
of a clean spectral gap makes it uniquely vulnerable to the choice of threshold, with direct consequences for quantum resource requirements and embedding accuracy examined subsequently.

\subsubsection{Analysis of $\varepsilon_{occ}$ within DMET-SQD}

\begin{figure}[htbp]
    \centering
    \includegraphics[width=\linewidth]{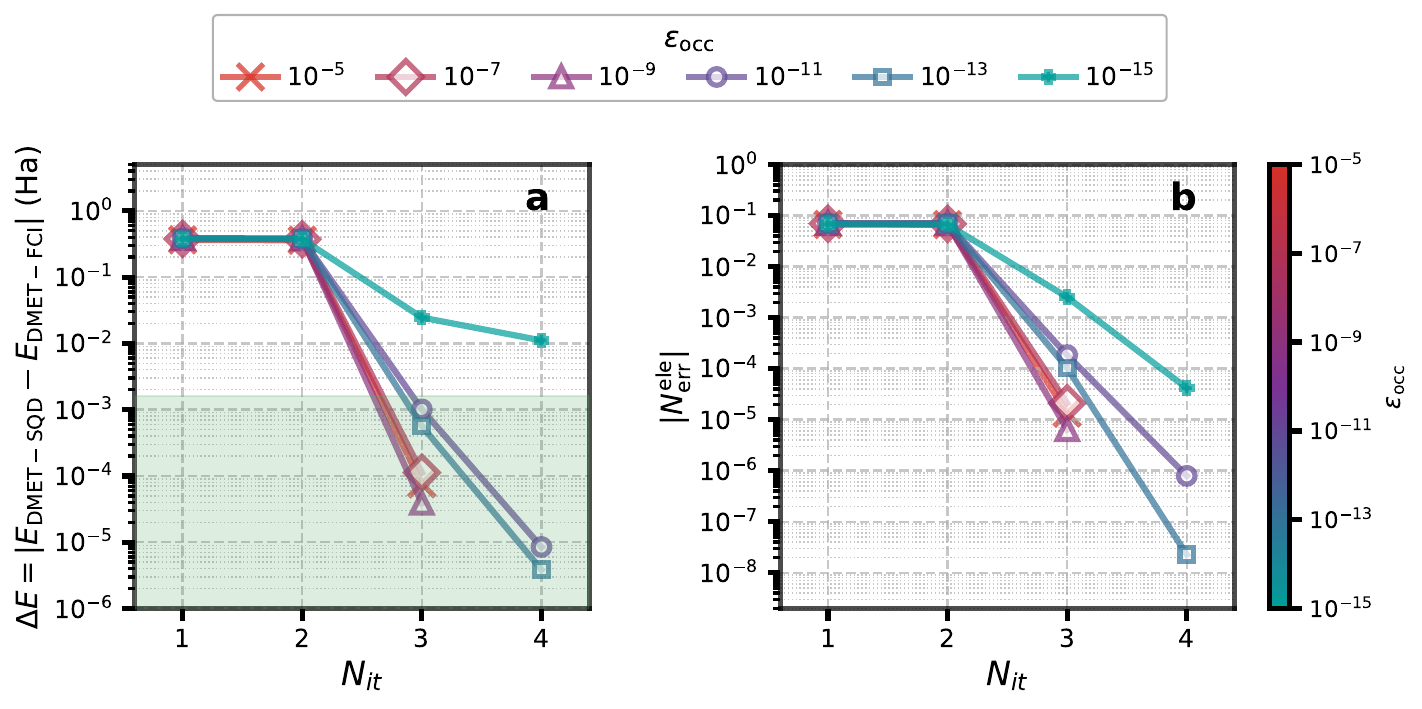}
    \caption{Convergence of the DMET-SQD energy error and global chemical-potential optimization for HOSCN at different occupation thresholds $\varepsilon_{\rm occ}$ on IBM Boston Heron R3 (13$^{th}$ June, 2026). (a) Energy error, $\Delta E = |E_{\rm DMET\text{-}SQD}-E_{\rm DMET\text{-}FCI}|$, and (b) electron-number error, $|N^{\rm ele}_{\rm err}|$, as functions of the DMET iteration number $N_{it}$. Marker styles distinguish different $\varepsilon_{\rm occ}$ values, while colours indicate $\log_{10}(\varepsilon_{\rm occ})$, with darker shades corresponding to smaller thresholds. The shaded region in panel (a) denotes the chemical-accuracy regime (1 kcal/mol $\approx 1.594\times10^{-3}$ Ha).}
    \label{fig:dmet_hoscn}
\end{figure}

Fig.~\ref{fig:dmet_hoscn} illustrates the convergence of the DMET-SQD workflow for HOSCN as a function of both $N_{it}$ and the occupation threshold $\varepsilon_{\rm occ}$, revealing that thresholds in the range $10^{-9}$ to $10^{-13}$ yield the fastest convergence and the lowest final $\Delta E$, reaching below $10^{-5}$~Ha by $N_{it} = 4$ while remaining within the chemical accuracy threshold throughout. At the largest threshold ($\varepsilon_{\rm occ} = 10^{-5}$) and the smallest ($\varepsilon_{\rm occ} = 10^{-15}$), the final $\Delta E$ is significantly larger, as shown in Fig.~\ref{fig:dmet_hoscn}.

\begin{figure}[H]
    \centering
    \includegraphics[width=\linewidth]{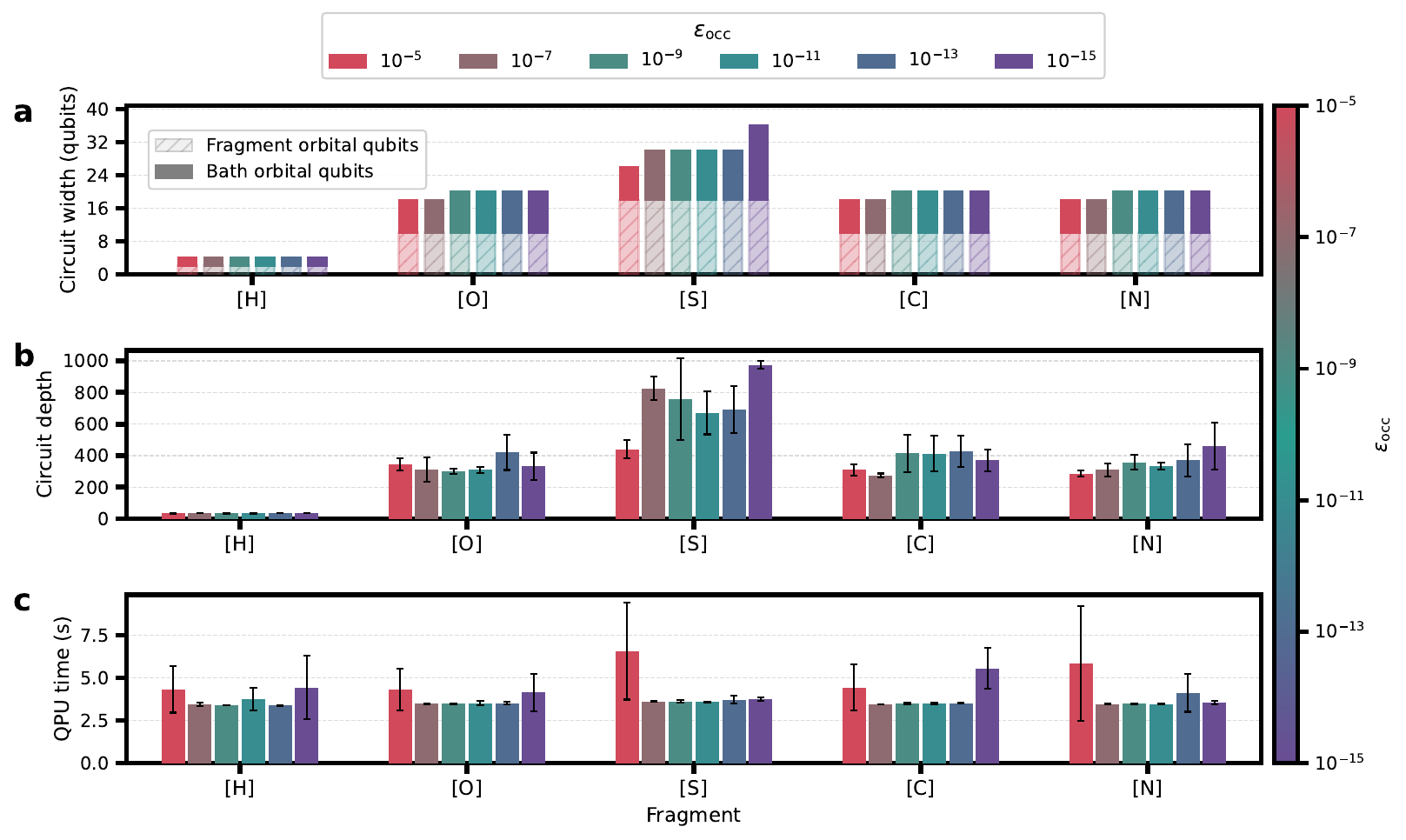}
    \caption{Quantum circuit resources for HOSCN/STO-3G across $\varepsilon_{\mathrm{occ}}$ thresholds Circuit width decomposed into fragment orbital qubits ($2|A_y|$, hatched) and bath orbital qubits ($2|B_y|$, solid)~(a), transpiled circuit depth~(b), and QPU execution time~(c) for each DMET fragment of HOSCN, executed on the IBM Heron R3 (\texttt{ibm\_boston}) quantum processor across six thresholds $\varepsilon_{\mathrm{occ}} \in \{10^{-5}, 10^{-7}, 10^{-9}, 10^{-11}, 10^{-13}, 10^{-15}\}$. Error bars denote $\pm 1\sigma$ over all $\mu_{\mathrm{glob}}$ convergence iterations.}
    \label{fig:resources_hoscn}
\end{figure}

Fig.~\ref{fig:resources_hoscn} summarizes the quantum circuit resources for HOSCN as a function of $\varepsilon_{\mathrm{occ}}$. As expected, the circuit width grows monotonically with decreasing threshold, since a finer $\varepsilon_{\mathrm{occ}}$ retains more bath orbitals and thereby enlarges the active space. The [S] fragment is the most sensitive to this truncation: owing to its large
fragment size ($|A_y| = 9$), it reaches a maximum of 36 qubits at $\varepsilon_{\mathrm{occ}} = 10^{-15}$. Circuit depth follows
a similar trend, with [S] exhibiting the deepest circuits across all thresholds.

Fig.~\ref{fig:hoscn_classical_resources} in Section~\ref{sec: classical_resources_HOSCN} of Supplementary Material presents the classical S-CoRe resource requirements for each DMET fragment of HOSCN as a function of $\varepsilon_{\mathrm{occ}}$. For the [H], [O], [C], and [N] fragments, the symmetry-space dimension $|S_{\mathrm{proj}}|$ remains largely stable across all thresholds, with sampling ratios consistently above 95\%, indicating that
the fixed shot budget of $10^4$ is sufficient to cover the projected subspace for these fragments at all thresholds. The [S] fragment presents a qualitatively different picture: $|S_{\mathrm{proj}}|$ grows by nearly two orders of magnitude from $\varepsilon_{\mathrm{occ}} = 10^{-5}$ to $10^{-15}$, accompanied by a collapse of the sampling ratio from
$\sim$83\% to $\sim$2\%. This reflects the growth of the [S] active space from 26 to 36 qubits as
$\varepsilon_{\mathrm{occ}}$ is tightened, driven by the bath truncation exception $|A_y| \neq |B_y|$ unique to this fragment (see Section~\ref{sec: orb_lists} in Supplementary Material). Taken together, these results establish that $\varepsilon_{\mathrm{occ}}$ functions as an
entanglement-aware embedding control parameter: for HOSCN, the threshold $10^{-13}$ achieves chemical accuracy while avoiding the unnecessary resource overhead incurred by admitting near-zero eigenvalues, providing a concrete and transferable criterion for threshold selection in low-symmetry molecular applications. However, an arbitrary lowering of the $\varepsilon_{occ}$ value without increasing the shot budget would lead to under exploration of the Hilbert space, resulting in lower accuracy.

The observations above suggest a practical preliminary analysis protocol for DMET-SQD calculations: prior to any quantum computation, performing the diagonalization of the environment block of the 1-RDM as part of a mean-field DMET-HF calculation and inspecting the resulting fractionality spectrum $\delta = \min(\varepsilon, 2-\varepsilon)$ per fragment provides a zeroth-order diagnostic for the selection of $\varepsilon_\text{occ}$. The spectral structure (specifically, the presence or absence of a clean gap separating genuinely fractional eigenvalues from the noise floor) directly quantifies the fragment--environment entanglement and anticipates the sensitivity of the impurity construction to the threshold choice. This is computationally inexpensive, requiring only a classical mean-field calculation, yet yields actionable guidance: fragments exhibiting sharp spectral gaps (such as [H], [O], [C], and [N] in HOSCN) are insensitive to $\varepsilon_\text{occ}$ and require no threshold tuning, whereas fragments with a gradual spectrum (such as [S]) signal that the threshold must be chosen with care to balance correlation completeness against quantum and classical resource constraints. In particular, a tighter $\varepsilon_\text{occ}$ admits more bath orbitals and captures more correlation, but enlarges the impurity Hamiltonian whose classical diagonalization and quantum circuit shot budget scale exponentially with active space size, rendering an overly stringent threshold computationally prohibitive.

Taken together, this preliminary spectral analysis elevates $\varepsilon_\text{occ}$ selection from an empirical numerical choice to an entanglement-aware decision: by grounding the threshold in the fractionality spectrum of the mean-field 1-RDM, the entire DMET-SQD workflow (from bath construction and impurity sizing through to quantum sampling and classical diagonalization) is guided by the true fragment-environment entanglement content of the system. Consequently, $\varepsilon_{\mathrm{occ}}$ is not merely a numerical convergence parameter but a physically transparent, entanglement-aware embedding control parameter whose optimal selection is essential for balancing correlation accuracy against qubit count, circuit depth, and shot budget in low-symmetry molecular systems on quantum hardware. This consideration extends naturally to recent embedding frameworks such as EWF~\cite{shajan2026_ewf_protein} that builds upon DMET as its foundational layer.

Moreover, SQD's reliance on recovering signal from noisy configurations, evidenced by the need for only $~2\%$ meaningful signal amid raw noisy data in benchmark cases~\cite{sqd_first_paper}. This reveals a deep hardware dependence that, while enabling NISQ-scale simulations, introduces inefficiencies tied to noise levels. In lower-noise regimes (higher signal configurations), the subspace dimension could remain fixed, but fewer invalid configurations would require correction via the iterative orbital occupancy updates, potentially shifting the burden to the quantum sampler itself. Sampling from this single ansatz distribution, however, introduces key caveats~\cite{critical_lims_sqd_2025}: notably, the risk of repetitive over-sampling of dominant configurations, leaving scant opportunity to capture rarer determinants that contribute non-negligibly to the true ground state.

Although lower hardware noise increases the fidelity of sampled configurations, it simultaneously suppresses the population of symmetry-breaking configurations. Configuration Recovery relies upon these noisy configurations to expand the sampled subspace using the orbital-occupancy distribution as a reference. In the extreme low-noise limit, this reduced exploration can narrow the recoverable determinant space and ultimately degrade the accuracy of SQD~\cite{AnuragKSV2026}.
Modern trapped-ion platforms achieve two-qubit gate fidelities of $\approx 99.92\%$ (Quantinuum)~\cite{ransford2025helios}, $\approx 99.99\%$ (IONQ)~\cite{hughes2025ionq}, neutral-atom arrays report two-qubit fidelities approaching $99.5\%$ (QuEra)~\cite{Evered2023}, and superconducting qubits have recently achieved $\approx 99.5\%$ (Rigetti)~\cite{du2025rigetti}, $\approx 99.88\%$ (IBM)~\cite{ibmquantum_heron_boston}, $\approx 99.93\%$ (IQM)~\cite{marxer2025_999_IQM_superconducting} two-qubit fidelities, all of which influence the balance between valid and recoverable configurations in SQD.

\section{Conclusion}\label{sec: conclusion}

In this work, we have successfully performed quantum simulations of a set of natural ligand-like molecules in the STO-3G basis set (with molecular weights between 40 Da and 76 Da). We employed DMET as a classical fragmentation technique followed by SQD as a high-level solver. The simulations were run on IBM's Eagle R3 (\texttt{ibm\_sherbrooke}) superconducting quantum computing hardware. All the results obtained were benchmarked against DMET-FCI and were found to be within chemical accuracy.

In addition, a detailed entanglement-aware analysis of the sample-based technique within the embedding framework was performed for the HOSCN molecule by systematically varying $\varepsilon_{\mathrm{occ}}$ across six orders of magnitude on IBM's Heron R3 (\texttt{ibm\_boston}). This analysis demonstrated that $\varepsilon_{\mathrm{occ}}$ functions as a physically meaningful, entanglement-aware control parameter rather than a simple numerical convergence threshold, with a fragment-dependent sensitivity that is particularly pronounced for strongly entangled fragments such as [S], and obtaining an optimal threshold range might be beneficial, which balances embedding completeness against quantum hardware feasibility.

Thereby, the current NISQ hardware seems to show potential to perform quantum sampling experiments, leading to classical sub-space diagonalization in post-processing with high accuracy. This paves the way for potential industrial use-cases in the field of quantum computer-aided drug design and materials design. In practice, an overly stringent threshold increases circuit depth and noise sensitivity without a comparable accuracy benefit, while an overly relaxed threshold risks neglecting non-negligible fragment-environment correlation. For HOSCN, the environment 1-RDM eigenvalue spectrum (Fig.~\ref{fig:hoscn_eigenvalue_spectrum}) provided a direct, per-fragment diagnostic for identifying this balance; extending such a diagnostic-driven selection of $\varepsilon_{occ}$ to other low-symmetry systems remains an important direction for future work.

Notably, this trade-off is not merely qualitative: our threshold sweep on HOSCN shows that the energy accuracy $\Delta E$ depends non-monotonically on $\varepsilon_{occ}$, with the tightest threshold ($\varepsilon_{occ}=10^{-15}$) yielding \emph{worse} accuracy than an intermediate value ($\varepsilon_{occ}=10^{-13}$) due to severe under-sampling of the enlarged [S] impurity. This finding cautions against the common assumption that tighter occupation thresholds are uniformly preferable, and instead establishes $\varepsilon_{occ}$ selection as an explicit accuracy--resource optimization problem for low-symmetry molecular systems on near-term hardware.

That being said, although SQD ensures that the noisy configurations are recovered into the symmetry space by using the average orbital occupancy distributions as reference, this does not guarantee that the noisy configurations are stochastically converted into determinants with a non-negligible contribution to the true ground state. The procedure still might lead to obtaining configurations from the symmetry space, which contribute little to the final energy estimate.

This highlights the need for approaches that explicitly promote compactness in the selected subspace, where the retained configurations are guided by user-defined criteria rather than by sampling alone. By constraining the search to a compact and size-controlled subspace, one can more reliably concentrate quantum and classical resources only on the determinants most relevant to the target state, leading to an overall reduction of computational cost for the procedure. 

This need is further underscored by the fact that the performance of S-CoRe is intrinsically tied to the noise level of the hardware: in ultra-low-noise regimes, the sampling distribution collapses onto a few high-weight determinants concentrated near the Hartree-Fock reference, severely limiting subspace diversity, whereas moderate noise increases the proportion of off-symmetry samples that, after configuration-recovery, can populate a richer and more correlation-relevant subspace. Relying solely on stochastic sampling and symmetry recovery, therefore, leaves subspace quality coupled to hardware noise characteristics outside the user's control, motivating more explicitly guided subspace construction strategies~\cite{hi_vqe_2024, Yoo2026extending_hi_vqe, Patra2026_PIGen, Patra2026mlcsg}.

A further promising direction is the incorporation of cumulant-based 
2-RDM reconstruction~\cite{Nusspickel2022} within the DMET-SQD framework, which has been shown to improve the accuracy of global observables and is in principle compatible with the RDM outputs of the SQD procedure. Additionally, extension of the present framework to larger basis sets and open-shell systems, as well as the systematic study of the effect of $\epsilon_{occ}$ across the full molecular set studied here, remain important avenues for future investigation.

\section*{Supplemetary Material}

The Supplementary Material provides comprehensive supporting data for the present study, including the optimized geometric coordinates of all ligand-like molecules considered, detailed comparisons of ground-state energies obtained using DMET-SQD and DMET-FCI (demonstrating good agreement with chemical accuracy), the calibration parameters of both the IBM Eagle R3 (\texttt{ibm\_sherbrooke}) and IBM Heron R3 (\texttt{ibm\_boston}) quantum processors employed in the simulations, the complete fragment orbital space decomposition for all studied molecules, and the classical reosurce cost for HOSCN molecule while varying the occupancy threshold.

\section*{Acknowledgements}

We acknowledge the use of IBM Quantum Credits via the IBM Quantum Startups Program for this work. The views expressed are those of the authors and do not reflect the official policy or position of IBM or the IBM Quantum Platform team. The authors would also like to extend appreciation to the advisors of Qclairvoyance Quantum Labs for their support, constructive discussions, and inspiration throughout the preparation of this work.

 \section*{Funding}
This research received no specific grant from any funding agency in the public, commercial, or not-for-profit sectors.

\section*{Competing interests}
R.M. and R.V. are paid consultants at Qclairvoyance Quantum Labs. The other authors declare no competing interests.

\bibliography{sample}

@ARTICLE{propane_fci_Gao2024,
  author    = {Gao, Hong and Imamura, Satoshi and Kasagi, Akihiko and Yoshida, Eiji},
  title     = {Distributed Implementation of Full Configuration Interaction for One Trillion Determinants},
  journal   = {Journal of Chemical Theory and Computation},
  volume    = {20},
  number    = {3},
  pages     = {1185--1192},
  year      = {2024},
  month     = {2},
  day       = {13},
  publisher = {American Chemical Society},
  issn      = {1549-9618},
  url       = {https://doi.org/10.1021/acs.jctc.3c01190},
  doi       = {10.1021/acs.jctc.3c01190}
}

@ARTICLE{kitaev1995abelian_qpe,
  author  = {A. Yu. Kitaev},
  title   = {Quantum Measurements and the Abelian Stabilizer Problem},
  journal = {arXiv:quant-ph/9511026},
  year    = {1995},
}

@ARTICLE{
sqd_first_paper,
author = {Javier Robledo-Moreno  and Mario Motta  and Holger Haas  and Ali Javadi-Abhari  and Petar Jurcevic  and William Kirby  and Simon Martiel  and Kunal Sharma  and Sandeep Sharma  and Tomonori Shirakawa  and Iskandar Sitdikov  and Rong-Yang Sun  and Kevin J. Sung  and Maika Takita  and Minh C. Tran  and Seiji Yunoki  and Antonio Mezzacapo },
title = {Chemistry beyond the scale of exact diagonalization on a quantum-centric supercomputer},
journal = {Science Advances},
volume = {11},
number = {25},
pages = {eadu9991},
year = {2025},
doi = {10.1126/sciadv.adu9991}}

@article{kanno2023qsci,
  author       = {Kanno, Keita and Kohda, Masaya and Imai, Ryosuke and Koh, Sho and Mitarai, Kosuke and Mizukami, Wataru and Nakagawa, Yuya O.},
  title        = {Quantum-Selected Configuration Interaction: Classical Diagonalization of Hamiltonians in Subspaces Selected by Quantum Computers},
  journal      = {arXiv},
  year         = {2023},
  publisher    = {arXiv},
  doi          = {10.48550/arXiv.2302.11320},
  url          = {https://doi.org/10.48550/arXiv.2302.11320}
}

@ARTICLE{Peruzzo2014_vqe,
  author = {Peruzzo, Alberto and McClean, Jarrod and Shadbolt, Peter and Yung, Man-Hong and Zhou, Xiao-Qi and Love, Peter J. and Aspuru-Guzik, Alán and O’Brien, Jeremy L.},
  title = {A variational eigenvalue solver on a photonic quantum processor},
  journal = {Nature Communications},
  year = {2014},
  volume = {5},
  number = {1},
  pages = {4213},
  issn = {2041-1723},
  doi = {10.1038/ncomms5213},
  url = {https://doi.org/10.1038/ncomms5213}
}

@article{Fock1930,
  author       = {Fock, V.},
  title        = {N{\"a}herungsmethode zur L{\"o}sung des quantenmechanischen Mehrk{\"o}rperproblems},
  journal      = {Zeitschrift f{\"u}r Physik},
  volume       = {61},
  number       = {1},
  pages        = {126--148},
  year         = {1930},
  doi          = {10.1007/BF01340294},
  url          = {https://doi.org/10.1007/BF01340294}
}

@article{hao2025largescaleefficientmoleculegeometry,
  author       = {Hao, Yajie and Ding, Qiming and Wang, Xiaoting and Yuan, Xiao},
  title        = {Large-scale Efficient Molecule Geometry Optimization with Hybrid Quantum-Classical Computing},
  journal      = {arXiv},
  year         = {2025},
  publisher    = {arXiv},
  doi          = {10.48550/arXiv.2509.07460},
  url          = {https://doi.org/10.48550/arXiv.2509.07460}
}

@ARTICLE{Belaloui2025_current_vqe,
  author = {Belaloui, Nacer Eddine and Tounsi, Abdellah and Khamadja, Abdelmouheymen Rabah and Louamri, Mohamed Messaoud and Benslama, Achour and Bernal Neira, David E. and Rouabah, Mohamed Taha},
  title = {Ground-State Energy Estimation on Current Quantum Hardware through the Variational Quantum Eigensolver: A Practical Study},
  journal = {Journal of Chemical Theory and Computation},
  year = {2025},
  volume = {21},
  number = {14},
  pages = {6777--6792},
  publisher = {American Chemical Society},
  issn = {1549-9618},
  doi = {10.1021/acs.jctc.4c01657},
  url = {https://doi.org/10.1021/acs.jctc.4c01657}
}

@ARTICLE{wiley_protein_ligand_interaction,
author = {Kirsopp, Josh J. M. and Di Paola, Cono and Manrique, David Zsolt and Krompiec, Michal and Greene-Diniz, Gabriel and Guba, Wolfgang and Meyder, Agnes and Wolf, Detlef and Strahm, Martin and Muñoz Ramo, David},
title = {Quantum computational quantification of protein–ligand interactions},
journal = {International Journal of Quantum Chemistry},
volume = {122},
number = {22},
pages = {e26975},
doi = {10.1002/qua.26975},
url = {https://onlinelibrary.wiley.com/doi/abs/10.1002/qua.26975},
year = {2022}
}

@ARTICLE{critical_lims_sqd_2025,
   title={Critical Limitations in Quantum-Selected Configuration Interaction Methods},
   volume={21},
   ISSN={1549-9626},
   url={http://dx.doi.org/10.1021/acs.jctc.5c00375},
   DOI={10.1021/acs.jctc.5c00375},
   number={14},
   journal={Journal of Chemical Theory and Computation},
   publisher={American Chemical Society (ACS)},
   author={Reinholdt, Peter and Ziems, Karl Michael and Kjellgren, Erik Rosendahl and Coriani, Sonia and Sauer, Stephan P. A. and Kongsted, Jacob},
   year={2025},
   month=jun, pages={6811–6822} }

@article{dmet_sqd,
  author       = {Shajan, Akhil and Kaliakin, Danil and Mitra, Abhishek and Robledo Moreno, Javier and Li, Zhen and Motta, Mario and Johnson, Caleb and Saki, Abdullah Ash and Das, Susanta and Sitdikov, Iskandar and Mezzacapo, Antonio and Merz, Kenneth M.},
  title        = {Toward Quantum-Centric Simulations of Extended Molecules: Sample-Based Quantum Diagonalization Enhanced with Density Matrix Embedding Theory},
  journal      = {Journal of Chemical Theory and Computation},
  volume       = {21},
  number       = {14},
  pages        = {6801--6810},
  year         = {2025},
  publisher    = {American Chemical Society},
  doi          = {10.1021/acs.jctc.5c00114},
  url          = {https://doi.org/10.1021/acs.jctc.5c00114}
}

@ARTICLE{lucj_ansatz,
author ="Motta, Mario and Sung, Kevin J. and Whaley, K. Birgitta and Head-Gordon, Martin and Shee, James",
title  ="Bridging physical intuition and hardware efficiency for correlated electronic states: the local unitary cluster Jastrow ansatz for electronic structure",
journal  ="Chem. Sci.",
year  ="2023",
volume  ="14",
issue  ="40",
pages  ="11213-11227",
publisher  ="The Royal Society of Chemistry",
doi  ="10.1039/D3SC02516K",
url  ="http://dx.doi.org/10.1039/D3SC02516K"}

@inproceedings{Gily_n_2019_qsvt,
  author       = {Gily{\'e}n, Andr{\'a}s and Su, Yuan and Low, Guang Hao and Wiebe, Nathan},
  title        = {Quantum Singular Value Transformation and Beyond: Exponential Improvements for Quantum Matrix Arithmetics},
  booktitle    = {Proceedings of the 51st Annual ACM SIGACT Symposium on Theory of Computing},
  series       = {STOC '19},
  pages        = {193--204},
  month        = jun,
  doi          = {10.1145/3313276.3316366},
  url          = {https://doi.org/10.1145/3313276.3316366}
}

@unpublished{AnuragKSV2026,
  author = {Anurag, K. S. V. and Ashish Kumar Patra and Manas Mukherjee and Ruchika Bhat and Sai Shankar, P. and Rahul Maitra and Jaiganesh, G.},
  title = {Bridging the NISQ and Fault-Tolerant Regimes: Generative-ML-Assisted Quantum Selected CI for Molecular Simulations},
  note = {Manuscript in preparation},
  year = {2026}
}

@article{senicourt2022tangeloopensourcepythonpackage,
  author       = {Senicourt, Valentin and Brown, James and Fleury, Alexandre and Day, Ryan and Lloyd, Erika and Coons, Marc P. and Bieniasz, Krzysztof and Huntington, Lee and Garza, Alejandro J. and Matsuura, Shunji and Plesch, Rudi and Yamazaki, Takeshi and Zaribafiyan, Arman},
  title        = {Tangelo: An Open-Source Python Package for End-to-End Chemistry Workflows on Quantum Computers},
  journal      = {arXiv},
  year         = {2022},
  publisher    = {arXiv},
  doi          = {10.48550/arXiv.2206.12424},
  url          = {https://doi.org/10.48550/arXiv.2206.12424}
}

@manual{ffsim,
author = {{The ffsim developers}},
title = {{ffsim: Faster simulations of fermionic quantum circuits, v0.0.56}},
organization = {qiskit-community},
year = {2025},
url = {https://github.com/qiskit-community/ffsim},
note = {Accessed: 2025-11-11},
}

@article{javadiabhari2024quantumcomputingqiskit,
  author       = {Javadi-Abhari, Ali and Treinish, Matthew and Krsulich, Kevin and Wood, Christopher J. and Lishman, Jake and Gacon, Julien and Martiel, Simon and Nation, Paul D. and Bishop, Lev S. and Cross, Andrew W. and Johnson, Blake R. and Gambetta, Jay M.},
  title        = {Quantum Computing with Qiskit},
  journal      = {arXiv},
  year         = {2024},
  publisher    = {arXiv},
  doi          = {10.48550/arXiv.2405.08810},
  url          = {https://doi.org/10.48550/arXiv.2405.08810}
}

@article{wouters2016dmets_guide,
  author       = {Wouters, Sebastian and Jiménez-Hoyos, Carlos A. and Sun, Qiming and Chan, Garnet K.-L.},
  title        = {A Practical Guide to Density Matrix Embedding Theory in Quantum Chemistry},
  journal      = {Journal of Chemical Theory and Computation},
  volume       = {12},
  number       = {6},
  pages        = {2706--2719},
  year         = {2016},
  month        = jun,
  publisher    = {American Chemical Society},
  doi          = {10.1021/acs.jctc.6b00316},
  url          = {https://doi.org/10.1021/acs.jctc.6b00316}
}

@ARTICLE{bartlett_cc_theory_2007,
  author={Bartlett, Rodney J. and Musiał, Monika},
  journal={Reviews of Modern Physics}, 
  title={Coupled-Cluster Theory in Quantum Chemistry}, 
  year={2007},
  volume={79},
  number={1},
  pages={291--352},
  doi={10.1103/RevModPhys.79.291},
  url={https://link.aps.org/doi/10.1103/RevModPhys.79.291},
  publisher={American Physical Society}
}

@article{Kawashima2021_dmet_vqe_ion,
  author       = {Kawashima, Yukio and Lloyd, Erika and Coons, Marc P. and Nam, Yunseong and Matsuura, Shunji and Garza, Alejandro J. and Johri, Sonika and Huntington, Lee and Senicourt, Valentin and Maksymov, Andrii O. and Nguyen, Jason H. V. and Kim, Jungsang and Alidoust, Nima and Zaribafiyan, Arman and Yamazaki, Takeshi},
  title        = {Optimizing electronic structure simulations on a trapped-ion quantum computer using problem decomposition},
  journal      = {Communications Physics},
  year         = {2021},
  volume       = {4},
  number       = {1},
  pages        = {245},
  doi          = {10.1038/s42005-021-00751-9},
  url          = {https://doi.org/10.1038/s42005-021-00751-9}
}

@article{sun2017pyscf,
  author       = {Sun, Qiming and Berkelbach, Timothy C. and Blunt, Nick S. and Booth, George H. and Guo, Sheng and Li, Zhendong and Liu, Junzi and McClain, James and Sayfutyarova, Elvira R. and Sharma, Sandeep and Wouters, Sebastian and Chan, Garnet Kin-Lic},
  title        = {The Python-Based Simulations of Chemistry Framework (PySCF)},
  journal      = {arXiv},
  year         = {2017},
  publisher    = {arXiv},
  doi          = {10.48550/arXiv.1701.08223},
  url          = {https://doi.org/10.48550/arXiv.1701.08223}
}

@book{szabo1996_modernqc,
  title        = {Modern Quantum Chemistry: Introduction to Advanced Electronic Structure Theory},
  author       = {Szabo, A. and Ostlund, N. S.},
  publisher    = {Dover Publications},
  year         = {1996},
  isbn         = {9780486691862},
  series       = {Dover Books on Chemistry},
  url          = {https://books.google.co.in/books?id=6mV9gYzEkgIC},
  note         = {See Section 2.4.1, pp.\ 89--95},
}

@article{papakonstantinou2013secant,
  title        = {Origin and Evolution of the Secant Method in One Dimension},
  author       = {Papakonstantinou, Joanna M. and Tapia, Richard A.},
  journal      = {The American Mathematical Monthly},
  volume       = {120},
  number       = {6},
  pages        = {500--518},
  year         = {2013},
  publisher    = {Mathematical Association of America},
  doi          = {10.4169/amer.math.monthly.120.06.500},
  url          = {http://www.jstor.org/stable/10.4169/amer.math.monthly.120.06.500}
}

@article{Eisert2025_QuantumAdvantage,
  author       = {Eisert, Jens and Preskill, John},
  title        = {Mind the Gaps: The Fraught Road to Quantum Advantage},
  journal      = {arXiv},
  volume       = {arXiv:2510.19928},
  year         = {2025},
  publisher    = {arXiv},
  doi          = {10.48550/arXiv.2510.19928},
  url          = {https://doi.org/10.48550/arXiv.2510.19928}
}

@article{Preskill_2018,
   title={Quantum Computing in the NISQ era and beyond},
   volume={2},
   ISSN={2521-327X},
   url={http://dx.doi.org/10.22331/q-2018-08-06-79},
   DOI={10.22331/q-2018-08-06-79},
   journal={Quantum},
   publisher={Verein zur Forderung des Open Access Publizierens in den Quantenwissenschaften},
   author={Preskill, John},
   year={2018},
   month=aug, pages={79} }

@article{sci_evangelista_1,
    author = {Evangelista, Francesco A.},
    title = {Adaptive multiconfigurational wave functions},
    journal = {The Journal of Chemical Physics},
    volume = {140},
    number = {12},
    pages = {124114},
    year = {2014},
    month = {03},
    issn = {0021-9606},
    doi = {10.1063/1.4869192},
    url = {https://doi.org/10.1063/1.4869192},
}

@ARTICLE{2020SciPy-NMeth,
  author  = {Virtanen, Pauli and Gommers, Ralf and Oliphant, Travis E. and
            Haberland, Matt and Reddy, Tyler and Cournapeau, David and
            Burovski, Evgeni and Peterson, Pearu and Weckesser, Warren and
            Bright, Jonathan and {van der Walt}, St{\'e}fan J. and
            Brett, Matthew and Wilson, Joshua and Millman, K. Jarrod and
            Mayorov, Nikolay and Nelson, Andrew R. J. and Jones, Eric and
            Kern, Robert and Larson, Eric and Carey, C J and
            Polat, {\.I}lhan and Feng, Yu and Moore, Eric W. and
            {VanderPlas}, Jake and Laxalde, Denis and Perktold, Josef and
            Cimrman, Robert and Henriksen, Ian and Quintero, E. A. and
            Harris, Charles R. and Archibald, Anne M. and
            Ribeiro, Ant{\^o}nio H. and Pedregosa, Fabian and
            {van Mulbregt}, Paul and {SciPy 1.0 Contributors}},
  title   = {{{SciPy} 1.0: Fundamental Algorithms for Scientific
            Computing in Python}},
  journal = {Nature Methods},
  year    = {2020},
  volume  = {17},
  pages   = {261--272},
  adsurl  = {https://rdcu.be/b08Wh},
  doi     = {10.1038/s41592-019-0686-2},
}

@article{Jordan1928,
  author       = {Jordan, P. and Wigner, E.},
  title        = {{\"U}ber das Paulische {\"A}quivalenzverbot},
  journal      = {Zeitschrift f{\"u}r Physik},
  volume       = {47},
  number       = {9},
  pages        = {631--651},
  year         = {1928},
  month        = sep,
  doi          = {10.1007/BF01331938},
  url          = {https://doi.org/10.1007/BF01331938}
}

@BOOK{jensen2006introduction,
  author    = {Jensen, Frank},
  title     = {Introduction to Computational Chemistry},
  publisher = {John Wiley \& Sons, Inc.},
  address   = {Hoboken, NJ, USA},
  year      = {2006},
  isbn      = {978-0470011874},
  note      = {Second Edition, see Section 4.9, pp. 169--174}
}

@ARTICLE{Barison2025_QCExcitedStates_SQD,
  author    = {Barison, Stefano and Robledo Moreno, Javier and Motta, Mario},
  title     = {Quantum-Centric Computation of Molecular Excited States with Extended Sample-Based Quantum Diagonalization},
  journal   = {Quantum Science and Technology},
  volume    = {10},
  number    = {2},
  pages     = {025034},
  year      = {2025},
  month     = {February},
  doi       = {10.1088/2058-9565/adb781},
  url       = {https://doi.org/10.1088/2058-9565/adb781},
  publisher = {IOP Publishing},
  issn      = {2058-9565},
}

@article{hi_vqe_2024,
  author       = {Pellow-Jarman, Aidan and McFarthing, Shane and Kang, Doo Hyung and Yoo, Pilsun and Elala, Eyuel Eshetu and Pellow-Jarman, Rowan and Nakliang, P. Mai and Kim, Jaewan and Rhee, June-Koo Kevin},
  title        = {HIVQE: Handover Iterative Variational Quantum Eigensolver for Efficient Quantum Chemistry Calculations},
  journal      = {arXiv},
  year         = {2025},
  publisher    = {arXiv},
  doi          = {10.48550/arXiv.2503.06292},
  url          = {https://doi.org/10.48550/arXiv.2503.06292}
}

@article{superconducting_quantum_engineer,
    author = {Krantz, P. and Kjaergaard, M. and Yan, F. and Orlando, T. P. and Gustavsson, S. and Oliver, W. D.},
    title = {A quantum engineer's guide to superconducting qubits},
    journal = {Applied Physics Reviews},
    volume = {6},
    number = {2},
    pages = {021318},
    year = {2019},
    month = {06},
    issn = {1931-9401},
    doi = {10.1063/1.5089550},
    url = {https://doi.org/10.1063/1.5089550}
}

@article{piccinelli2025_sqdrift,
  author       = {Piccinelli, Samuele and Baiardi, Alberto and Rossmannek, Max and Carrera Vazquez, Almudena and Tacchino, Francesco and Mensa, Stefano and Altamura, Edoardo and Alavi, Ali and Motta, Mario and Robledo-Moreno, Javier and Kirby, William and Sharma, Kunal and Mezzacapo, Antonio and Tavernelli, Ivano},
  title        = {Quantum Chemistry with Provable Convergence via Randomized Sample-Based Quantum Diagonalization},
  journal      = {arXiv},
  year         = {2025},
  publisher    = {arXiv},
  doi          = {10.48550/arXiv.2508.02578},
  url          = {https://doi.org/10.48550/arXiv.2508.02578}
}

@ARTICLE{Matsuzawa2020_JastrowLowDepth,
  author    = {Matsuzawa, Yuta and Kurashige, Yuki},
  title     = {Jastrow-Type Decomposition in Quantum Chemistry for Low-Depth Quantum Circuits},
  journal   = {Journal of Chemical Theory and Computation},
  volume    = {16},
  number    = {2},
  pages     = {944--952},
  year      = {2020},
  month     = {February},
  day       = {11},
  doi       = {10.1021/acs.jctc.9b00963},
  url       = {https://doi.org/10.1021/acs.jctc.9b00963},
  publisher = {American Chemical Society},
  issn      = {1549-9618}
}

@article{Yu2025_SBKrylov,
  author       = {Yu, Jeffery and Robledo Moreno, Javier and Iosue, Joseph T. and Bertels, Luke and Claudino, Daniel and Fuller, Bryce and Groszkowski, Peter and Humble, Travis S. and Jurcevic, Petar and Kirby, William and Maier, Thomas A. and Motta, Mario and Pokharel, Bibek and Seif, Alireza and Shehata, Amir and Sung, Kevin J. and Tran, Minh C. and Tripathi, Vinay and Mezzacapo, Antonio and Sharma, Kunal},
  title        = {Quantum-Centric Algorithm for Sample-Based Krylov Diagonalization},
  journal      = {arXiv},
  year         = {2025},
  publisher    = {arXiv},
  doi          = {10.48550/arXiv.2501.09702},
  url          = {https://doi.org/10.48550/arXiv.2501.09702}
}

@ARTICLE{Alexeev2024_QCSMaterials_sqd,
  author    = {Alexeev, Yuri and Amsler, Maximilian and Barroca, Marco Antonio and Bassini, Sanzio and Battelle, Torey and Camps, Daan and Casanova, David and Choi, Young Jay and Chong, Frederic T. and Chung, Charles and Codella, Christopher and C{\'o}rcoles, Antonio D. and Cruise, James and Di Meglio, Alberto and Duran, Ivan and Eckl, Thomas and Economou, Sophia and Eidenbenz, Stephan and Elmegreen, Bruce and Fare, Clyde and Faro, Ismael and Sanz Fern{\'a}ndez, Cristina and Neumann Barros Ferreira, Rodrigo and Fuji, Keisuke and Fuller, Bryce and Gagliardi, Laura and Galli, Giulia and Glick, Jennifer R. and Gobbi, Isacco and Gokhale, Pranav and de la Puente Gonzalez, Salvador and Greiner, Johannes and Gropp, Bill and Grossi, Michele and Gull, Emanuel and Healy, Burns and Hermes, Matthew R. and Huang, Benchen and Humble, Travis S. and Ito, Nobuyasu and Izmaylov, Artur F. and Javadi-Abhari, Ali and Jennewein, Douglas and Jha, Shantenu and Jiang, Liang and Jones, Barbara and de Jong, Wibe Albert and Jurcevic, Petar and Kirby, William and Kister, Stefan and Kitagawa, Masahiro and Klassen, Joel and Klymko, Katherine and Koh, Kwangwon and Kondo, Masaaki and K{\"u}rk{\c{c}}uog{\u{l}}u, Do{\u{g}}a Murat and Kurowski, Krzysztof and Laino, Teodoro and Landfield, Ryan and Leininger, Matt and Leyton-Ortega, Vicente and Li, Ang and Lin, Meifeng and Liu, Junyu and Lorente, Nicolas and Luckow, Andre and Martiel, Simon and Martin-Fernandez, Francisco and Martonosi, Margaret and Marvinney, Claire and Castaneda Medina, Arcesio and Merten, Dirk and Mezzacapo, Antonio and Michielsen, Kristel and Mitra, Abhishek and Mittal, Tushar and Moon, Kyungsun and Moore, Joel and Mostame, Sarah and Motta, Mario and Na, Young-Hye and Nam, Yunseong and Narang, Prineha and Ohnishi, Yu-ya and Ottaviani, Daniele and Otten, Matthew and Pakin, Scott and Pascuzzi, Vincent R. and Pednault, Edwin and Piontek, Tomasz and Pitera, Jed and Rall, Patrick and Ravi, Gokul Subramanian and Robertson, Niall and Rossi, Matteo A.C. and Rydlichowski, Piotr and Ryu, Hoon and Samsonidze, Georgy and Sato, Mitsuhisa and Saurabh, Nishant and Sharma, Vidushi and Sharma, Kunal and Shin, Soyoung and Slessman, George and Steiner, Mathias and Sitdikov, Iskandar and Suh, In-Saeng and Switzer, Eric D. and Tang, Wei and Thompson, Joel and Todo, Synge and Tran, Minh C. and Trenev, Dimitar and Trott, Christian and Tseng, Huan-Hsin and Tubman, Norm M. and Tureci, Esin and Garc{\'i}a Vali{\~n}as, David and Vallecorsa, Sofia and Wever, Christopher and Wojciechowski, Konrad and Wu, Xiaodi and Yoo, Shinjae and Yoshioka, Nobuyuki and Yu, Victor Wen-zhe and Yunoki, Seiji and Zhuk, Sergiy and Zubarev, Dmitry},
  title     = {Quantum-centric supercomputing for materials science: A perspective on challenges and future directions},
  journal   = {Future Generation Computer Systems},
  year      = {2024},
  volume    = {160},
  pages     = {666--710},
  month     = {04},
  doi       = {10.1016/j.future.2024.04.060},
  url       = {https://doi.org/10.1016/j.future.2024.04.060},
  issn      = {0167-739X},
  publisher = {Elsevier}
}

@ARTICLE{MacDonald1933_RayleighRitz,
  author    = {MacDonald, J. K. L.},
  title     = {Successive Approximations by the Rayleigh–Ritz Variation Method},
  journal   = {Phys. Rev.},
  year      = {1933},
  volume    = {43},
  number    = {10},
  pages     = {830--833},
  month     = {May},
  doi       = {10.1103/PhysRev.43.830},
  url       = {https://link.aps.org/doi/10.1103/PhysRev.43.830},
  publisher = {American Physical Society}
}

@ARTICLE{AbuGhanem2025_heavy_hex_eagle,
  author = {AbuGhanem, Muhammad},
  title = {IBM Quantum Computers: Evolution, Performance, and Future Directions},
  journal = {The Journal of Supercomputing},
  volume = {81},
  number = {5},
  pages = {687--715},
  year = {2025},
  month = {4},
  day = {1},
  doi = {10.1007/s11227-025-07047-7},
  url = {https://doi.org/10.1007/s11227-025-07047-7},
  publisher = {Springer},
  issn = {1573-0484}
}

@ARTICLE{Knizia2012_DMET,
  author    = {Knizia, Gerald and Chan, Garnet Kin-Lic},
  title     = {Density Matrix Embedding: A Simple Alternative to Dynamical Mean-Field Theory},
  journal   = {Physical Review Letters},
  volume    = {109},
  number    = {18},
  pages     = {186404},
  year      = {2012},
  month     = {November},
  doi       = {10.1103/PhysRevLett.109.186404},
  url       = {https://doi.org/10.1103/PhysRevLett.109.186404},
  publisher = {American Physical Society},
  issn      = {0031-9007}
}

@ARTICLE{Georges1996_DMFT,
  author    = {Georges, Antoine and Kotliar, Gabriel and Krauth, Werner and Rozenberg, Marcelo J.},
  title     = {Dynamical Mean-Field Theory of Strongly Correlated Fermion Systems and the Limit of Infinite Dimensions},
  journal   = {Reviews of Modern Physics},
  volume    = {68},
  number    = {1},
  pages     = {13--125},
  year      = {1996},
  month     = {January},
  doi       = {10.1103/RevModPhys.68.13},
  url       = {https://doi.org/10.1103/RevModPhys.68.13},
  publisher = {American Physical Society},
  issn      = {0034-6861}
}

@ARTICLE{Hehre1969_GaussianSTO,
  author    = {Hehre, W. J. and Stewart, R. F. and Pople, J. A.},
  title     = {Self-Consistent Molecular-Orbital Methods. I. Use of Gaussian Expansions of Slater-Type Atomic Orbitals},
  journal   = {The Journal of Chemical Physics},
  volume    = {51},
  number    = {6},
  pages     = {2657--2664},
  year      = {1969},
  month     = {September},
  doi       = {10.1063/1.1672392},
  url       = {https://doi.org/10.1063/1.1672392},
  publisher = {AIP Publishing},
  issn      = {0021-9606}
}

@ARTICLE{Davidson1975_iterative_eigensolver,
  author    = {Davidson, Ernest R.},
  title     = {The iterative calculation of a few of the lowest eigenvalues and corresponding eigenvectors of large real-symmetric matrices},
  journal   = {Journal of Computational Physics},
  year      = {1975},
  volume    = {17},
  number    = {1},
  pages     = {87--94},
  doi       = {10.1016/0021-9991(75)90065-0},
  url       = {https://doi.org/10.1016/0021-9991(75)90065-0},
  issn      = {0021-9991},
  publisher = {Elsevier}
}

@inbook{Levine2014Ch16,
  author       = {Levine, Ira N.},
  title        = {Molecular Orbital Theory},
  booktitle    = {Quantum Chemistry},
  chapter      = {16},
  pages        = {653--710},
  year         = {2014},
  publisher    = {Pearson Education},
  address      = {Boston},
  edition      = {7},
  isbn         = {978-0-321-80345-0}
}

@ARTICLE{Tilly2022_VQE_Review,
  author    = {Tilly, Jules and Chen, Hongxiang and Cao, Shuxiang and Picozzi, Dario and Setia, Kanav and Li, Ying and Grant, Edward and Wossnig, Leonard and Rungger, Ivan and Booth, George H. and Tennyson, Jonathan},
  title     = {The Variational Quantum Eigensolver: A Review of Methods and Best Practices},
  journal   = {Physics Reports},
  year      = {2022},
  volume    = {986},
  pages     = {1--128},
  month     = {November},
  doi       = {10.1016/j.physrep.2022.08.003},
  url       = {https://doi.org/10.1016/j.physrep.2022.08.003},
  publisher = {Elsevier BV},
  issn      = {0370-1573}
}

@software{RDKit_2025_09_1,
  title        = {RDKit: Open‐source cheminformatics software(2025)},
  author       = {Greg Landrum and Paolo Tosco and Brian Kelley and Ricardo Rodriguez and David Cosgrove and Riccardo Vianello and Eisuke Kawashima and Nadine Schneider and Dan Nealschneider and Andrew Dalke and Matt Swain and Samo Turk and Aleksandr Savelev and Alain Vaucher and Maciej Wójcikowski and Hussein Faara and Ichiru Take and Niels Maeder and Rachel Walker and Vincent F. Scalfani and Daniel Probst and Kazuya Ujihara and Axel Pahl and Guillaume Godin and Juuso Lehtivarjo},
  version      = {2025_09_1},
  date         = {2025-09-30},
  publisher    = {Zenodo},
  doi          = {10.5281/zenodo.17232453},
  url          = {https://doi.org/10.5281/zenodo.17232453}
}

@ARTICLE{Caimi2017_Methoxyamine_Fludarabine,
  author    = {Caimi, Paolo F. and Cooper, Brenda W. and William, Basem M. and Dowlati, Afshin and Barr, Paul M. and Fu, Pingfu and Pink, John and Xu, Yan and Lazarus, Hillard M. and de Lima, Marcos and Gerson, Stanton L.},
  title     = {Phase I Clinical Trial of the Base Excision Repair Inhibitor Methoxyamine in Combination with Fludarabine for Patients with Advanced Hematologic Malignancies},
  journal   = {Oncotarget},
  year      = {2017},
  volume    = {8},
  number    = {45},
  pages     = {79864--79875},
  month     = {October},
  doi       = {10.18632/oncotarget.20094},
  url       = {https://doi.org/10.18632/oncotarget.20094},
  publisher = {Impact Journals},
  issn      = {1949-2553},
  pmid      = {29108368}
}

@ARTICLE{Eads2021_Temozolomide_Methoxyamine,
  author    = {Eads, Jennifer R. and Krishnamurthi, Smitha S. and Saltzman, Joel and Bokar, Joseph A. and Savvides, Panos and Meropol, Neal J. and Gibbons, Joseph and Koon, Henry and Sharma, Neelesh and Rogers, Lisa and Pink, John J. and Xu, Yan and Beumer, Jan H. and Riendeau, John and Fu, Pingfu and Gerson, Stanton L. and Dowlati, Afshin},
  title     = {Phase I Clinical Trial of Temozolomide and Methoxyamine (TRC-102), an Inhibitor of Base Excision Repair, in Patients with Advanced Solid Tumors},
  journal   = {Investigational New Drugs},
  year      = {2021},
  volume    = {39},
  number    = {1},
  pages     = {142--151},
  month     = {February},
  doi       = {10.1007/s10637-020-00962-x},
  url       = {https://doi.org/10.1007/s10637-020-00962-x},
  publisher = {Springer},
  issn      = {1573-0646}
}

@ARTICLE{Ghosh2020_Urea_DrugDiscovery,
  author    = {Ghosh, Arun K. and Brindisi, Margherita},
  title     = {Urea Derivatives in Modern Drug Discovery and Medicinal Chemistry},
  journal   = {Journal of Medicinal Chemistry},
  year      = {2020},
  volume    = {63},
  number    = {6},
  pages     = {2751--2788},
  month     = {March},
  doi       = {10.1021/acs.jmedchem.9b01541},
  url       = {https://doi.org/10.1021/acs.jmedchem.9b01541},
  publisher = {American Chemical Society},
  issn      = {0022-2623},
  pmid      = {31789518}
}

@ARTICLE{Listro2022_UreaAnticancerReview,
  author    = {Listro, Roberta and Rossino, Giacomo and Piaggi, Federica and Sonekan, Falilat Folasade and Rossi, Daniela and Linciano, Pasquale and Collina, Simona},
  title     = {Urea-Based Anticancer Agents: Exploring 100 Years of Research with an Eye to the Future},
  journal   = {Frontiers in Chemistry},
  year      = {2022},
  volume    = {10},
  doi       = {10.3389/fchem.2022.995351},
  url       = {https://doi.org/10.3389/fchem.2022.995351},
  publisher = {Frontiers Media SA},
  issn      = {2296-2646}
}

@article{Li2019SABRE,
  author       = {Li, Gushu and Ding, Yufei and Xie, Yuan},
  title        = {Tackling the Qubit Mapping Problem for NISQ-Era Quantum Devices},
  journal      = {arXiv},
  year         = {2019},
  publisher    = {arXiv},
  doi          = {10.48550/arXiv.1809.02573},
  url          = {https://doi.org/10.48550/arXiv.1809.02573}
}

@article{Campbell2019RandomCompiler,
  title        = {Random Compiler for Fast Hamiltonian Simulation},
  author       = {Campbell, Earl},
  journal      = {Physical Review Letters},
  volume       = {123},
  number       = {7},
  pages        = {070503},
  year         = {2019},
  month        = aug,
  doi          = {10.1103/PhysRevLett.123.070503},
  publisher    = {American Physical Society}
}

@article{Yoshioka2025KrylovDiagonalization,
  title        = {Krylov Diagonalization of Large Many-Body Hamiltonians on a Quantum Processor},
  author       = {Yoshioka, Nobuyuki and Amico, Mirko and Kirby, William and Jurcevic, Petar and Dutt, Arkopal and Fuller, Bryce and Garion, Shelly and Haas, Holger and Hamamura, Ikko and Ivrii, Alexander and Majumdar, Ritajit and Minev, Zlatko and Motta, Mario and Pokharel, Bibek and Rivero, Pedro and Sharma, Kunal and Wood, Christopher J. and Javadi-Abhari, Ali and Mezzacapo, Antonio},
  journal      = {Nature Communications},
  volume       = {16},
  number       = {1},
  pages        = {5014},
  year         = {2025},
  month        = jun,
  doi          = {10.1038/s41467-025-59716-z},
  issn         = {2041-1723},
  url          = {https://doi.org/10.1038/s41467-025-59716-z}
}

@article{aspuruguzik2005_simulated,
  author       = {Aspuru-Guzik, Al{\'a}n and Dutoi, Austin D. and Love, Peter J. and Head-Gordon, Martin},
  title        = {Simulated quantum computation of molecular energies},
  journal      = {Science},
  year         = {2005},
  volume       = {309},
  pages        = {1704--1707},
  doi          = {10.1126/science.1113479},
}

@article{cao2019_quantumchemistry,
  author       = {Cao, Yudong and Romero, Jonathan and Olson, Jonathan P. and Degroote, Matthias and Johnson, Peter D. and Kieferov{\'a}, M{\'a}ria and Kivlichan, Ian D. and Menke, Tim and Peropadre, Borja and Sawaya, Nicolas P. D. and Sim, Sukin and Veis, Libor and Aspuru-Guzik, Al{\'a}n},
  title        = {Quantum chemistry in the age of quantum computing},
  journal      = {Chemical Reviews},
  year         = {2019},
  volume       = {119},
  pages        = {10856--10915},
  doi          = {10.1021/acs.chemrev.8b00803},
}

@article{schuch2009_qmahard,
  author       = {Schuch, Norbert and Verstraete, Frank},
  title        = {Computational complexity of interacting electrons and fundamental limitations of density functional theory},
  journal      = {Nature Physics},
  year         = {2009},
  volume       = {5},
  pages        = {732--735},
  doi          = {10.1038/nphys1370},
}

@ARTICLE{born_oppie_1927,
       author = {{Born}, M. and {Oppenheimer}, R.},
        title = "{Zur Quantentheorie der Molekeln}",
      journal = {Annalen der Physik},
         year = 1927,
        month = jan,
       volume = {389},
       number = {20},
        pages = {457-484},
          doi = {10.1002/andp.19273892002},
       adsurl = {https://ui.adsabs.harvard.edu/abs/1927AnP...389..457B}
}

@article{ransford2025helios,
  author       = {Ransford, Anthony and Allman, M. S. and Arkinstall, Jake and Campora III, J. P. and Cooper, Samuel F. and Delaney, Robert D. and Dreiling, Joan M. and Estey, Brian and Figgatt, Caroline and Hall, Alex and Husain, Ali A. and Isanaka, Akhil and Kennedy, Colin J. and Kotibhaskar, Nikhil and Madjarov, Ivaylo S. and Mayer, Karl and Milne, Alistair R. and Park, Annie J. and Reed, Adam P. and Ancona, Riley and Andersen, Molly P. and Andres-Martinez, Pablo and Angenent, Will and Argueta, Liz and Arkin, Benjamin and Ascarrunz, Leonardo and Baker, William and Barnes, Corey and Bartolotta, John and Berg, Jordan and Besand, Ryan and Bjork, Bryce and Blain, Matt and Blanchard, Paul and Blume-Kohout, Robin and Bohn, Matt and Borgna, Agustin and Botamanenko, Daniel Y. and Boutelle, Robert and Brown, Natalie and Buckingham, Grant T. and Burdick, Nathaniel Q. and Burton, William Cody and Carey, Varis and Carron, Christopher J. and Chambers, Joe and Children, John and Colussi, Victor E. and Crepinsek, Steven and Cureton, Andrew and Davies, Joe and Davis, Daniel and DeCross, Matthew and Deen, David and Delaney, Conor and DelVento, Davide and DeSalvo, B. J. and Dominy, Jason and Duncan, Ross and Eccles, Vanya and Edgington, Alec and Erickson, Neal and Erickson, Stephen and Ertsgaard, Christopher T. and Evans, Bruce and Evans, Tyler and Fabrikant, Maya I. and Fischer, Andrew and Foltz, Cameron and Foss-Feig, Michael and Francois, David and Freyberg, Brad and Gao, Charles and Garay, Robert and Garvin, Jane and Gaudiosi, David M. and Gilbreth, Christopher N. and Giles, Josh and Glynn, Erin and Graves, Jeff and Hansen, Azure and Hayes, David and Heidemann, Lukas and Higashi, Bob and Hilbun, Tyler and Hines, Jordan and Hlavaty, Ariana and Hoffman, Kyle and Hoffman, Ian M. and Holliman, Craig and Hooper, Isobel and Horning, Bob and Hostetter, James and Hothem, Daniel and Houlton, Jack and Hout, Jared and Hutson, Ross and Jacobs, Ryan T. and Jacobs, Trent and Johannsen, Melf and Johansen, Jacob and Jones, Loren and Julian, Sydney and Jung, Ryan and Keay, Aidan and Klein, Todd and Koch, Mark and Kondo, Ryo and Kong, Chang and Kosto, Asa and Lawrence, Alan and Liefer, David and Lollie, Michelle and Lucchetti, Dominic and Lysne, Nathan K. and Lytle, Christian and MacPherson, Callum and Malm, Andrew and Mather, Spencer and Mathewson, Brian and Maxwell, Daniel and McCaffrey, Lauren and McDougall, Hannah and Mendoza, Robin and Mills, Michael and Morrison, Richard and Narmour, Louis and Nguyen, Nhung and Nugent, Lora and Olson, Scott and Ouellette, Daniel and Parks, Jeremy and Peters, Zach and Petricka, Jessie and Pino, Juan M. and Polito, Frank and Preidl, Matthias and Price, Gabriel and Proctor, Timothy and Pugh, McKinley and Ratcliff, Noah and Raymondson, Daisy and Rhodes, Peter and Roman, Conrad and Roy, Craig and Ryan-Anderson, Ciaran and Sanchez, Fernando Betanzo and Sangiolo, George and Sawadski, Tatiana and Schaffer, Andrew and Schow, Peter and Sedlacek, Jon and Semenenko, Henry and Shevchuk, Peter and Shore, Susan and Siegfried, Peter and Singhal, Kartik and Sivarajah, Seyon and Skripka, Thomas and Sletten, Lucas and Spaun, Ben and Sprenkle, R. Tucker and Stoufer, Paul and Tader, Mariel and Taylor, Stephen F. and Thompson, Travis H. and Tobey, Raanan and Tran, Anh and Tran, Tam and Vittorini, Grahame and Volin, Curtis and Walker, Jim and White, Sam and Wilson, Douglas and Wolf, Quinn and Wringe, Chester and Young, Kevin and Zheng, Jian and Zuraski, Kristen and Baldwin, Charles H. and Chernoguzov, Alex and Gaebler, John P. and Sanders, Steven J. and Neyenhuis, Brian and Stutz, Russell and Bohnet, Justin G.},
  title        = {Helios: A 98-qubit trapped-ion quantum computer},
  journal      = {arXiv},
  year         = {2025},
  publisher    = {arXiv},
  doi          = {10.48550/arXiv.2511.05465},
  url          = {https://doi.org/10.48550/arXiv.2511.05465}
}

@article{Evered2023,
  author    = {Evered, Simon J. and Bluvstein, Dolev and Kalinowski, Marcin and Ebadi, Sepehr
               and Manovitz, Tom and Zhou, Hengyun and Li, Sophie H. and Geim, Alexandra A.
               and Wang, Tout T. and Maskara, Nishad and Levine, Harry and Semeghini, Giulia
               and Greiner, Markus and Vuleti{\'c}, Vladan and Lukin, Mikhail D.},
  title     = {High-fidelity parallel entangling gates on a neutral-atom quantum computer},
  journal   = {Nature},
  volume    = {622},
  number    = {7982},
  pages     = {268--272},
  year      = {2023},
  doi       = {10.1038/s41586-023-06481-y},
  url       = {https://doi.org/10.1038/s41586-023-06481-y},
  issn      = {1476-4687}
}

@article{marxer2025_999_IQM_superconducting,
  author       = {Marxer, Fabian and Mro{\.z}ek, Jakub and Andersson, Joona and Abdurakhimov, Leonid and Adam, Janos and Bergholm, Ville and Beriwal, Rohit and Chan, Chun Fai and Dahl, Saga and Das, Soumya Ranjan and Deppe, Frank and Fedorets, Olexiy and Gao, Zheming and Gomez Frieiro, Alejandro and Gusenkova, Daria and Guthrie, Andrew and Hiltunen, Tuukka and Hsu, Hao and Hyypp{\"a}, Eric and Ikonen, Joni and Inel, Sinan and Jolin, Shan W. and Karis, Azad and Kim, Seung-Goo and Kindel, William and Komlev, Anton and Koistinen, Miikka and Kokkoniemi, Roope and Kumar, Snigdha and Ku, Hsiang-Sheng and Lamprich, Julia and Laine, Sami and Landra, Alessandro and Lee, Lan-Hsuan and Lethif, Nizar and Liebermann, Per and Liu, Wei and Mitra, Kunal and Myll{\"a}ri, Tuomas and Ockeloen-Korppi, Caspar and Orell, Tuure and Plyshch, Alexander and R{\"a}bin{\"a}, Jukka and Rebello, Arthur and Renger, Michael and Reentil{\"a}, Outi and Ritvas, Jussi and Saarinen, Sampo and Salmenkivi, Otto and Sarsby, Matthew and Savytskyi, Mykhailo and Selinmaa, Ville and Steggles, Matthew and Takala, Eelis and Takmakov, Ivan and Tarasinski, Brian and Tuorila, Jani and V{\"a}lim{\"a}{\"a}, Alpo and Verjauw, Jeroen and Wesdorp, Jaap and Wurz, Nicola and Qiu, Wei and Zhu, Lihuang and Hassel, Juha and Heinsoo, Johannes and Geresdi, Attila and Veps{\"a}l{\"a}inen, Antti},
  title        = {Above 99.9\% Fidelity Single-Qubit Gates, Two-Qubit Gates, and Readout in a Single Superconducting Quantum Device},
  journal      = {arXiv},
  year         = {2025},
  publisher    = {arXiv},
  doi          = {10.48550/arXiv.2508.16437},
  url          = {https://doi.org/10.48550/arXiv.2508.16437}
}

@article{du2025rigetti,
  author       = {Du, Zefan and Chumpitaz Flores, Pedro and Wei, Wenqi and Chen, Juntao and Hua, Kaixun and Mao, Ying},
  title        = {Optimizing Inter-chip Coupler Link Placement for Modular and Chiplet Quantum Systems},
  journal      = {arXiv},
  year         = {2025},
  publisher    = {arXiv},
  doi          = {10.48550/arXiv.2509.10409},
  url          = {https://doi.org/10.48550/arXiv.2509.10409}
}

@misc{ibmquantum_heron_boston,
  author       = {{IBM Quantum}},
  title        = {IBM Quantum — Heron \& IBM Boston Processor Listings},
  howpublished = {\url{https://quantum.cloud.ibm.com/computers?processorType=Heron&system=ibm_boston}},
  note         = {Accessed: 28 January 2026},
  year         = {2026}
}

@article{hughes2025ionq,
  author       = {Hughes, A. C. and Srinivas, R. and L{\"o}schnauer, C. M. and Knaack, H. M. and Matt, R. and Ballance, C. J. and Malinowski, M. and Harty, T. P. and Sutherland, R. T.},
  title        = {Trapped-ion two-qubit gates with >99.99\% fidelity without ground-state cooling},
  journal      = {arXiv},
  year         = {2025},
  publisher    = {arXiv},
  doi          = {10.48550/arXiv.2510.17286},
  url          = {https://doi.org/10.48550/arXiv.2510.17286}
}

@Article{sapt_pro_lig,
author ="Malone, Fionn D. and Parrish, Robert M. and Welden, Alicia R. and Fox, Thomas and Degroote, Matthias and Kyoseva, Elica and Moll, Nikolaj and Santagati, Raffaele and Streif, Michael",
title  ="Towards the simulation of large scale protein–ligand interactions on NISQ-era quantum computers",
journal  ="Chem. Sci.",
year  ="2022",
volume  ="13",
issue  ="11",
pages  ="3094-3108",
publisher  ="The Royal Society of Chemistry",
doi  ="10.1039/D1SC05691C",
url  ="http://dx.doi.org/10.1039/D1SC05691C"}

@article{Bowling2025_ProteinLigand_FragQC,
  author  = {Bowling, Paige E. and Broderick, Dustin R. and Herbert, John M.},
  title   = {Convergent Protocols for Computing Protein--Ligand Interaction Energies Using Fragment-Based Quantum Chemistry},
  journal = {Journal of Chemical Theory and Computation},
  year    = {2025},
  volume  = {21},
  number  = {2},
  pages   = {951--966},
  doi     = {10.1021/acs.jctc.4c01429},
  url     = {https://doi.org/10.1021/acs.jctc.4c01429},
  issn    = {1549-9618},
  publisher = {American Chemical Society}
}

@article{Anurag2025_ButanePES,
  author       = {Anurag, K. S. V. and Patra, Ashish Kumar and Anand, Chinmay and Ghevade, Vikas Dattatraya and Raghavendra, V. and Bhat, Ruchika and Jaiganesh, G},
  title        = {Potential Energy Surface Scan of \emph{n}-butane Using Various Quantum Chemistry Software},
  journal      = {Authorea},
  year         = {2025},
  month        = {July},
  doi          = {10.22541/au.174624847.78353339/v2},
  url          = {https://doi.org/10.22541/au.174624847.78353339/v2}
}

@article{scalable_qc_qpe_vqe,
  title = {Scalable Quantum Simulation of Molecular Energies},
  author = {O'Malley, P. J. J. and Babbush, R. and Kivlichan, I. D. and Romero, J. and McClean, J. R. and Barends, R. and Kelly, J. and Roushan, P. and Tranter, A. and Ding, N. and Campbell, B. and Chen, Y. and Chen, Z. and Chiaro, B. and Dunsworth, A. and Fowler, A. G. and Jeffrey, E. and Lucero, E. and Megrant, A. and Mutus, J. Y. and Neeley, M. and Neill, C. and Quintana, C. and Sank, D. and Vainsencher, A. and Wenner, J. and White, T. C. and Coveney, P. V. and Love, P. J. and Neven, H. and Aspuru-Guzik, A. and Martinis, J. M.},
  journal = {Phys. Rev. X},
  volume = {6},
  issue = {3},
  pages = {031007},
  numpages = {13},
  year = {2016},
  month = {Jul},
  publisher = {American Physical Society},
  doi = {10.1103/PhysRevX.6.031007},
  url = {https://link.aps.org/doi/10.1103/PhysRevX.6.031007}
}

@Article{pes_scan_qchem_ml,
author ="Mancini, Giordano and Fusè, Marco and Lazzari, Federico and Barone, Vincenzo",
title  ="Fast exploration of potential energy surfaces with a joint venture of quantum chemistry{,} evolutionary algorithms and unsupervised learning",
journal  ="Digital Discovery",
year  ="2022",
volume  ="1",
issue  ="6",
pages  ="790-805",
publisher  ="RSC",
doi  ="10.1039/D2DD00070A",
url  ="http://dx.doi.org/10.1039/D2DD00070A"}

@Article{barton_1956_pes,
author ="Barton, D. H. R. and Cookson, R. C.",
title  ="The principles of conformational analysis",
journal  ="Q. Rev. Chem. Soc.",
year  ="1956",
volume  ="10",
issue  ="1",
pages  ="44-82",
publisher  ="The Royal Society of Chemistry",
doi  ="10.1039/QR9561000044",
url  ="http://dx.doi.org/10.1039/QR9561000044"}

@inbook{horn2012matrix,
title={Matrix Analysis},
author={Horn, R.A. and Johnson, C.R.},
chapter={4.3},
pages={242--248},
edition={2nd},
isbn={9781139788885},
url={https://books.google.co.in/books?id=O7sgAwAAQBAJ},
year={2012},
publisher={Cambridge University Press}
}

@Article{x2c_TZV_basis_set,
author ="Franzke, Yannick J. and Treß, Robert and Pazdera, Tobias M. and Weigend, Florian",
title  ="Error-consistent segmented contracted all-electron relativistic basis sets of double- and triple-zeta quality for NMR shielding constants",
journal  ="Phys. Chem. Chem. Phys.",
year  ="2019",
volume  ="21",
issue  ="30",
pages  ="16658-16664",
publisher  ="The Royal Society of Chemistry",
doi  ="10.1039/C9CP02382H",
url  ="http://dx.doi.org/10.1039/C9CP02382H"}

@article{Shayit2025_HBrTe,
  author  = {Shayit, Agam and Liao, Can and Upadhyay, Shiv and Hu, Hang and Zhang, Tianyuan and DePrince III, A. Eugene and Yang, Chao and Li, Xiaosong},
  title   = {Numerically exact configuration interaction at quadrillion-determinant scale},
  journal = {Nature Communications},
  year    = {2025},
  volume  = {16},
  number  = {1},
  pages   = {11016},
  issn    = {2041-1723},
  doi     = {10.1038/s41467-025-65967-7},
  url     = {https://doi.org/10.1038/s41467-025-65967-7}
}

@article{cho2025quemb,
  author = {Cho, Minsik and Meitei, Oinam Romesh and Weisburn, Leah P. and Weser, Oskar and Weatherly, Shaun and Alexiu, Alexandra and Hanscam, Rebecca and Tran, Henry K. and Ye, Hong-Zhou and Welborn, Matthew and Ricke, Nathan and Tsuchimochi, Takashi and Trofimov, Aleksandr and Orkhon, Temujin and Whelpley, Noah and Luo, Carina and Van Voorhis, Troy},
  title = {QuEmb: A Toolbox for Bootstrap Embedding Calculations of Molecular and Periodic Systems},
  journal = {J. Phys. Chem. A},
  year = {2025},
  month = {July},
  volume = {129},
  number = {28},
  pages = {6538--6551},
  doi = {10.1021/acs.jpca.5c02983},
  url = {https://doi.org/10.1021/acs.jpca.5c02983},
  publisher = {American Chemical Society}
}

@article{ye2020bootstrap,
  author = {Ye, Hong-Zhou and Tran, Henry K. and Van Voorhis, Troy},
  title = {Bootstrap Embedding For Large Molecular Systems},
  journal = {J. Chem. Theory Comput.},
  year = {2020},
  month = {June},
  volume = {16},
  number = {8},
  pages = {5035--5046},
  doi = {10.1021/acs.jctc.0c00438},
  url = {https://doi.org/10.1021/acs.jctc.0c00438},
  publisher = {American Chemical Society}
}

@article{Liu2023Bootstrap,
  author  = {Liu, Yuan and Meitei, Oinam R. and Chin, Zachary E. and Dutt, Arkopal and Tao, Max and Chuang, Isaac L. and Van Voorhis, Troy},
  title   = {Bootstrap Embedding on a Quantum Computer},
  journal = {Journal of Chemical Theory and Computation},
  year    = {2023},
  volume  = {19},
  number  = {8},
  pages   = {2230--2247},
  publisher = {American Chemical Society},
  issn    = {1549-9618},
  doi     = {10.1021/acs.jctc.3c00012},
  url     = {https://doi.org/10.1021/acs.jctc.3c00012}
}

@inproceedings{Hardikar2024,
author = {Hardikar, Tarini and Heitritter, Kenneth and Brown, James and D'Cunha, Ruhee and Mitra, Abhishek and Weatherly, Shaun and Liu, Yuan and Otten, Matthew and Voorhis, Troy and Gagliardi, Laura and Setia, Kanav},
year = {},
month = {09},
booktitle = {2024 IEEE International Conference on Quantum Computing and Engineering (QCE)},
pages = {538-544 (2024)},
title = {Quanta-Bind: A Quantum Computing Pipeline for Modeling Strongly Correlated Metal-Protein Interactions},
doi = {10.1109/QCE60285.2024.00069}
}

@article{Otten2022Localized_laura,
  author    = {Otten, Matthew and Hermes, Matthew R. and Pandharkar, Riddhish and Alexeev, Yuri and Gray, Stephen K. and Gagliardi, Laura},
  title     = {Localized Quantum Chemistry on Quantum Computers},
  journal   = {Journal of Chemical Theory and Computation},
  year      = {2022},
  volume    = {18},
  number    = {12},
  pages     = {7205--7217},
  publisher = {American Chemical Society},
  issn      = {1549-9618},
  doi       = {10.1021/acs.jctc.2c00388},
  url       = {https://doi.org/10.1021/acs.jctc.2c00388}
}

@article{Pham2018DMET_laura,
  author    = {Pham, Hung Q. and Bernales, Varinia and Gagliardi, Laura},
  title     = {Can Density Matrix Embedding Theory with the Complete Active Space Self-Consistent Field Solver Describe Single and Double Bond Breaking in Molecular Systems?},
  journal   = {Journal of Chemical Theory and Computation},
  year      = {2018},
  volume    = {14},
  number    = {4},
  pages     = {1960--1968},
  publisher = {American Chemical Society},
  issn      = {1549-9618},
  doi       = {10.1021/acs.jctc.7b01248},
  url       = {https://doi.org/10.1021/acs.jctc.7b01248}
}

@article{Verma2025DMETPDFT_laura,
  author    = {Verma, Shreya and Hermes, Matthew R. and Gagliardi, Laura},
  title     = {Density Matrix Embedding Pair-Density Functional Theory for Molecules},
  journal   = {The Journal of Physical Chemistry Letters},
  year      = {2025},
  volume    = {16},
  number    = {21},
  pages     = {5348--5357},
  publisher = {American Chemical Society},
  doi       = {10.1021/acs.jpclett.5c00829},
  url       = {https://doi.org/10.1021/acs.jpclett.5c00829}
}

@article{kowalski_cc_downfolding,
    author = {Bauman, Nicholas P. and Kowalski, Karol},
    title = {Coupled cluster downfolding methods: The effect of double commutator terms on the accuracy of ground-state energies},
    journal = {The Journal of Chemical Physics},
    volume = {156},
    number = {9},
    pages = {094106},
    year = {2022},
    month = {03},
    issn = {0021-9606},
    doi = {10.1063/5.0076260},
    url = {https://doi.org/10.1063/5.0076260},
}

@article{bauman2025coupledclusterdownfoldingtheory,
  author       = {Bauman, Nicholas P. and Zheng, Muqing and Liu, Chenxu and Myers, Nathan M. and Panyala, Ajay and Peng, Bo and Li, Ang and Kowalski, Karol},
  title        = {Coupled Cluster Downfolding Theory in Simulations of Chemical Systems on Quantum Hardware},
  journal      = {arXiv},
  year         = {2025},
  publisher    = {arXiv},
  doi          = {10.48550/arXiv.2507.01199},
  url          = {https://doi.org/10.48550/arXiv.2507.01199}
}

@article{Nakai2023DC,
  title   = {Divide-and-Conquer Linear-Scaling Quantum Chemical Computations},
  author  = {Nakai, Hiromi and Kobayashi, Masato and Yoshikawa, Takeshi and Seino, Junji and Ikabata, Yasuhiro and Nishimura, Yoshifumi},
  journal = {The Journal of Physical Chemistry A},
  volume  = {127},
  number  = {3},
  pages   = {589--618},
  year    = {2023},
  doi     = {10.1021/acs.jpca.2c06965},
  pmid    = {36630608},
}

@article{Hermes2020VLASSCF,
  title   = {Variational Localized Active Space Self-Consistent Field Method},
  author  = {Hermes, Matthew R. and Pandharkar, Riddhish and Gagliardi, Laura},
  journal = {Journal of Chemical Theory and Computation},
  volume  = {16},
  number  = {8},
  pages   = {4923--4937},
  year    = {2020},
  doi     = {10.1021/acs.jctc.0c00222},
}

@article{ewf_embedding_2022,
  title = {Systematic Improvability in Quantum Embedding for Real Materials},
  author = {Nusspickel, Max and Booth, George H.},
  journal = {Phys. Rev. X},
  volume = {12},
  issue = {1},
  pages = {011046},
  numpages = {15},
  year = {2022},
  month = {Mar},
  publisher = {American Physical Society},
  doi = {10.1103/PhysRevX.12.011046},
  url = {https://link.aps.org/doi/10.1103/PhysRevX.12.011046}
}

@article{blumenthal2025_dmft_intuition,
  author       = {Blumenthal, Emmy},
  title        = {Building Intuition for Dynamical Mean-Field Theory: A Simple Model and the Cavity Method},
  journal      = {arXiv},
  year         = {2025},
  publisher    = {arXiv},
  doi          = {10.48550/arXiv.2507.16654},
  url          = {https://doi.org/10.48550/arXiv.2507.16654}
}

@article{Xu2024MBE_VQE_Deflation,
  title   = {Many-Body-Expansion Based on Variational Quantum Eigensolver and Deflation for Dynamical Correlation},
  author  = {Xu, Enhua and Shimomoto, Yuma and Ten-no, Seiichiro L. and Tsuchimochi, Takashi},
  journal = {The Journal of Physical Chemistry A},
  year    = {2024},
  volume  = {128},
  number  = {12},
  pages   = {2507--2521},
  doi     = {10.1021/acs.jpca.4c00351},
  publisher = {American Chemical Society}
}

@Article{fmo_fedorov_2012,
author ="Fedorov, Dmitri G. and Nagata, Takeshi and Kitaura, Kazuo",
title  ="Exploring chemistry with the fragment molecular orbital method",
journal  ="Phys. Chem. Chem. Phys.",
year  ="2012",
volume  ="14",
issue  ="21",
pages  ="7562-7577",
publisher  ="The Royal Society of Chemistry",
doi  ="10.1039/C2CP23784A",
url  ="http://dx.doi.org/10.1039/C2CP23784A"}

@article{FMO_VQE_Lim_2024,
  title   = {Fragment Molecular Orbital-Based Variational Quantum Eigensolver for Quantum Chemistry in the Age of Quantum Computing},
  author  = {Lim, Hocheol and Kang, Doo Hyung and Kim, Jeonghoon and Pellow-Jarman, Aidan and McFarthing, Shane and Pellow-Jarman, Rowan and Jeon, Hyeon-Nae and Oh, Byungdu and Rhee, June-Koo Kevin and No, Kyoung Tai},
  journal = {Scientific Reports},
  volume  = {14},
  number  = {1},
  pages   = {2422},
  year    = {2024},
  doi     = {10.1038/s41598-024-52926-3},
  issn    = {2045-2322},
}

@article{yamazaki2018_practical_appl,
  author       = {Yamazaki, Takeshi and Matsuura, Shunji and Narimani, Ali and Saidmuradov, Anushervon and Zaribafiyan, Arman},
  title        = {Towards the Practical Application of Near-Term Quantum Computers in Quantum Chemistry Simulations: A Problem Decomposition Approach},
  journal      = {arXiv},
  year         = {2018},
  publisher    = {arXiv},
  doi          = {10.48550/arXiv.1806.01305},
  url          = {https://doi.org/10.48550/arXiv.1806.01305}
}

@article{DCStatePrep_2021_Araujo,
  title   = {A Divide-and-Conquer Algorithm for Quantum State Preparation},
  author  = {Araujo, Israel F. and Park, Daniel K. and Petruccione, Francesco and da Silva, Adenilton J.},
  journal = {Scientific Reports},
  volume  = {11},
  number  = {1},
  pages   = {6329},
  year    = {2021},
  doi     = {10.1038/s41598-021-85474-1},
  issn    = {2045-2322},
}

@article{wang2025randomizedQSVT,
  author       = {Wang, Xinzhao and Zhang, Yuxin and Hazra, Soumyabrata and Li, Tongyang and Shao, Changpeng and Chakraborty, Shantanav},
  title        = {Randomized Quantum Singular Value Transformation},
  journal      = {arXiv},
  year         = {2025},
  publisher    = {arXiv},
  doi          = {10.48550/arXiv.2510.06851},
  url          = {https://doi.org/10.48550/arXiv.2510.06851}
}

@article{fujii_2022_deep_vqe,
  title = {Deep Variational Quantum Eigensolver: A Divide-And-Conquer Method for Solving a Larger Problem with Smaller Size Quantum Computers},
  author = {Fujii, Keisuke and Mizuta, Kaoru and Ueda, Hiroshi and Mitarai, Kosuke and Mizukami, Wataru and Nakagawa, Yuya O.},
  journal = {PRX Quantum},
  volume = {3},
  issue = {1},
  pages = {010346},
  numpages = {12},
  year = {2022},
  month = {Mar},
  publisher = {American Physical Society},
  doi = {10.1103/PRXQuantum.3.010346},
  url = {https://link.aps.org/doi/10.1103/PRXQuantum.3.010346}
}

@Article{jinlong_2023_mbe,
author ="Ma, Huan and Liu, Jie and Shang, Honghui and Fan, Yi and Li, Zhenyu and Yang, Jinlong",
title  ="Multiscale quantum algorithms for quantum chemistry",
journal  ="Chem. Sci.",
year  ="2023",
volume  ="14",
issue  ="12",
pages  ="3190-3205",
publisher  ="The Royal Society of Chemistry",
doi  ="10.1039/D2SC06875C",
url  ="http://dx.doi.org/10.1039/D2SC06875C"}

@article{negre2025_new_dmet,
  author       = {Negre, Alicia and Faulstich, Fabian and Kim, Raehyun and Ayral, Thomas and Lin, Lin and Canc{\`e}s, Eric},
  title        = {New perspectives on Density-Matrix Embedding Theory},
  journal      = {arXiv},
  year         = {2025},
  publisher    = {arXiv},
  doi          = {10.48550/arXiv.2503.09881},
  url          = {https://doi.org/10.48550/arXiv.2503.09881}
}

@article{asthana2025quantumkrylovalgorithmusing,
  author       = {Asthana, Ayush},
  title        = {Quantum Krylov algorithm using unitary decomposition for exact eigenstates of fermionic systems using quantum computers},
  journal      = {arXiv},
  year         = {2025},
  publisher    = {arXiv},
  doi          = {10.48550/arXiv.2512.11788},
  url          = {https://doi.org/10.48550/arXiv.2512.11788}
}

@article{shajan2026_ewf_protein,
  author       = {Shajan, Akhil and Kaliakin, Danil and Liang, Fangchun and Pellegrini, Thaddeus and Doga, Hakan and Bhowmik, Subhamoy and Das, Susanta and Mezzacapo, Antonio and Motta, Mario and Merz Jr., Kenneth M.},
  title        = {Molecular Quantum Computations on a Protein},
  journal      = {arXiv},
  year         = {2026},
  publisher    = {arXiv},
  doi          = {10.48550/arXiv.2512.17130},
  url          = {https://doi.org/10.48550/arXiv.2512.17130}
}

@article{kowalski2024resourceadaptivequantumflowalgorithms,
  author       = {Kowalski, Karol and Bauman, Nicholas P.},
  title        = {Resource-adaptive quantum flow algorithms for quantum simulations of many-body systems: sub-flow embedding procedures},
  journal      = {arXiv},
  year         = {2024},
  publisher    = {arXiv},
  doi          = {10.48550/arXiv.2410.11992},
  url          = {https://doi.org/10.48550/arXiv.2410.11992}
}

@article{Ai_2025_DMET_ROHF_Lanthanide,
  author  = {Ai, Yuhang and Li, Ze-Wei and Guan, Zhe-Bin and Jiang, Hong},
  title   = {Density Matrix Embedding Theory-Based Multiconfigurational Quantum Chemistry Approach to Lanthanide Single-Ion Magnets},
  journal = {Journal of Chemical Theory and Computation},
  year    = {2025},
  volume  = {21},
  number  = {19},
  pages   = {9631--9640},
  doi     = {10.1021/acs.jctc.5c01336},
  url     = {https://doi.org/10.1021/acs.jctc.5c01336},
  issn    = {1549-9618},
  publisher = {American Chemical Society}
}

@article{Ai_2022_DMET_ROHF_SIM,
  author    = {Ai, Yuhang and Sun, Qiming and Jiang, Hong},
  title     = {Efficient Multiconfigurational Quantum Chemistry Approach to Single-Ion Magnets Based on Density Matrix Embedding Theory},
  journal   = {The Journal of Physical Chemistry Letters},
  year      = {2022},
  volume    = {13},
  number    = {45},
  pages     = {10627--10634},
  doi       = {10.1021/acs.jpclett.2c02890},
  url       = {https://doi.org/10.1021/acs.jpclett.2c02890},
  publisher = {American Chemical Society}
}

@article{Mitra_2021_DMET_ROHF_Defects,
  author    = {Mitra, Abhishek and Pham, Hung Q. and Pandharkar, Riddhish and Hermes, Matthew R. and Gagliardi, Laura},
  title     = {Excited States of Crystalline Point Defects with Multireference Density Matrix Embedding Theory},
  journal   = {The Journal of Physical Chemistry Letters},
  year      = {2021},
  volume    = {12},
  number    = {48},
  pages     = {11688--11694},
  doi       = {10.1021/acs.jpclett.1c03229},
  url       = {https://doi.org/10.1021/acs.jpclett.1c03229},
  publisher = {American Chemical Society}
}

@article{Patra2026_PIGen,
  author       = {Patra, Chayan and Mondal, Dibyendu and Halder, Sonaldeep and Halder, Dipanjali and Laskar, Mostafizur Rahaman and Goel, Richa and Maitra, Rahul},
  title        = {Physics-Informed Generative Machine Learning for Accelerated Quantum-centric Supercomputing},
  journal      = {arXiv},
  year         = {2026},
  publisher    = {arXiv},
  doi          = {10.48550/arXiv.2512.06858},
  url          = {https://doi.org/10.48550/arXiv.2512.06858}
}

@article{Yoo2026extending_hi_vqe,
  author    = {Yoo, Pilsun and Kim, Kyungmin and Elala, Eyuel E. and McFarthing, Shane and Pellow, Aidan and Fuks, Johanna I. and Kang, Doo Hyung and Nakliang, Pratanphorn and Kim, Jaewan and Pathak, Himadri and Shirakawa, Tomonori and Yunoki, Seiji and Rhee, June-Koo Kevin},
  title     = {Extending the Handover-Iterative VQE to Challenging Strongly Correlated Systems: $N_2$ and Fe-S Cluster},
  journal   = {arXiv},
  year      = {2026},
  publisher = {arXiv},
  doi       = {10.48550/arXiv.2601.06935},
  url       = {https://doi.org/10.48550/arXiv.2601.06935}
}

@article{cances_2025_analysis_dmet,
author = {Cancès, Eric and Faulstich, Fabian M. and Kirsch, Alfred and Letournel, Eloïse and Levitt, Antoine},
title = {Analysis of density matrix embedding theory around the non-interacting limit},
journal = {Communications on Pure and Applied Mathematics},
volume = {78},
number = {8},
pages = {1359-1410},
doi = {https://doi.org/10.1002/cpa.22244},
url = {https://onlinelibrary.wiley.com/doi/abs/10.1002/cpa.22244},
eprint = {https://onlinelibrary.wiley.com/doi/pdf/10.1002/cpa.22244},
year = {2025}
}

@unpublished{Patra2026mlcsg,
  author = {Ashish Kumar Patra and  Anurag, K. S. V. and Rahul Maitra and
            Ruchika Bhat and Sai Shankar, P. and Jaiganesh, G.},
  title = {Machine-Learned Compact Subspace Generation for Quantum Selected Configuration Interaction within Density Matrix Embedding Framework},
  note = {Manuscript in preparation},
  year = {2026}
}

@Article{Bierman_2026_QBE_SQD,
author ="Bierman, Joel and Liu, Yuan",
title  ="Towards utility-scale electronic structure with sample-based quantum bootstrap embedding",
journal  ="Digital Discovery",
year  ="2026",
pages  ="-",
publisher  ="RSC",
doi  ="10.1039/D5DD00416K",
url  ="http://dx.doi.org/10.1039/D5DD00416K"}

@misc{iijima2023accuratequantumchemicalcalculations,
      title={Towards Accurate Quantum Chemical Calculations on Noisy Quantum Computers}, 
      author={Naoki Iijima and Satoshi Imamura and Mikio Morita and Sho Takemori and Akihiko Kasagi and Yuhei Umeda and Eiji Yoshida},
      year={2023},
      eprint={2311.09634},
      archivePrefix={arXiv},
      primaryClass={quant-ph},
      url={https://arxiv.org/abs/2311.09634}, 
}

@article{Sun_2016_quantum_embedding_theories,
   title={Quantum Embedding Theories},
   volume={49},
   ISSN={1520-4898},
   url={http://dx.doi.org/10.1021/acs.accounts.6b00356},
   DOI={10.1021/acs.accounts.6b00356},
   number={12},
   journal={Accounts of Chemical Research},
   publisher={American Chemical Society (ACS)},
   author={Sun, Qiming and Chan, Garnet Kin-Lic},
   year={2016},
   month=nov, pages={2705–2712} 
}

@article{Nusspickel2022,
  title = {Systematic Improvability in Quantum Embedding for Real Materials},
  author = {Nusspickel, Max and Booth, George H.},
  journal = {Phys. Rev. X},
  volume = {12},
  issue = {1},
  pages = {011046},
  numpages = {15},
  year = {2022},
  month = {Mar},
  publisher = {American Physical Society},
  doi = {10.1103/PhysRevX.12.011046},
  url = {https://link.aps.org/doi/10.1103/PhysRevX.12.011046}
}

\newpage

\appendix
\thispagestyle{empty}

\vspace*{\fill}
\begin{center}
    {\Huge\bfseries SUPPLEMENTARY MATERIAL}
\end{center}
\vspace*{\fill}

\section{Geometric Coordinates of the Studied Molecules}\label{sec:coords_supp}
\vspace{-1.5em}

\setcounter{table}{0}
\renewcommand{\thetable}{A\arabic{table}}

\setcounter{figure}{0}
\renewcommand{\thefigure}{A\arabic{figure}}

\setcounter{equation}{0}
\renewcommand{\theequation}{A\arabic{equation}}

\begin{table}[H]
\centering
\begin{tabular}{|p{3.8cm}|c|c|c|c|}
\hline
\multirow{2}{*}{\textbf{Molecule}} & \multicolumn{4}{c|}{\textbf{Geometric Coordinates}} \\ \cline{2-5}
 & \textbf{Atom} & \textbf{x} & \textbf{y} & \textbf{z} \\
\hline

\multirow{4}{*}{\makecell[l]{Cyanic Acid \\ ($HOCN$)}} 
 & H & -1.458587 & -0.272810 & 0.065495 \\
 & O & -0.600578 & 0.099568 & -0.321249 \\
 & C & 0.544407 & 0.090463 & 0.422709 \\
 & N & 1.514757 & 0.082779 & 1.053115 \\
\hline

\multirow{6}{*}{\makecell[l]{Formaldehyde Oxime \\ ($CH_3NO$)}} 
 & C & -0.910484 & -0.020915 & -0.332484 \\
 & H & 2.170747  & -0.116917 & 0.612439 \\
 & H & -1.005526 & 1.043453  & -0.507441 \\
 & H & -1.760983 & -0.667161 & -0.500891 \\
 & N & 0.207852  & -0.523560 & 0.074740 \\
 & O & 1.298394  & 0.285100  & 0.293422 \\
\hline

\multirow{8}{*}{\makecell[l]{Methoxyamine \\ ($CH_5NO$)}} 
 & C & -0.966149 & 0.201296  & 0.149956 \\
 & H & -1.000265 & 0.965450  & -0.658496 \\
 & H & -1.667205 & 0.516700  & 0.949826 \\
 & H & -1.306933 & -0.785003 & -0.237043 \\
 & H & 1.882536  & 0.441132  & -0.409602 \\
 & H & 1.601346  & -1.163838 & -0.023188 \\
 & N & 1.132081  & -0.281427 & -0.328214 \\
 & O & 0.324588  & 0.105691  & 0.692069 \\
\hline

\multirow{7}{*}{\makecell[l]{Methyl Isocyanate \\ ($C_2H_3NO$)}} 
 & C & -0.872655 & -0.048579 & -0.012732 \\
 & C & 1.498758  & 0.200579  & -0.121172 \\
 & H & -0.658520 & -0.731912 & 0.839315 \\
 & H & -1.419752 & -0.612203 & -0.796487 \\
 & H & -1.514077 & 0.782782  & 0.345983 \\
 & N & 0.351223  & 0.500017  & -0.581535 \\
 & O & 2.615024  & -0.090684 & 0.326629 \\
\hline

\multirow{9}{*}{\makecell[l]{Acetaldehyde Oxime \\ ($C_2H_5NO$)}} 
 & C & -1.321954 & 0.079847  & -0.059915 \\
 & C & 0.125140  & 0.194559  & 0.278307 \\
 & H & -1.848828 & -0.492848 & 0.731931 \\
 & H & -1.772167 & 1.091808  & -0.134570 \\
 & H & -1.440397 & -0.441345 & -1.033016 \\
 & H & 0.567039  & 1.175157  & 0.409184 \\
 & H & 2.659033  & 0.076802  & 0.861245 \\
 & N & 0.845312  & -0.873562 & 0.413794 \\
 & O & 2.186822  & -0.810420 & 0.726078 \\
\hline

\multirow{8}{*}{\makecell[l]{Carbamide / Urea \\ ($CH_4N_2O$)}} 
 & C & 0.023609  & 0.174787  & 0.475047 \\
 & H & 2.152264  & -0.041841 & 0.244225 \\
 & H & 1.187776  & -0.518005 & -1.199231 \\
 & H & -2.122930 & 0.259006  & 0.346006 \\
 & H & -1.304030 & -0.342648 & -1.139910 \\
 & N & -1.243207 & 0.016569  & -0.161562 \\
 & N & 1.224229  & -0.157069 & -0.220304 \\
 & O & 0.082288  & 0.609201  & 1.655728 \\
\hline

\multirow{3}{*}{\makecell[l]{Nitrosyl Chloride \\ ($NOCl$)}} 
 & N  & -0.222631 & 0.458624  & 0.000000 \\
 & O  & -1.157638 & -0.281711 & 0.000000 \\
 & Cl & 1.380269  & -0.176914 & 0.000000 \\
\hline

\multirow{5}{*}{\makecell[l]{Hydroxythiocyanate \\ ($HOSCN$)}} 
 & H & -1.576204 & 0.567554  & -0.304395 \\
 & O & -1.272167 & 0.443194  & 0.630206 \\
 & S & -0.421261 & -1.048230 & 0.616500 \\
 & C & 1.126988  & -0.244750 & 0.353656 \\
 & N & 2.142645  & 0.282231  & 0.181157 \\
\hline

\end{tabular}
\caption{Geometric co-ordinates of the molecules used for simulation(geometry-optimized upto force-field level)~\cite{RDKit_2025_09_1}.}
\end{table}

\section{Energy Results} \label{sec: tab:nat_mols_energy}

\setcounter{table}{0}
\renewcommand{\thetable}{B\arabic{table}}

\setcounter{figure}{0}
\renewcommand{\thefigure}{B\arabic{figure}}

\setcounter{equation}{0}
\renewcommand{\theequation}{B\arabic{equation}}

\begin{table}[H]
\centering
\begin{tabular}{|p{4.2cm}|p{3.8cm}|p{3.8cm}|}
\hline
\textbf{Molecule} & \textbf{$E_{\mathrm{DMET\text{-}SQD}}$ (Ha)} 
& \textbf{$E_{\mathrm{DMET\text{-}FCI}}$ (Ha)}\\
\hline

\makecell[l]{Cyanic Acid \\ ($HOCN$)} 
& -165.80748733 
& -165.80748835  \\
\hline

\makecell[l]{Formaldehyde Oxime \\ ($CH_3NO$)} 
& -166.89443901 
& -166.89443875  \\
\hline

\makecell[l]{Methoxyamine \\ ($CH_5NO$)} 
& -168.04584566 
& -168.04584180  \\
\hline

\makecell[l]{Methyl Isocyanate \\ ($C_2H_3NO$)} 
& -204.57166716 
& -204.57166689 \\
\hline

\makecell[l]{Acetaldehyde Oxime \\ ($C_2H_5NO$)} 
& -205.56816508 
& -205.56816496 \\
\hline

\makecell[l]{Carbamide / Urea \\ ($CH_4N_2O$)} 
& -221.44130237 
& -221.44130142 \\
\hline

\makecell[l]{Nitrosyl Chloride \\ ($NOCl$)} 
& -582.38458086 
& -582.38458236 \\
\hline

\makecell[l]{Hydroxythiocyanate \\ ($HOSCN$)} 
& -559.05055857 
& -559.05056231 \\
\hline

\end{tabular}
\caption{Energy results obtained from the DMET-SQD simulations for the studied molecules along with the energy reference $E_{\mathrm{DMET\text{-}FCI}}$ in Hartree (Ha).}
\label{tab:nat_mols_energy}
\end{table}


\section{Quantum Computing Hardware Calibration Data - IBM Sherbrooke} \label{sec: sherbrooke_calib}

\setcounter{table}{0}
\renewcommand{\thetable}{C\arabic{table}}

\setcounter{figure}{0}
\renewcommand{\thefigure}{C\arabic{figure}}

\setcounter{equation}{0}
\renewcommand{\theequation}{C\arabic{equation}}

\begin{table}[H] 
\centering
\begin{tabular}{|c|c|}
\hline
\textbf{Parameter} & \textbf{Value} \\
\hline
T1 & 270.447138 $\mu s$\\
\hline
T2 & 211.026521 $\mu s$\\
\hline
Frequency & 4.794027 GHz\\
\hline
Anharmonicity & -0.310939 GHz\\
\hline
Readout assignment error & 0.021240 \\
\hline
P(0 $\mid$ 1) & 0.020507 \\
\hline
P(1 $\mid$ 0) & 0.019042 \\
\hline
Readout length & 1216.0 ns\\
\hline
Identity gate error & 0.000244 \\
\hline
$R_z$ gate error & 0.0 \\
\hline
$\sqrt{X}$ ($S_x$) gate error & 0.000244 \\
\hline
Pauli-X gate error & 0.000244 \\
\hline
ECR gate error & 0.006767 \\
\hline
Gate time & 533.333333 ns \\
\hline
\end{tabular}
\caption{Calibration metrics for the Eagle R3 processor (IBM Sherbrooke, containing 127 qubits) recorded on 4 June 2025 at 01:48:22 EST. These values correspond to the device’s pre-execution calibration snapshot, including coherence times ($T_1$, $T_2$), qubit frequency, anharmonicity, readout assignment errors, gate error rates, and gate durations~\cite{superconducting_quantum_engineer}. Please note that the quantum hardware is usually calibrated every 24 hours.}
\label{tab:sherbrooke_calib}
\end{table}

\section{Quantum Computing Hardware Calibration Data - IBM Boston} \label{sec: boston_calib}

\setcounter{table}{0}
\renewcommand{\thetable}{D\arabic{table}}

\setcounter{figure}{0}
\renewcommand{\thefigure}{D\arabic{figure}}

\setcounter{equation}{0}
\renewcommand{\theequation}{D\arabic{equation}}

\begin{table}[H] 
\centering
\begin{tabular}{|c|c|}
\hline
\textbf{Parameter} & \textbf{Value} \\
\hline
T1 & 272.208074 $\mu s$\\
\hline
T2 & 326.176439 $\mu s$\\
\hline
Readout assignment error & 0.003418 \\
\hline
P(0 $\mid$ 1) & 0.005615 \\
\hline
P(1 $\mid$ 0) & 0.001465 \\
\hline
Readout length & 2180.0 ns\\
\hline
Identity gate error & 0.000157 \\
\hline
Identity gate time & 32.0 ns \\
\hline
$R_z$ gate error & 0.0 \\
\hline
$\sqrt{X}$ ($S_x$) gate error & 0.000157 \\
\hline
$\sqrt{X}$ ($S_x$) gate time & 32.0 ns \\
\hline
Pauli-X gate error & 0.000157 \\
\hline
Pauli-X gate time & 32.0 ns \\
\hline
CZ gate error & 0.001180 \\
\hline
CZ gate time & 68.0 ns \\
\hline
\end{tabular}
\caption{Calibration metrics for the Heron R3 processor (\texttt{ibm\_boston})~\cite{ibmquantum_heron_boston} recorded on 13 June 2026 at 13:18:06 IST. These values correspond to the device's pre-execution calibration snapshot, including coherence times ($T_1$, $T_2$), readout assignment errors, gate error rates, and gate durations. Median values are reported across all qubits and gate instances on the device. Please note that the quantum hardware is usually calibrated every 24 hours.}
\label{tab:boston_calib}
\end{table}

\newpage

\section{Fragment orbital space decomposition for DMET-SQD simulations.} \label{sec: orb_lists}

\setcounter{table}{0}
\renewcommand{\thetable}{E\arabic{table}}

\setcounter{figure}{0}
\renewcommand{\thefigure}{E\arabic{figure}}

\setcounter{equation}{0}
\renewcommand{\theequation}{E\arabic{equation}}

\begin{table*}[!ht]
\centering
\small
\begin{tabular}{llccccccc}
\hline\hline
Molecule & Fragment & $(o,\,e)$ & $N_{\mathrm{orb}}$ & $|A_y|$ & $|B_y|$ & $|\mathrm{Cor}_y|$ & $|\mathrm{Vir}_y|$ & $|\mathrm{Env}_y|$ \\
\hline

\multirow{4}{*}{HCNO}
 & {[H]} & $(2,\;2)$   & \multirow{4}{*}{16} & 1 & 1 & 10 & 4 & 15 \\
 & {[C]} & $(10,\;10)$ &                     & 5 & 5 &  6 & 0 & 11 \\
 & {[N]} & $(10,\;10)$ &                     & 5 & 5 &  6 & 0 & 11 \\
 & {[O]} & $(10,\;10)$ &                     & 5 & 5 &  6 & 0 & 11 \\
\hline

\multirow{6}{*}{CH$_3$NO}
 & {[C]} & $(10,\;10)$ & \multirow{6}{*}{18} & 5 & 5 &  7 &  1 & 13 \\
 & {[H]} & $(2,\;2)$   &                     & 1 & 1 & 11 &  5 & 17 \\
 & {[H]} & $(2,\;2)$   &                     & 1 & 1 & 11 &  5 & 17 \\
 & {[H]} & $(2,\;2)$   &                     & 1 & 1 & 11 &  5 & 17 \\
 & {[N]} & $(10,\;10)$ &                     & 5 & 5 &  7 &  1 & 13 \\
 & {[O]} & $(10,\;10)$ &                     & 5 & 5 &  7 &  1 & 13 \\
\hline

\multirow{8}{*}{CH$_5$NO}
 & {[C]} & $(10,\;10)$ & \multirow{8}{*}{20} & 5 & 5 &  8 &  2 & 15 \\
 & {[H]} & $(2,\;2)$   &                     & 1 & 1 & 12 &  6 & 19 \\
 & {[H]} & $(2,\;2)$   &                     & 1 & 1 & 12 &  6 & 19 \\
 & {[H]} & $(2,\;2)$   &                     & 1 & 1 & 12 &  6 & 19 \\
 & {[H]} & $(2,\;2)$   &                     & 1 & 1 & 12 &  6 & 19 \\
 & {[H]} & $(2,\;2)$   &                     & 1 & 1 & 12 &  6 & 19 \\
 & {[O]} & $(10,\;10)$ &                     & 5 & 5 &  8 &  2 & 15 \\
 & {[N]} & $(10,\;10)$ &                     & 5 & 5 &  8 &  2 & 15 \\
\hline

\multirow{7}{*}{C$_2$H$_3$NO}
 & {[C]} & $(10,\;10)$ & \multirow{7}{*}{23} & 5 & 5 & 10 &  3 & 18 \\
 & {[C]} & $(10,\;10)$ &                     & 5 & 5 & 10 &  3 & 18 \\
 & {[H]} & $(2,\;2)$   &                     & 1 & 1 & 14 &  7 & 22 \\
 & {[H]} & $(2,\;2)$   &                     & 1 & 1 & 14 &  7 & 22 \\
 & {[H]} & $(2,\;2)$   &                     & 1 & 1 & 14 &  7 & 22 \\
 & {[N]} & $(10,\;10)$ &                     & 5 & 5 & 10 &  3 & 18 \\
 & {[O]} & $(10,\;10)$ &                     & 5 & 5 & 10 &  3 & 18 \\
\hline

\multirow{9}{*}{C$_2$H$_5$NO}
 & {[C]} & $(10,\;10)$ & \multirow{9}{*}{25} & 5 & 5 & 11 &  4 & 20 \\
 & {[C]} & $(10,\;10)$ &                     & 5 & 5 & 11 &  4 & 20 \\
 & {[H]} & $(2,\;2)$   &                     & 1 & 1 & 15 &  8 & 24 \\
 & {[H]} & $(2,\;2)$   &                     & 1 & 1 & 15 &  8 & 24 \\
 & {[H]} & $(2,\;2)$   &                     & 1 & 1 & 15 &  8 & 24 \\
 & {[H]} & $(2,\;2)$   &                     & 1 & 1 & 15 &  8 & 24 \\
 & {[H]} & $(2,\;2)$   &                     & 1 & 1 & 15 &  8 & 24 \\
 & {[N]} & $(10,\;10)$ &                     & 5 & 5 & 11 &  4 & 20 \\
 & {[O]} & $(10,\;10)$ &                     & 5 & 5 & 11 &  4 & 20 \\
\hline

\multirow{8}{*}{CH$_4$N$_2$O}
 & {[C]} & $(10,\;10)$ & \multirow{8}{*}{24} & 5 & 5 & 11 &  3 & 19 \\
 & {[N]} & $(10,\;10)$ &                     & 5 & 5 & 11 &  3 & 19 \\
 & {[H]} & $(2,\;2)$   &                     & 1 & 1 & 15 &  7 & 23 \\
 & {[H]} & $(2,\;2)$   &                     & 1 & 1 & 15 &  7 & 23 \\
 & {[H]} & $(2,\;2)$   &                     & 1 & 1 & 15 &  7 & 23 \\
 & {[H]} & $(2,\;2)$   &                     & 1 & 1 & 15 &  7 & 23 \\
 & {[N]} & $(10,\;10)$ &                     & 5 & 5 & 11 &  3 & 19 \\
 & {[O]} & $(10,\;10)$ &                     & 5 & 5 & 11 &  3 & 19 \\
\hline

\multirow{3}{*}{NOCl}
 & {[N]}  & $(8,\;10)$  & \multirow{3}{*}{19} & 5 & 3 & 11 & 0 & 14 \\
 & {[O]}  & $(8,\;10)$  &                     & 5 & 3 & 11 & 0 & 14 \\
 & {[Cl]} & $(12,\;18)$ &                     & 9 & 3 &  7 & 0 & 10 \\
\hline

\multirow{5}{*}{HOSCN}
 & {[H]} & $(2,\;2)$   & \multirow{5}{*}{25} & 1 &  1 & 18 &  5 & 24 \\
 & {[O]} & $(10,\;10)$ &                     & 5 &  5 & 14 &  1 & 20 \\
 & {[S]} & $(15,\;18)$ &                     & 9 &  6 & 10 &  0 & 16 \\
 & {[C]} & $(10,\;10)$ &                     & 5 &  5 & 14 &  1 & 20 \\
 & {[N]} & $(10,\;10)$ &                     & 5 &  5 & 14 &  1 & 20 \\
\hline\hline
\end{tabular}
\caption{%
    For each fragment: impurity active space $(o,\,e)$ (spatial orbitals and electrons); $|A_y|$, fragment orbitals; $|B_y|$, bath orbitals ($|B_y|\leq|A_y|$); $|\mathrm{Cor}_y|$, fully occupied environment orbitals; $|\mathrm{Vir}_y|$, empty environment orbitals; $|\mathrm{Env}_y|=|B_y|+|\mathrm{Cor}_y|+|\mathrm{Vir}_y|$, total environment orbitals at $\varepsilon_{occ} = 10^{-13}$.
}
\label{tab:orbital_spaces}
\end{table*}

\section{Classical Resources for HOSCN molecule for varying $\varepsilon_{occ}$} \label{sec: classical_resources_HOSCN}

\setcounter{table}{0}
\renewcommand{\thetable}{F\arabic{table}}

\setcounter{figure}{0}
\renewcommand{\thefigure}{F\arabic{figure}}

\setcounter{equation}{0}
\renewcommand{\theequation}{F\arabic{equation}}

\begin{figure}[H]
    \centering
    \includegraphics[width=\linewidth]{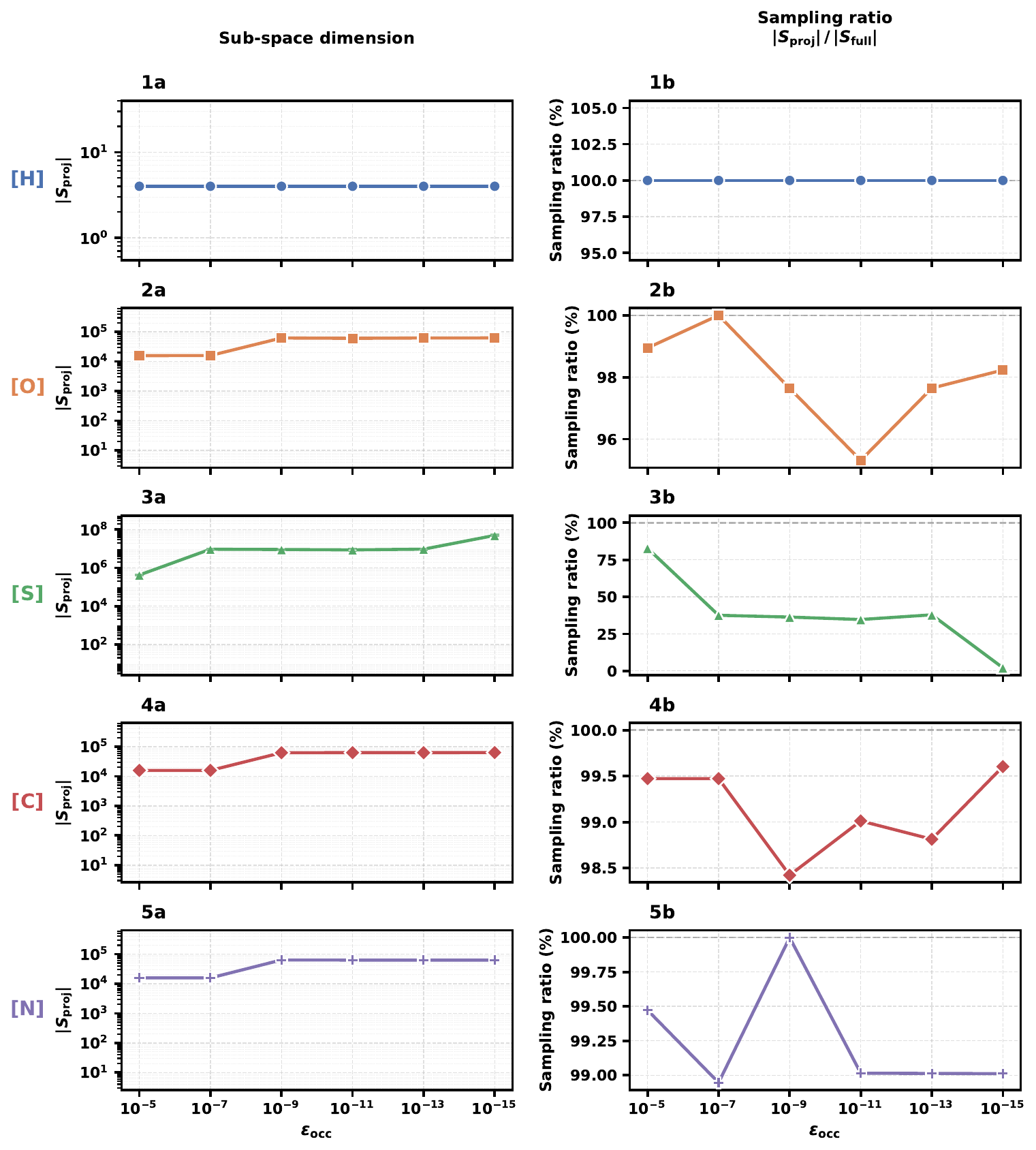}
    \caption{Classical S-CoRe resources for HOSCN/STO-3G as a function of $\varepsilon_{\mathrm{occ}}$ : Sub-space dimension $|S_{\mathrm{proj}}|$~(a), and sampling ratio $|S_{\mathrm{proj}}|/|S_{\mathrm{full}}|$~(b) for each DMET fragment of HOSCN across different $\epsilon_{occ}$, executed on IBM Heron R3 Boston.}
    \label{fig:hoscn_classical_resources}
\end{figure}
\end{document}